\documentclass[a4paper,11pt]{article}
\usepackage{iftex}
\ifPDFTeX\pdfoutput=1\fi 
\usepackage{jheppub} 
\usepackage[T1]{fontenc} 
\usepackage{mathrsfs} 
\usepackage{shuffle}
\usepackage{musicography}
\usepackage{empheq}
\usepackage{mathrsfs}
\usepackage{bm,bbm}
\usepackage{verbatim}
 \usepackage{makecell}
 \usepackage{float} 
 \usepackage[makeroom]{cancel}

\usepackage{amsmath, amssymb, amsthm}
\usepackage{mathrsfs}
\let\mathscrRsfs\mathscr
\usepackage{mathtools}  
\usepackage{dsfont}
\usepackage[mathscr]{eucal}
 
\usepackage{tikz-cd}

\usepackage{musicography}
\usepackage{empheq}

\usepackage{xcolor}
\usepackage{slashed}
\usepackage{float}
 
\usepackage[
]{hyperref}

\DeclareRobustCommand{\frl}[1]{\url{#1}\hspace{0.5em} {\bf\color{orange}   \rule{0.5em}{.81em}  } \hspace{0.5em}}

\theoremstyle{definition}
\newtheorem{theorem}{Theorem}[section]  
\newtheorem{lemma}[theorem]{Lemma}       
\newtheorem{proposition}[theorem]{Proposition}

\newtheorem{conjecture}[theorem]{Conjecture}

\theoremstyle{definition}
\newtheorem{definition}{Definition}[section]   
\newtheorem{example}[definition]{Example}

\theoremstyle{remark}
\newtheorem{remark}{Remark}[section]

\newcommand{\bedef}{\begin{definition}}
\newcommand{\eedef}{\end{definition}}

\newcommand{\bethm}{\begin{theorem}}
\newcommand{\eethm}{\end{theorem}}

\newcommand{\bermk}{\begin{remark}}
\newcommand{\eermk}{\end{remark}}

\newcommand{\belma}{\begin{lemma}}
\newcommand{\eelma}{\end{lemma}}

\newcommand{\beconj}{\begin{conjecture}}
\newcommand{\eeconj}{\end{conjecture}}

\newcommand{\beprop}{\begin{proposition}}
\newcommand{\eeprop}{\end{proposition}}

\newcommand{\becor}{\begin{Corollary}}
\newcommand{\eecor}{\end{Corollary}}

\newcommand{\beex}{\begin{example}}
\newcommand{\eeex}{\end{example}}

       \newcommand{\ceff}{c^\text{eff}}
        
       \newcommand{\prin}{ \text{prin}}

\usepackage{xparse}

\newcommand{\Def}[1][]{%
  \par\medskip\noindent \textbf{Definition%
  \if\relax\detokenize{#1}\relax
  \else\ (\textbf{#1})%
  \fi. } \ignorespaces
}
\newcommand{\Thm}[1][]{%
  \par\medskip\noindent \textbf{Theorem%
  \if\relax\detokenize{#1}\relax
  \else\ (\textbf{#1})%
  \fi. } \ignorespaces
}
\newcommand{\Lem}[1][]{%
  \par\noindent\textbf{Lemma%
  \if\relax\detokenize{#1}\relax
  \else\ (\textbf{#1})%
  \fi. }\ignorespaces
}
\newcommand{\Prop}[1][]{%
  \par\noindent\textbf{Proposition%
  \if\relax\detokenize{#1}\relax
  \else\ (\textbf{#1})%
  \fi. }\ignorespaces
}

\newcommand{\ket}[1]{\ensuremath{\left|#1\right\rangle}}

\newcommand{\be}{\begin{equation}}
\newcommand{\ee}{\end{equation}}
\newcommand{\bpm}{\begin{pmatrix}}
\newcommand{\epm}{\end{pmatrix}}

\newcommand{\beqn}{\begin{eqnarray}}
\newcommand{\eeqn}{\end{eqnarray}}

 \newcommand{\C}{\mathbb C}
\newcommand{\Z}{\mathbb Z}

\newcommand{\id}{\mathrm{id}}

\usepackage{pgffor}
\foreach \x in {A,...,Z}{
  \expandafter\xdef\csname b\x\endcsname{\noexpand\mathbb{\x}}
}
\foreach \x in {A,...,Z}{
  \expandafter\xdef\csname c\x\endcsname{\noexpand\mathcal{\x}}
}
\foreach \x in {A,...,Z}{
  \expandafter\xdef\csname sc\x\endcsname{\noexpand\mathscr{\x}}
}
\foreach \x in {A,...,Z}{
  \expandafter\xdef\csname s\x\endcsname{\noexpand\mathscr{\x}}
}
\foreach \x in {A,...,Z}{
  \expandafter\xdef\csname sf\x\endcsname{\noexpand\mathsf{\x}}
}
\foreach \x in {a,...,z}{
  \expandafter\xdef\csname sf\x\endcsname{\noexpand\mathsf{\x}}
}
\foreach \x in {a,...,z}{
  \expandafter\xdef\csname ds\x\endcsname{\noexpand\mathds{\x}}
}
\foreach \x in {A,...,Z}{
  \expandafter\xdef\csname  fk\x\endcsname{\noexpand\frak{\x}}
}
\foreach \x in {a,...,z}{
  \expandafter\xdef\csname  fk\x\endcsname{\noexpand\frak{\x}}
}
\foreach \x in {A,...,Z}{
  \expandafter\xdef\csname ss\x\endcsname{\noexpand\mathscrRsfs{\x}}
}

\usepackage{etoolbox}

\newcommand{\DeclareOperator}[1]{%
  \expandafter\DeclareMathOperator\csname #1\endcsname{#1}%
}

\newcommand{\DeclareText}[1]{%
  \expandafter\providecommand\csname #1\endcsname{\ensuremath{\mathrm{#1}}}%
}

\newcommand{\DeclareBoldface}[1]{%
  \expandafter\providecommand\csname #1\endcsname{\ensuremath{\mathbf{#1}}}%
}

\newcommand{\DeclareSerif}[1]{%
  \expandafter\providecommand\csname #1\endcsname{\ensuremath{\mathsf{#1}}}%
}

\forcsvlist{\DeclareOperator}
{gr, Sym, Fun, End, Aut, Inn, Out, Pic, Tor, Bock, Ext, Hom, Spec, FPdim, Dim, Ker, Cok, PE, PL, rk,Diag,diag, arccosh, sgn, Real, Res, Imag, Tr, tr, lcm, Span, GL, SL, PGL, PSL,   Sq, Arf, ABK, HH, HC, PV, Map
}

\forcsvlist{\DeclareText}{Strip,Tube,Dome,Tub,Dom,id,Id,ev,coev, 
 Spin,Pin,HB,CB,HS,SNF,AdS,Irr, OS,
Ch,CFT,QFT,YM,Ver,SU,SO,Sp,Vol,Vect,sVec,sVect,Reps,Grp,Bord,op,loc 
 }

\forcsvlist{\DeclareBoldface}{Mod,Bimod,Alg}
\forcsvlist{\DeclareSerif}{Cats}

\author[a,b ]{Hongliang Jiang} 
\affiliation[a]{Center for Mathematics and Interdisciplinary Sciences, Fudan University, Shanghai 200433, China }
\affiliation[b]{Shanghai Institute for Mathematics and Interdisciplinary Sciences (SIMIS), Shanghai 200433, China} 

\emailAdd{jianghongliang@fudan.edu.cn }

\title{\Large Infinite    Index Identities for Argyres--Douglas Theories and  
Graded Vector Space Isomorphisms for VOAs }

\abstract{Supersymmetric indices provide stringent tests of dualities between superconformal field theories (SCFTs). We establish infinite families of Schur-index identities among four-dimensional $\mathcal N=2$ Argyres--Douglas (AD) theories, with the logarithms of their fugacities related linearly. These identities connect theories with markedly different physical properties without requiring infrared duality,   while still implying simple relations among their central charges and flavor levels. Through the SCFT/VOA correspondence, they become character identities for the associated affine and $W$-algebras. One broad family relates $D_p(G)$ and $(A_{(h+1)(p-1)},G)$ theories, corresponding to affine and principal $W$-algebras at boundary admissible levels. At the level of the underlying state spaces, these character identities suggest that the corresponding VOAs are related by graded vector space isomorphisms, which are weaker than full VOA isomorphisms. We explicitly construct such a graded isomorphism between the VOAs associated with $(A_1,A_{4p-3})$ and $(A_1,D_{2p})$. We also show the correspondence for universal affine and principal $W$-algebras and test its compatibility with their simple quotients by matching the lowest-degree null states. We establish analogous character identities for certain non-vacuum modules in concrete examples and conjecture a broader extension. Finally, beyond the SCFT/VOA correspondence, we discover similar identities between level-one affine and Virasoro algebras, whose further generalizations point toward a new  {spectral level-rank  duality}  in WZW models.
}

\begin{document} 
\maketitle
\raggedbottom

\section{Introduction and summary}

Superconformal indices    play a central role in the study of supersymmetric quantum field theories. They encode infinitely many protected quantities and provide powerful tools for studying dualities, renormalization-group flows, moduli spaces, and, through the AdS/CFT correspondence, black-hole entropy. In particular, they furnish stringent tests of infrared (IR) dualities: distinct ultraviolet descriptions that flow to the same IR fixed point are expected to yield identical superconformal indices. Such agreement provides strong evidence for a proposed duality by matching an infinite collection of protected quantities. When the indices admit contour-integral or infinite sum representations, these equalities also yield highly nontrivial   identities of independent mathematical interest.

In this paper, we investigate a weaker notion of index matching: indices of theories that need not be IR dual can coincide after simple but nontrivial redefinitions and specializations of fugacities. Focusing on the Schur indices of four-dimensional $\mathcal{N}=2$ superconformal field theories (SCFTs), we systematically construct and prove infinitely many such identities. Through the SCFT/VOA correspondence, these results translate into identities among specialized   characters of vertex operator algebras (VOAs).

Our primary examples are Argyres--Douglas (AD) theories with class $\mathcal{S}$ realizations involving an irregular puncture and, in some cases, an additional regular puncture \cite{Xie:2012hs}. In suitable cases, the associated VOAs are affine vertex algebras or $W$-algebras at admissible levels \cite{Song:2017oew}. These VOAs have a well-developed mathematical structure, including systematic constructions via Drinfeld--Sokolov (DS) reduction and uniform formulas for their characters \cite{Bouwknegt:1992wg,FrenkelKacWakimoto1992,KacWakimotoBoundary2017}. They therefore provide a broad, mathematically tractable setting in which to identify and establish such relations.

A precursor to our construction is the identity $\mathcal I_{(A_1,A_{6n})}(q)=\mathcal I_{(A_1,D_{2n+1})}(q^3;q^{1/2})$ for $n\ge 1$, established in \cite{Adamovic:2025efe}. The associated VOAs are Virasoro and affine vertex algebras, respectively. This identity was itself motivated by a relation between the indices of AD theories and $\cN=4$ super-Yang--Mills theories \cite{Buican:2020moo,Jiang:2024baj}. Our construction extends these results to a broad range of SCFTs and VOAs.

For illustration, we present a few simple, representative examples:
\beqn\nonumber
\mathcal I_{(A_{N-1},A_{(p-1)(N+1)})}(q)
&=&
\mathcal I_{D_p(SU(N))}
\left(q^{N+1};\{q^{s-(N+1)/2}\}_{s=1}^N\right),
\qquad \gcd(p,N)=1,
\\ \nonumber
\mathcal I_{(A_1,A_{4p-3})}(q;w)
&=&
\mathcal I_{(A_1,D_{2p})}
\left(q^2;y=q^{1/2},w\right),
\quad p\ge2,
\\ \nonumber
\cI_{(A_2,A_4)}(q)
&=&
\cI_{(A_1,A_5)}(q^2;1).
\eeqn
These examples illustrate different features of our construction. The first generalizes the identity of \cite{Adamovic:2025efe} to higher rank, relating affine vertex algebras to principal $W$-algebras. The second retains nontrivial flavor dependence on both sides, whereas the third relates unflavored indices. Combining the second and third examples further shows that these identities can be composed into chains connecting more than two SCFTs.

The index identities also imply relations among the anomaly coefficients of the SCFTs, including their $a$ and $c$ central charges and flavor levels. These relations follow by comparing the high-temperature asymptotics of the indices. In particular, the effective central charges obey a universal rescaling relation, with the rescaling fixed by the power of $q$ appearing in the index identity.

These exact index identities also suggest structural connections between the underlying VOAs. More precisely, they imply graded vector space isomorphisms (GVISOs),   with gradings defined by suitable combinations of conformal weights and flavor charges.~\footnote{More generally, one may also need to include the grading by fermion parity in the presence of fermionic operators.
Due to boson-fermion cancellation, the index identity itself may not lead to a GVISO. 
 }
This was first studied in \cite{Buican:2020moo}. Such isomorphisms are weaker than VOA isomorphisms, which must also preserve operator product expansions (OPEs) and normally ordered products. We seek a direct realization of these GVISOs in terms of VOA states and their relations. For a particular pair of VOAs, we explicitly construct the map and prove that it defines a GVISO. We also examine the obstacles to a general construction. For the universal affine and principal $W$-algebras considered here, the correspondence follows from a one-to-one matching of modes. Passing to the simple quotients, however, requires compatibility with null relations, making the construction substantially more subtle. Nevertheless, we explicitly demonstrate the matching of the lowest-degree null states on the two sides.

We further extend our discussion to non-vacuum VOA modules and their interpretation in terms of surface-defect insertions in the SCFTs. Analogous character and index identities also arise in this setting. The problem is more intricate than for vacuum modules: VOA modules generally have complicated structures, simple character formulas are not always available, and the numbers of modules on the two sides can differ substantially. In an example involving three AD theories, we analyze these issues in detail and establish character identities relating selected modules of the corresponding VOAs. We conjecture a broader extension to suitable non-vacuum modules and outline its form for affine and principal $W$-algebras at boundary admissible levels. In this setting, we express the module characters in plethystic exponential (PE) form, obtaining formulas useful for studying SCFTs with surface defects. A detailed treatment of the general extension, including its proof, is left to future work.

Finally, we extend the discussion beyond the SCFT/VOA correspondence. We establish a GVISO between a level-one affine VOA and the Ising VOA, together with explicit character identities that extend to non-vacuum modules. A free-field realization in terms of Majorana fermions provides a direct explanation of this correspondence. Generalizing further, this example   leads us to character relations between pairs of affine VOAs whose ranks and levels are exchanged. We refer to these relations as \emph{spectral level-rank dualities}, emphasizing the correspondence between graded state spaces under an exchange of rank and level. This terminology distinguishes them from the conventional topological level-rank dualities in  Wess--Zumino--Witten (WZW)  models, or Chern--Simons (CS)   theories  via the CS/WZW correspondence.

The paper is organized as follows. Section \ref{scftvoa} reviews the SCFT/VOA correspondence and the high-temperature asymptotics of Schur indices, and explains how index identities constrain anomaly coefficients.
 Section \ref{ANDNSFs} presents the basic family of index identities relating ${(A_{N-1},A_{(p-1)(N+1)})}$ and ${D_p(SU(N))}$.
 Section \ref{AGDpG} further extends this construction to identities relating ${(A_{(p-1)(h+1)},G)}$ and ${D_p(G)}$.
 Section \ref{gwsecgeneral} develops the general construction and establishes further families of index identities for a  broader class of AD theories.
 Section \ref{GVISOsec} investigates explicit GVISOs between the underlying VOAs, going beyond character matching.
 Section \ref{charmodules} explores extensions to non-vacuum modules.
 Section \ref{IsingnoSCFT} extends the analysis beyond the SCFT/VOA correspondence.
 Section \ref{conssec} concludes with a discussion of open questions. The appendices collect further technical details.

 \section{SCFT/VOA correspondence and index identity }\label{scftvoa}
 
 \subsection{Review of the SCFT/VOA correspondence     }
 
 Every 4d $\cN=2$ SCFT can be associated with a vertex operator  algebra or chiral algebra~\cite{Beem:2013sza}  
 \be
 \bV: \text{4d $\cN=2$ SCFT} \to \text{ VOA}.
 \ee
 The SCFT/VOA correspondence captures the Schur sector of the SCFTs. 
 
 Two  important  entries in the SCFT/VOA dictionary are 
     \be
c_{2d}=-12c_{4d}, \qquad k_{2d}=-\frac12k_{4d}.
 \ee

  The Schur index is defined as
  \be
\cI=\Tr(-1)^F q^{E-R}  x^{J}, \qquad q=e^{2\pi i \tau}, \qquad x=e^{2\pi i z},
\ee 
where the trace is taken over all Schur operators, $J$  is the flavor symmetry generator and    $x$ is its fugacity. A generalization to multiple flavor fugacities is straightforward.

 In particular, the SCFT/VOA correspondence implies that  the Schur index
 of  a 4d $\cN=2$ SCFT $\cT$ is identified with the vacuum (super)character of the associated VOA~\cite{Beem:2013sza}  
 \be
 \cI_\cT=\chi_{\bV[\cT]}.
 \ee

 In this paper,    $\chi$   denotes the normalized character, and $\widehat \chi=q^{-c/24}\chi$   denotes the full character.   The former has the nice feature that it starts with 1 for the unique vacuum and coincides with the index, while the latter transforms  covariantly under modular transformations.  

\subsection{Cardy limit of the Schur index}

In this subsection, we would like to make full use of the modular property in order to study  the high temperature or Cardy limit of the Schur index. See \cite{Beem:2017ooy,ArabiArdehali:2023bpq} for previous discussions.

 The VOAs associated with 4d SCFTs are expected to be quasi-lisse in general. For a quasi-lisse VOA, its full vacuum character satisfies an MLDE \cite{Arakawa:2016hkg,Beem:2017ooy}.  The solution space of MLDE gives a vector valued modular form, which includes the   vacuum   and  non-vacuum module characters. Note that the modular vector  may also need logarithmic or pseudotrace solutions in addition to ordinary module characters. 

 As a result,  under the $S$ transformation, the character vectors formed by the solutions of MLDEs  transform  as
 \footnote{For half-integer-graded VOAs obeying $\Gamma^0(2)$-MLDEs, one needs to enlarge the character vectors by including the solutions to the $S$-transformed MLDEs.}
  \be\label{Schis}
  \widehat {\bm\chi} (-1/\tau )
= 
S \widehat {\bm\chi}(\tau ).
  \ee
 
Taking into account possible logarithmic solutions, each character   can be chosen to admit  an expansion of the  following     form    
\footnote{More generally, one may also generalize and replace $A_i (-\log q/(2\pi))^{d_i}$ with a polynomial  of degree $d_i$ such that $P_i(x) =A_i \; x^{d_i}+\cdots$.   }
 \be\label{chiiS}
\widehat\chi_i(\tau)=q^{h_i-c/24} A_i (-\log q/(2\pi))^{d_i}\bigl[1+O(q^\epsilon)\bigr],\qquad d_i\in\mathbb N, \qquad A_i \neq 0, \qquad\epsilon>0.
\ee

  Taking the limit  $\tau\to   +\infty i  $,  we learn from the $S$ transformation \eqref{Schis}  that
  \be\label{chi0hat}
  \widehat\chi_0(-1/\tau)=\sum_i S_{0i}e^{2\pi i\tau(h_i-c/24)}\left[A_i (-\log q/(2\pi))^{d_i}+O(e^{2\pi i\epsilon\tau})\right].\qquad  
    \ee 
   The  dominant contribution comes from  the one with the smallest $h_i$ and largest exponent $d_i$~\footnote{One can    figure out the candidate $ h_{\text{min}}$ by looking at the roots of the indicial equation of  MLDE. For twisted $\Gamma^0(2)$-MLDE, which happens for Schur index in series of $q^{1/2}$, one needs to consider the $S$-transformed MLDE and its indicial equation.  One example is the  Bershadsky-Polyakov algebra for $(A_1,A_5)$, as discussed in \cite{Beem:2017ooy}. }
    \be
 h_{\text{min}}=\min\{h_i | S_{0i}\neq 0 \},\qquad  d_{\text{max}}=\max\{d_i | h_i =h_{\text{min}}, S_{0i}\neq 0\}.
 \ee

We can then rewrite \eqref{chi0hat}  and get the following asymptotic growth    in the limit $\tau \to i 0^+$ 
 \be
 \chi (\tau) =e^{2\pi i \tau c_{2d}/24} e^{-2\pi i  (h_ \text{min}-c_{2d}/24)/\tau} (-i\tau)^{-d_{\max}}
 \Big[\sum_{i:h_i=h_ {\min}}    A_i S_{0i}   (-i\tau)^{d_\text{max}-d_i} +O(e^{-\#/\text{Im }   \tau})\Big] ,
 \ee
 where we omit the subscript 0 for the vacuum. 
 
   The universal expansion of the index is given by
  
 \be\label{asymI}
\log\cI=\log\chi= {\frac{\pi i}{12\tau}}c^\text{eff}+\gamma \log(-i \tau)+\log \cS+\frac{\pi i \tau\cC}{12}+\cdots,
\ee
where the dots represent the exponentially small terms.  By comparison, we find
\be
\ceff=c_{2d}- 24h_{\text{min}}, \qquad \gamma=-d_{\text{max}}, \quad 
\cS(\tau)= \sum_{i:h_i=h_{\text{min}}}  A_i S_{0i}   (-i\tau)^{d_{\text{max}}-d_i}, \qquad
\cC=c_{2d},
\ee
where $\cS$ is a finite polynomial of $\tau$, and only the minimal-exponent sector belongs in this polynomial. \footnote{If the leading contributions cancel, one then needs to modify and consider the first surviving exponent and degree. 
}

 The logarithmic term $\gamma \log(-i \tau)$ is absent if the MLDE has no  logarithmic solution and the $T$ matrix is diagonalizable, which holds for    rational VOAs. In such a case  $d_i=d_{\text{max}}=0$, and $\cS$ is just a constant, which is related to the modular $S$-matrix.  
 \footnote{If $\cS$ is not a constant, isolating $\cC$ from the linear contribution  of $\cS$ requires  non-trivial care in \eqref{asymI} .}

 Note that the above equations apply to the asymptotic growth  of the unflavored index in the high temperature limit $\tau \to 0$.   If we also turn on the flavor fugacity and scale its additive   coordinate as $z=\alpha\tau\to0$,   $\cC$ also receives a contribution from the  flavor symmetry, which is proportional to the corresponding  level. In this case, we also need to include the flavor anomaly contribution $\exp( \pi i k_J z^2/\tau )$ to the $S$ transformation in \eqref{Schis}. Then the general relation is
 \be\label{cCsJ}
  \mathcal C=c_{2d}-12k_J\alpha^2,
 \ee 
  where $k_J$ is the   level for  the effective U(1) symmetry, namely  $J(z)J(w)\sim k_J/(z-w)^2$.  For several currents we need to replace $k_J\alpha^2$ by their level quadratic form.  
  
  In principle, the effective central charge could also be corrected by the flavor contribution, but the extra flavor contribution  turns out to be absent for the specializations considered here.    The effective central charge is also related to the 4d   central charges   \cite{Beem:2017ooy}
  \footnote{This relation assumes that the leading Schur Cardy coefficient is the anomaly contribution which might fail, in particular if $c<a$. The examples in this paper do  satisfy this assumption.} 
\be\label{ceffac}
\ceff=48(c - a).
\ee
 This makes the effective central charge particularly useful.

  \subsection{AD theory and associated VOA }
  
  Our main examples of SCFTs are Argyres--Douglas (AD) theories  \cite{Argyres:1995jj}. We focus on theories admitting a class $\cS$ realization by compactifying six-dimensional $\mathcal N=(2,0)$ SCFTs on a sphere with one irregular puncture and, possibly, one regular puncture \cite{Xie:2012hs}. This construction can be extended by introducing outer-automorphism twists \cite{Wang:2018gvb}. We mainly focus on the untwisted case, in which $G$ is of ADE type. We denote its complexified Lie algebra by $\mathfrak g$ and its Coxeter and dual Coxeter numbers by $h_G$ and $h_G^\vee$, respectively; for simply-laced $G$, these coincide.

One family of interest is $D_p(G)$, with $p\ge2$. These theories arise from the six-dimensional $\mathcal N=(2,0)$ theory of type $G$ compactified on a sphere with one full regular puncture and an irregular puncture of type $G^{h_G}[p-h_G]$ \cite{Xie:2012hs}. The full puncture contributes a $G$ flavor symmetry, while the irregular puncture may contribute additional flavor symmetry; the combined symmetry may also exhibit enhancement. When $\gcd(p,h_G)=1$, the irregular puncture contributes no additional flavor symmetry, and the flavor algebra of $D_p(G)$ is precisely $\mathfrak g$.

More general regular punctures can be obtained by partially closing the full puncture through nilpotent Higgsing, whenever the corresponding Higgsing is allowed. The resulting theory is denoted by $D_p(G,Y)$, where $Y$ labels the Nahm nilpotent orbit in $\mathfrak g$ associated with the regular puncture.

For $G=SU(N)$, the two commonly used notations are related by
\be
D_p(SU(N),Y)=\left(A_{N-1}^{N}[p-N],Y\right).
\ee
The puncture $Y$ is labeled by a partition $[\lambda_1,\ldots,\lambda_t]$ of $N$, with $\lambda_1\ge\cdots\ge\lambda_t>0$ and $\sum_{i=1}^t\lambda_i=N$. Equivalently, grouping equal parts, we may write
$ Y=[\mu_1^{s_1},\ldots,\mu_k^{s_k}]$ subject to   $\mu_1>\cdots>\mu_k>0$, $  s_I>0$ and $\sum_{I=1}^k\mu_I s_I=N$, where $s_I$ is the multiplicity of the part $\mu_I$ and $t=\sum_{I=1}^k s_I$. The continuous flavor symmetry associated with the regular puncture is conventionally denoted  by \footnote{Here $S[\cdots]$ removes the overall $U(1)$, and  denotes the determinant-one condition inherited from the embedding into $SU(N)$. In terms of flavor fugacities, the constraint is $\prod_{i=1}^t\prod_{s=1}^{\lambda_i}x_{i,s}= \prod_{i=1}^t y_i^{\lambda_i}=1$,
where $y_i$ is the   fugacity attached to the $\lambda_i$ block, and  $x_{i,s}$  is the  fugacity      induced from the affine torus, related by 
$ x_{i,s} =y_iq^{(\lambda_i+1-2s)/2} $.
} \footnote{Note that in this paper, we do not distinguish global forms of the flavor symmetry. 
}
\be
F_Y=S\!\left[U(s_1)\times\cdots\times U(s_k)\right],
\qquad
\operatorname{rank}F_Y=\sum_{I=1}^k s_I-1=t-1.
\ee

Some frequently encountered punctures are
\beqn
\text{full puncture:}\quad Y
&=&[1^N]
\quad\longrightarrow\quad SU(N),
\\
\text{simple (subregular) puncture:}\quad Y
&=&[N-1,1]
\quad\longrightarrow\quad U(1),
\qquad N\ge3,
\\
\text{completely closed puncture:}\quad Y
&=&[N]
\quad\longrightarrow\quad \text{none},
\eeqn
where we also indicate the  flavor symmetry contributed by the regular puncture. In the Nahm labeling, the full and completely closed punctures correspond to the trivial and principal nilpotent orbits, respectively.

Another large class     of AD theories  also  admits a geometric engineering description in type IIB string theory on noncompact Calabi--Yau threefolds with suitable isolated quasihomogeneous hypersurface singularities. A distinguished family constructed in this way is denoted by $(G,G')$, where $G$ and $G'$ label ADE types and the defining polynomial is the sum of the corresponding two ADE singularity polynomials \cite{Cecotti:2010fi}.

The class $\cS$ and type IIB constructions overlap and can yield the same SCFT. In particular,
\beqn
(A_{N-1},A_{k-1})
&=&D_{N+k}(SU(N),[N]), \qquad N,k\ge2,
\\
(A_1,D_k)
&=&D_k(SU(2))
=D_k(SU(2),[1,1]),\qquad k\ge 3,
\\
D_3(SU(2))
&=&(A_1,D_3)
=(A_1,A_3).
\eeqn
 
The class $\cS$ description is particularly useful for identifying the associated VOAs. Under the coprimality assumption $\gcd(p,h_G)=1$, the VOA associated with $D_p(G)$ is identified with the simple affine vertex algebra \cite{Xie:2016evu,Song:2017oew}
\be
\mathbb V[D_p(G)]
=
L_{k}(\mathfrak g),
\qquad
k=-h_G^\vee+\frac{h_G^\vee}{p}.
\ee
Here $k$ is the affine level, which is boundary admissible. The theory $D_p(G)$ corresponds to the special case of $D_p(G,Y)$ in which $Y$ is the full puncture.

More generally, for a regular puncture $Y$ compatible with the irregular data, the associated  VOA   is  \cite{Xie:2016evu,Song:2017oew}
\be
\mathbb V[D_p(G,Y)]
=
\mathcal W_{k}(\mathfrak g,f_Y),
\qquad
k=-h_G^\vee+\frac{h_G^\vee}{p},
\ee
where $f_Y$ is a     nilpotent  element in the   Nahm orbit determined by the puncture $Y$, and $\mathcal W_{k}(\mathfrak g,f_Y)$ denotes the simple $W$-algebra. The construction uses quantum Drinfeld--Sokolov (DS) reduction associated with $f_Y$. Thus, partial closure of the regular puncture is reflected by DS reduction on the VOA side, with the relevant reduction understood to be nonzero. For the full puncture, $f_Y=0$, and the description reduces to the affine vertex algebra above. We will discuss affine and $W$-algebras in more detail below. See  section~\ref{gwsecgeneral}.

The construction can also be extended to twisted class $\cS$ setups \cite{Wang:2018gvb}, leading to non-simply-laced flavor symmetries and associated affine and $W$-algebras. Such extensions require specifying the outer-automorphism twist, the relevant Lie algebra, and the level normalization. We will not discuss these additional subtleties here.

 \subsection{A recap on known   index identities}
 
 The  prototypical example of an index identity  is given by  \cite{Buican:2020moo}
 \be\label{A2D4indx}
 \cI_{(A_2,D_4)}(q)=\cI_{\mathcal{N}=4\;SU(2)  \text{  SYM}}(q^{3}, q^{\frac{1}{2} }),
 \ee
where the second argument for the $SU(2) $ flavor fugacity is specialized. 

More generally,   there is a class of AD theories which are obtained by the conformal gauging of a set of $D_{p_i}(G)$ theories,   denoted as $\cT_{(p,N)}$ with  $p\in\{2,3,4,6\}$, $N\geq2$ and $\gcd(p,N)=1$ \cite{Buican:2020moo,Jiang:2024baj}. Their Schur indices are also related to those of $\cN=4$ SYM 
\be\label{TpNindex}
\cI_{\cT_{(p,N)}}(q)=\cI_{\mathcal{N}=4 \; SU(N)  \text{  SYM}}(q^{p}, q^{\frac{p}{2}-1}).
\ee
 Note that $(A_2,D_4)=\cT_{(3,2)}$. Another similarity between the two SCFTs above is that they both   have the feature $a=c$.

As shown in \cite{Jiang:2026keq,Yoshida:2026ugm}, there is   another kind of character identity relating   three theories
  \be
\mathcal{I} _{(A_2,D_{3m+1})}(q) = \cI_{(A_1,A_{2m-2})} (q)\; \cI_{\mathcal{N}=4\;SU(2)\text{ SYM}}(q^{2m+1}, q^{\frac{1}{2}}),\qquad m\ge1,
\ee 
where the prefactor  is trivial if $m=1$, which reduces to \eqref{A2D4indx}.

 This kind of relation is further generalized to a pair of $\cN=2$ AD theories. In \cite{Adamovic:2025efe}, it was found that 
 \footnote{Since the index is invariant under the inversion of $SU(2)$ flavor fugacity, we also have $\cI_{(A_{1},D_{2n+1})}(q^3,q^{-\frac12})=\cI_{(A_1,A_{6n})} (q)$. Similar statements hold for the remaining index identities.}
 \be
\cI_{(A_{1},D_{2n+1})}(q^3,q^{\frac12})=\cI_{(A_1,A_{6n})} (q),
\ee
where the  $(A_{1},D_{2n+1})$ AD theory has flavor symmetry $SU(2)$ whose fugacity is specialized. 
This translates to a  vacuum character identity. 
More non-trivially, it was also shown in     \cite{Adamovic:2025efe}   that similar identities can be generalized to the non-vacuum modules of VOAs.

These index and character identities motivate us to ask whether this is just a coincidence, or a general feature. Also, it is natural to wonder what  they imply beyond the matching of indices.

This is the goal of this paper. We systematically generalize these index identities, which
occur widely among AD theories and $W$-algebras. We also show that they imply simple
relations between characteristic anomaly coefficients.

 \subsection{Anomaly relations  from  index identity}
 
 The   general   index identities  take  the  form  of
 \be
\prod_{i=1}^{N_L}  {\mathcal{I}}_{\mathfrak{L}_i}(Q_i, \bm X_i)
=\prod_{j=1}^{N_R}  {\mathcal{I}}_{\mathfrak{R}_j}(Q_j',\bm X_j'),
 \ee
 where the fugacities on the LHS and RHS are all related by a specific  map. This   includes all the   identities we reviewed before.

 Our main goal is to study in detail the simplest case of $N_L=N_R=1$:   
 \be\label{ISinds}
   {\mathcal{I}}_{\mathfrak{L} }(q,\bm x  )=  {\mathcal{I}}_{\mathfrak{R} }(q^\ell, \bm X _*(q,\bm x)),
 \ee
where $X_* $ is specialized to a specific  homogeneous function of $q, \bm x$. Such identities can relate AD theories whose
associated VOAs are not isomorphic.

As discussed before, the leading asymptotic growth of the index is controlled by the effective central charge  \eqref{asymI}. 
 The index identity in \eqref{ISinds} thus     implies that 
 \footnote{For the specializations used below, the leading transformed flavored coefficient survives, so the stated effective-central-charge relation holds. This flavor independence is not an unconditional statement for arbitrary specializations.  }
 \be\label{LRcc}
\ceff_\fkL=\ceff_\fkR/\ell, \qquad \cC_\fkL=\ell \; \cC_\fkR.
 \ee
  This is particularly useful because these   central charges  are characteristic quantities, computable from the  4d central charges and levels in a simple way \eqref{ceffac}.  It also provides a natural way to understand  why both theories in \eqref{TpNindex} feature $a=c$.

  In case the flavor fugacities on the two sides can be set to one simultaneously, we have the  simpler relation $   {\mathcal{I}}_{\mathfrak{L} }(q  )=  {\mathcal{I}}_{\mathfrak{R} }(q^\ell)$. The asymptotic  expansion \eqref{asymI} gives  the   stronger and simpler relations   
  \be\label{ceffecL}
     \ceff_\fkL=\ceff_\fkR/\ell,\qquad
c_{\fkL}=\ell \; c_{\fkR} .
  \ee
Note that the  two central charges above scale oppositely, which is related to the factor $\tau$ appearing in the denominator and numerator, respectively  \eqref{asymI}. 

For the more general case \eqref{ISinds}, the $\tau$-linear   coefficient $\cC$ receives a correction from the flavor fugacity, as in \eqref{cCsJ}, and the $c$ central charges on the two sides are therefore related
in a more complicated but controlled way.  
One can also try to match other coefficients in \eqref{asymI}, in particular the  constant term $\cS$.  For $\gamma$,   it actually always vanishes in our boundary admissible class of AD theories below.  \footnote{This   does not extend automatically to all AD theories. For example, \eqref{TpNindex} has $\ceff=0$ and  $\gamma\neq 0$.}

%
%

As a side remark, we also find many examples with index identities of the form  
  \be
\cI_{\mathfrak{L}_1}(Q_1 , \bm X _1)\cI_{\mathfrak{L}_2}(Q _2, \bm X _2)=\cI_{\mathfrak{R}_3}(Q', \bm X'),
 \ee
 where the fugacities are   related and specialized as well. 
 
\section{Index identity between $(A_{N-1},A_{(N+1)(p-1)}) $ and $D_p(SU(N))$} \label{ANDNSFs}
In this section, we generalize the result in  \cite{Adamovic:2025efe} and establish that
\be
\mathcal{I}_{(A_{N-1},A_{(N+1)(p-1)})}(q) = \mathcal{I}_{D_p(SU(N))}(Q = q^{N+1}, \bm{x} = q^\rho), \qquad
 \gcd(p,N)=1,
\ee
where we use   $Q,\bm x$ to represent the fugacities of the flavored theory on the RHS. In particular, the  flavor fugacities are set  to the    principal  specialization 
 \be\label{prinspecial}
x_i =q^{\rho_i}= q^{\frac{N+1-2i}{2}}, \qquad i = 1,\dots,N, \qquad \prod_{i=1}^N x_i = 1,
\ee 
where $\rho$ is the Weyl vector.

The Schur index of $D_p(SU(N))$ is given by
\be
\mathcal{I}_{D_p(SU(N))}(Q, \bm x) = \mathrm{PE} \left[ \frac{Q - Q^p}{(1 - Q)(1 - Q^p)} \chi^{SU(N)}_{\mathrm{adj}}(\bm x) \right],
\ee
where the plethystic exponential  is  defined as
\be
\mathrm{PE}[f(q,x)] = \exp\left[ \sum_{n \ge 1} \frac{f(q^n, x^n)}{n} \right].
\ee
The inverse is denoted as $ \mathrm{PL}$.

 Under \eqref{prinspecial}, the adjoint character becomes 
\be
\chi^{SU(N)}_{\mathrm{adj}}(q^\rho) = \sum_{1\le i, j\le N} \frac{x_i}{x_j}-1= \sum_{1\le i, j\le N} \frac{q^j}{q^i}-1
=  \frac{(1 - q^{N-1})(1 - q^{N+1})}{q^{N-1}(1 - q)^2}  .
\ee
Setting $ Q = q^{N+1}$ gives

\beqn
\mathcal{I}_{D_p}(q^{N+1}, q^\rho) &=& \mathrm{PE} \left[ \frac{q^{N+1}(1 - q^{(N+1)(p-1)})}{(1 - q^{N+1})(1 - q^{(N+1)p})} \chi_{\mathrm{adj}}(q^\rho) \right]
\nonumber\\
 &=& \mathrm{PE} \left[ \frac{q^{N+1}(1 - q^{(N+1)(p-1)})}{(1 - q^{N+1})(1 - q^{(N+1)p})} \frac{(1 - q^{N-1})(1 - q^{N+1})}{q^{N-1}(1 - q)^2}  \right]
\nonumber\\
 &=& \mathrm{PE} \left[ \frac{q^{2}(1 - q^{(N+1)(p-1)})(1 - q^{N-1})}{ (1 - q)^2(1 - q^{(N+1)p})}  \right] .\nonumber
\eeqn

Meanwhile, the Schur index of  $(A_{N-1}, A_{k-1})$ is given by  
\be\label{AkANIdx}
\mathcal{I}_{(A_{N-1}, A_{k-1})} = \mathrm{PE}\left[ \frac{q^2 (1 - q^{k-1}) (1 - q^{N-1})}{(1 - q)^2 (1 - q^{k+N})} \right],
\qquad \gcd(k,N)=1.
\ee
By comparing the two expressions above, it is easy to see that they coincide if we set $k=(N+1)(p-1)+1=pN+p-N$ and $\gcd(k,N)=\gcd(p,N)=1$ as required. Therefore 
\be
\mathcal{I}_{(A_{N-1},A_{(N+1)(p-1)})}(q) = \mathcal{I}_{D_p(SU(N))}(Q = q^{N+1}, \bm x = q^\rho).
\ee

Using \eqref{ceffac}, one can easily compute   the effective central charges of the two theories  
\be \ceff_{D_p(SU(N))} = \frac{(N^2 - 1)(p - 1)}{p}, \qquad
\ceff_{(A_{N-1},A_{k-1})} = \frac{(N - 1)(k - 1)}{N + k}. \ee
As a result, one can verify 
\be \ceff_ {(A_{N-1},A_{(N+1)(p-1)})} =
   \frac{(N - 1)(p - 1)}{p} 
 = \frac{1}{N + 1} \ceff_{D_p(SU(N))}.
  \ee

\section{Index identity between $( A_{(h+1)(p-1)},G)$ and $D_p(G)$} \label{AGDpG}
 
In this section, we generalize the identity in the previous section  and prove that
\be\label{DpGA}
\mathcal I_{D_p(G)}
(q^{h+1},q^\rho)
=
\mathcal I_{( A_{(h+1)(p-1)},G)}(q), \qquad \gcd(h,p)=1,
\ee
for any \emph{simply laced } $G$ with  Coxeter number $h$, which coincides with the  dual Coxeter number $h^\vee$.  Throughout the section, we always assume that  $p\ge 2$ is an integer and $\gcd(p,h)=1$.

For  $ D_p(G)$ with  no additional  flavor symmetry from  the irregular puncture, the flavored Schur index is \cite{Song:2017oew}
\be\label{DpGindex}
\mathcal I_{D_p(G)}(Q,\bm x)
=
\operatorname{PE}\!\left[
\frac{Q-Q^p}{(1-Q)(1-Q^p)}
\chi_{\rm adj}^{G}(\bm x)
\right].
\ee
This is precisely the boundary-admissible affine vacuum character.  
 
Let $G$ be simply laced, with rank $r$, Coxeter number
$ h=h^\vee $, 
exponents
$
e_1,\ldots,e_r,
$
and Casimir degrees
$
d_i=e_i+1.
$
Under the principal specialization
$
\bm x=q^\rho,
$
we have
\be \chi_{\rm adj}(q^\rho)
=
\sum_{i=1}^r\sum_{m=-e_i}^{e_i}q^m
=
\sum_{i=1}^r\sum_{m=1 -d_i }^{d_i-1}q^m. \ee
Using
$
e_i+e_{r+1-i}=h,
$
one gets the following  crucial identity
\be \frac{q^{h+1}}{1-q^{h+1}}\,
\chi_{\rm adj}(q^\rho)
=
\frac{\displaystyle\sum_{i=1}^r q^{d_i}}{1-q}. \ee
Therefore setting $ Q=q^{h+1} $ in \eqref{DpGindex} 
gives
\be \mathcal I_{D_p(G)}
(q^{h+1},q^\rho)
=
\operatorname{PE}\!\left[
\frac{
\left(\sum_iq^{d_i}\right)
\left(1-q^{(h+1)(p-1)}\right)
}
{(1-q)(1-q^{(h+1)p})}
\right]. \ee
This immediately implies that 
\be\label{DpGAGrel}
\mathcal I_{D_p(G)}
(q^{h+1},q^\rho)
=
\mathcal I_{( A_{(h+1)(p-1)},G)}(q),
\ee
 by recalling the following Schur-index formula for $(A_{k-1},G)$ with $k\ge2$ and $\gcd(k,h)=1$ 
\be \label{IkG}
\mathcal{I}_{(A_{k-1}, G)}(q) = \mathrm{PE} \left[ \frac{\sum_{i=1}^r \left( q^{d_i} - q^{h+k+1-d_i} \right)}{(1 - q)(1 - q^{h+k})} \right].
\ee

More explicitly, the two related theories are 
 \be
\begin{array}{c|c|c|c}
G & \ell & D_p(G) &  ( A_{(h+1)(p-1)},G) \\ \hline
A_{N-1} &  N+1 & D_p(SU(N)) & (A_{N-1},A_{(N+1)(p-1)})\\
D_N &  2N-1 & D_p(D_N) & (D_N,A_{(2N-1)(p-1)})\\
E_6 &  13 & D_p(E_6) & (E_6,A_{13(p-1)})\\
E_7 &  19 & D_p(E_7) & (E_7,A_{19(p-1)})\\
E_8 &  31 & D_p(E_8) & (E_8,A_{31(p-1)})
\end{array}
\ee

The associated VOAs of the two SCFTs are     
\beqn
\mathbb V[D_p(G)] &=& L_{-h+h/p}(\mathfrak g),
\nonumber\\
\mathbb V[(A_{(h+1)(p-1)},G)] &=& \mathcal W_{-h+h/((h+1)p)}(\mathfrak g,f_{\prin})
 =  W_G(h,(h+1)p),
\eeqn
where $ {W}_G(h, h+k)$ is the simple principal $W$-algebra minimal model. 

Since the Schur index of an SCFT equals   the vacuum  (super)character of the associated VOA, the index relation \eqref{DpGAGrel}  of  SCFTs   translates to the character relation of VOAs
\be
\chi_{\mathcal W_{-h+h/((h+1)p)}(\mathfrak g,f_{\prin})}(q)  =  \chi_{L_{-h+h/p}(\mathfrak g)}(q^{h+1},q^\rho).
\ee

So the rescaling factor is generally given by $
\ell={h_G+1}.
$
We next check the anomaly relations.

For $ \gcd(p,h)=1$, the  4d central charges of $D_p(G)$ are given by     \cite{Cecotti:2013lda}
\be
a_{D_p(G)}
=
\frac{(4p-1)(p-1)}{48p}\dim G,
\qquad
c_{D_p(G)}
=
\frac{p-1}{12}\dim G, \label{cDpg}
\ee
and the effective central charge is \eqref{ceffac}
\be\label{DpGceff}
 \ceff _{D_p(G)}  = \frac{p-1}{p} \dim G =\frac{r(h+1) (p-1)}{p} ,
\ee 
where we used
 $ \dim G=r(h+1)$ with $r$ the rank of $G$.
 
Meanwhile, the  central charges of $(A_{k-1},G)$ in the coprime case   $\gcd(k,h)=1$  are
\be\label{AGca}
c_{(A_{k-1}, G)} = \frac{r(k - 1)(kh + k + h)}{12(k + h)}, \qquad
a_{(A_{k-1}, G)} = \frac{r(k - 1)(4kh + 4k + 4h - 1)}{48(k + h)},
\ee
which can be computed from the type IIB realization. These expressions  imply the following effective central charge
\be\label{cAkG}
\ceff_{(A_{k-1}, G)} 
  = \frac{r(k - 1)}{k + h}.
\ee
 This can also be confirmed from the direct   high temperature limit of the index  \eqref{IkG}
\be
\mathcal I_{(A_{k-1},G)}(e^{-\beta})
 =\mathcal S_{(A_{k-1},G)}
 \exp\!\left(\frac{\pi^2r(k-1)}{6(h+k)\beta}
              -\frac{\beta c_{2d}}{24}\right)
 \left[1+O(e^{-\#/\beta})\right] ,
\ee
where
\be \cS_{(A_{k-1}, G)} = (2\pi)^{-r(k-1)/2} \prod_{i=1}^r \prod_{s=0}^{k-2} \Gamma\left( \frac{d_i + s}{h + k} \right). 
\ee

 By comparing \eqref{DpGceff} and \eqref{cAkG}, it is easy to see that 
\be
\ceff_{( A_{(h+1)(p-1)},G)}= \frac{r(p-1)}{p}= \frac{1}{h+1}\ceff_{D_p(G)}.
\ee

We can also match the  $\tau$-linear term $\cC$  in \eqref{asymI} once the flavor contribution is taken into account.  
On the affine side, we have 
\be
\tau_{\mathrm{aff}}=\ell\tau,\qquad \bm z_{\mathrm{aff}}=\tau\rho=\tau_{\mathrm{aff}}\frac{\rho}{\ell},\qquad
k_{J^\rho}= k_{\mathrm{aff}}(\rho,\rho) ,\qquad
(\rho,\rho)=\frac{h\dim\mathfrak g}{12},
\ee
where $\ell=h+1$. 
Therefore,  \eqref{cCsJ}  gives
\be
\cC_{D_p(G)}=c_{2d,D_p(G)}-\frac{12k_{\mathrm{aff}}(\rho,\rho)}{\ell^2}
=(1-p)\dim\fkg\big( 1-\frac{h^2}{p\ell^2}\big) 
=-\frac{r(p-1) }{\ell p}\left[p(h+1)^2-h^2\right],
\ee
where  we used \eqref{cDpg} and $k_{\mathrm{aff}}=-h+h/p$.

On the principal $W$ side which has no flavor symmetry, we have
\be
\cC_{(A_{(h+1)(p-1)}, G)} =c_{2d,(A_{(h+1)(p-1)}, G)}  =-12c_{ (A_{(h+1)(p-1)}, G)} = -\frac{r(p-1) \big[p(h+1)^2 - h^2\big]}{p},
\ee
where we used \eqref{AGca}.  By comparison,  we obtain
\be
\cC_{(A_{(h+1)(p-1)}, G)} =\ell\;\cC_{D_p(G)}.
\ee
This confirms \eqref{LRcc}. 

Let us conclude this section by comparing the Coulomb-branch spectra
of the two theories, which probe a protected sector different from
the Schur sector.

The Coulomb spectra of the two theories are given by   \cite{Cecotti:2013lda}
\begin{align}\label{SCBADpG}
 \text{CB}_{D_p(G)}
 &=\bigsqcup_{i=1}^{r}
 \Big\{
 d_i-\frac{hl}{p} \;\Big|\;
 l\in\mathbb Z,\quad
 1\le l<\frac{p(d_i-1)}h
 \Big\},\\ \label{SCBAG}
 \text{CB}_{(A_{(h+1)(p-1)},G)}
 &=\bigsqcup_{i=1}^{r}
 \Big\{
 d_i-\frac{hj}{(h+1)p}\;\Big|\;
 j\in\mathbb Z,\quad
 d_i\le j<\frac{(h+1)p(d_i-1)}h
 \Big\},
\end{align}
where the second formula can be derived from the type IIB construction \cite{Wang_2016}. Here  $d_i$'s are the degrees of Casimirs of $\fkg$, and all elements in the set above should be counted with multiplicity. In particular, the upper bounds on $l$ and $j$ select Coulomb-branch operators with         $\Delta>1$.

 Consequently, one can verify that  
\be
 \text{CB}_{D_p(G)}
 =
 \Big\{
 \Delta\in \text{CB}_{(A_{(h+1)(p-1)},G)}\;\Big|\;
 p\Delta\in\mathbb Z
 \Big\}
 \subsetneq
 \text{CB}_{(A_{(h+1)(p-1)},G)}.
\ee
Indeed, starting from the spectrum in \eqref{SCBAG}, we have
\be
 p\Delta=pd_i-\frac{hj}{h+1}\in\mathbb Z
 \quad\Longleftrightarrow\quad
 l=\frac{j}{ h+1} \in\mathbb Z
  \quad\Longrightarrow\quad
  1\le l<\frac{p(d_i-1)}{h},
\ee
where we have used the fact that  $2\le d_i\le h$, and the range of $j$ in \eqref{SCBAG}. The resulting $l$ agrees exactly with that in \eqref{SCBADpG}.

The ranks, which are dimensions of the Coulomb branches, of the two theories are also related: 
\be 
\dim \text{CB}_{D_p(G)} 
  =\frac{r(p-1)}2, \qquad
\dim \text{CB}_{(A_{(h+1)(p-1)},G)} =\frac{r(h+1)(p-1)}2
 =\ell\, \dim  \text{CB}_{D_p(G)} .
 \ee

\section{Index identities for  general AD theories  }
\label{gwsecgeneral}
We now turn to more general AD theories and further extend the identities established previously. The resulting character identities involve $W$-algebras exhibiting a range of new features, possibly involving  non-simply laced algebras, non-principal nilpotent elements, non-boundary admissible  and even non-admissible levels.

\subsection{$W$ algebra at boundary admissible level  }
\label{gw-subsec-domain}

Let $\mathfrak g$ be a finite-dimensional simple complex Lie algebra.
Write $h$, $h^\vee$, $r$, and $r^\vee$ for its Coxeter number, dual
Coxeter number, rank, and lacing number, respectively.  See Table~\ref{Liedata} for   details.

In this paper, we mainly consider the  principal boundary-admissible family, whose level is given by
  \footnote{The simple affine vertex algebra $ L_k(\mathfrak{g})$ is called admissible when its level $k$ is admissible, namely when \cite{KacWakimoto1988} 
\be
k + h^\vee_{\mathfrak{g}} = \frac{n}{p}\qquad \text{ with } n,p \in \mathbb{Z}_{\ge 1}, \qquad (p, n) = 1,\qquad n \ge 
\begin{cases} 
h^\vee_{\mathfrak{g}} &   \text{if } (r^\vee, p) = 1, \\  
h_{\mathfrak{g}} & \text{if } (r^\vee, p) = r^\vee,
\end{cases}\nonumber
\ee
where $ r^\vee$ is the lacing number. 
If $ k$ is admissible with $ (r^\vee, p) = 1$, we say that $ k$ is  \emph{principal}. If $ k$ is admissible with $ (r^\vee, p) = r^\vee$, we say that $ k$ is  \emph{coprincipal}.  We say $k$ is \emph{boundary admissible } if $k$ is admissible and $ n=h^\vee_\fkg$. 
\label{admiscond}
}
\be\label{klevel}
 k(u)=k=-h^\vee+\frac{h^\vee}{u}, \qquad \qquad u =\kappa +h^\vee,
\ee
 subject to the conditions  
\be
 u\in\mathbb Z_{\geq1},\qquad
 \gcd(u,h^\vee)=\gcd(u,r^\vee)=1.
 \label{gw-eq-admissibility}
\ee
   
   The  corresponding simple affine  vertex algebra will be denoted as $L_k(\fkg)$, which is the simple quotient of the universal affine vertex algebra  $V^k(\fkg)$.  
The quotient removes all the null states. On the 4d SCFT side,   this is the VOA associated with $D_u(G)$, constructed by compactifying a six-dimensional $(2,0)$ theory on a Riemann sphere with an irregular puncture and a full puncture.~\footnote{One may also need to consider twisted punctures and specify extra data if $G$ is not simply laced. } 

Performing a further DS reduction on the affine vertex algebra $L_k(\fkg)$ gives a $W$-algebra.   On the 4d SCFT side, this corresponds to partially closing the full puncture.  To specify the DS reduction, one also needs to choose a nilpotent element.  The nilpotent element 
is denoted as $f$, and can be completed to an $\mathfrak{sl}_2$-triple
$(e_f,h_f,f)$. We also set $\sfh_f=h_f/2$.
In particular, $[\sfh_f,e_f]=e_f$ and $[\sfh_f,f]=-f$.
Let $\mathfrak g^\natural$ denote the centralizer of the triple. 
Under  the commuting action of $ \mathfrak{sl}_2\oplus\mathfrak g^\natural\subset \fkg $, the adjoint representation of $G$, which is identified with    $\fkg$,  can be decomposed as
\be
 \mathfrak g=\bigoplus_{j\in\frac12\mathbb Z_{\geq0}}
 V_j\otimes R_j,\qquad
 \chi_{V_j}(q)=\sum_{\ell=-j}^{j}q^\ell
 =\frac{q^{-j}-q^{j+1}}{1-q}.
 \label{gw-eq-decomposition}
\ee
Here $V_j$ is the spin-$j$ representation of $\fks\fku(2)$, namely $\dim V_j=2j+1$,  and $R_j$ is a representation of
$\mathfrak g^\natural$.  We will use  $\bm x$ to denote the flavor fugacity of   $\mathfrak g^\natural$.

 The corresponding simple  $W$-algebra after DS reduction specified by nilpotent element $f$ is denoted as $\mathcal W_k(\mathfrak g,f)$, \footnote{More precisely, there is also a nonvanishing condition on reduction \cite{Arakawa:2010ni}: $u>j_{\text{max}}$ in order to ensure that    $f\in X_{L_{k(u)}(\mathfrak g)}$. Here  $ X_{L_{k(u)}(\mathfrak g)}$ is the associated variety  of the simple affine algebra, $u$ appears in the level \eqref{klevel}, and $j_{\text{max}}=\max\{j:R_j\neq0\}$, which implies that the largest Jordan block of  $\operatorname{ad}f$ has size $2j_{\text{max}}+1$. 
For $\mathfrak{sl}_N$, the nilpotent element is specified by  a partition  $\lambda=[\lambda_1,\lambda_2, \cdots]$ subject to  $\lambda_1\ge \lambda_2\ge\cdots > 0$ and $\sum_i \lambda_i=N$. As a result, the   nonvanishing reduction condition is
  $u\geq \lambda_1$, as $j_{\max}=\lambda_1-1$. We will always assume that this nonvanishing     reduction condition is satisfied.  
  We also assume certain simplicity conditions are satisfied in this paper, so that the DS reduction  of a simple affine vertex algebra  yields  a simple $W$ algebra.  
  } which should be distinguished from the  universal algebra $\mathcal W^k(\mathfrak g,f)$.

\subsection{The character and index formula}

 We now present the vacuum-character formula for the W-algebra above. At boundary-
admissible levels, these formulas are particularly simple.

The character is defined as \footnote{In the presence of fermionic operators in the VOA, to compare with the Schur index, the trace must be replaced by a supertrace, and the corresponding supercharacter should be considered.}
\be
 \chi_{\mathcal W_{k }(\mathfrak g,f)}(q,\bm x)
 =\operatorname{Tr}_{\mathcal W_{k }(\mathfrak g,f)} 
 q^{L_0}\bm x^{J_0},
 \label{gw-eq-normalized}
\ee
where  we  omit  the factor $q^{-c/24}$, and normalize the character such that it starts with 1 for the vacuum.

Let us introduce
\beqn \label{ABfeq}
 A_f(q,\bm x) &=& \sum_jq^{j+1}\chi_{R_j}(\bm x), 
\\  \label{ABfeq2}
 B_f(q,\bm x) &=& \chi_{\mathrm{adj}}(\bm x q^{\sfh_f})
= \sum_j\chi_{V_j}(q)\chi_{R_j}(\bm x). 
\eeqn
The multiplicative notation $\bm x q^{\sfh_f}$ means that a weight
$\lambda$ contributes $\bm x^\lambda q^{\lambda(\sfh_f)}$.

Then the  character of the $W$-algebra is   given by  \cite[eq.~(36)]{Xie:2019zlb}
\be 
   \chi_{\mathcal W_{k(u) }(\mathfrak g,f)}(q,\bm x)
 =\PE\!\left[
 \frac{A_f(q,\bm x)}{1-q}
 -\frac{q^u}{1-q^u}B_f(q,\bm x)
 \right],
 \label{gw-eq-product}
\ee
where $A_f,B_f$ are given by  \eqref{ABfeq} and  \eqref{ABfeq2}.

Using the formula in \eqref{gw-eq-decomposition}, this  can be simplified further \cite[eq.~(39)]{Xie:2019zlb}  
\be 
 \chi_{ \mathcal W_{k(u)}(\mathfrak g,f)}(q,\bm x)
 =\PE\!\left[
 \frac{\sum_j(q^{j+1}-q^{u-j})\chi_{R_j}(\bm x)}
 {(1-q)(1-q^u)}\right]
  =\PE\!\left[
 \frac{A_f(q,\bm x) -q^{u +1}A_f(q^{-1},\bm x)}
 {(1-q)(1-q^u)}\right]. 
 \label{gw-eq-XY39}
\ee
 
These character formulas  allow  us to derive  the effective central charge
\footnote{From the asymptotic growth, one can show that each factor of  $\PE[m q^a/(1-q^u)] = \frac1{(q^a;q^u)^m_\infty}$ with $a,u>0$ contributes effective central charge $m/u $,   independent of $a$.
}
 \be
\ceff_{\mathcal W_{k(u) }(\mathfrak g,f)} = \dim\mathfrak g^f- \frac{\dim \mathfrak{g}}{u}, \qquad\qquad
\dim\mathfrak g^f=\sum_j\dim R_j ,
\ee 
  where $\mathfrak g^f $ is the centralizer of $f$, whereas $\mathfrak g^\natural\cong R_0\subset\mathfrak g^f$ is the centralizer of the full $\mathfrak{sl}_2$-triple and has dimension $\dim R_0$.

If $f$ is    the principal nilpotent element, then $\mathfrak g^\natural$ is trivial and  we have the decomposition 
\be
 \mathfrak g=\bigoplus_{ i=1}^r  V_{d_i-1},
 \ee
 where the $d_i$  are the degrees of the Casimirs of $\fkg$.  See Table~\ref{Liedata} for   explicit values.
  In this case the character simplifies  to
\be \label{chiprin}
 \chi_{\mathcal W_{k(u) }(\mathfrak g,f_\prin)}(q )
 =\PE\!\left[
 \frac{\sum_{i=1}^r (q^{d_i}-q^{u+1-d_i }) }
 {(1-q)(1-q^u)}\right].
\ee

If $f$ is a trivial nilpotent element,  we have $\mathfrak g=\mathfrak g^\natural$ and  only $V_0$ survives in \eqref{gw-eq-decomposition}.  In this case, we get the original affine vertex algebra, whose character is given by
\be  \label{chiaff}
  \chi_{  L_{k(u) }(\mathfrak g )}( Q,\bm x) 
 =\PE\!\left[
 \frac{Q-Q^u}{(1-Q)(1-Q^u)}\chi_{\mathrm{adj}}(\bm x)
 \right],
\ee
 where we use the symbol  $Q$ in place of $q$ for later convenience.

 In the two cases above, the effective central charges are
\be \label{ceffprins}
\ceff_{   L_{k(u) }(\mathfrak g )}
=
\left(1-\frac{1}{ u}\right)\dim\mathfrak g,
\qquad
\ceff_{\mathcal W_{k(u) }(\mathfrak g,f_\prin)}
=
r -\frac{ \dim\mathfrak g}{ u}=
\frac{ \dim\mathfrak g}{ h+1} -\frac{ \dim\mathfrak g}{ u},
\ee
where we used the formula  $\dim\mathfrak g=r(h+1)$.
As a consistency check, if $u=h+1$, then the second formula gives  vanishing effective central charge, which is expected as the corresponding principal $W$ algebra is trivial.

\subsection{Master equation}
Now we show a very useful identity: 

\be
\chi_{\mathcal W_{-h^\vee+\frac{h^\vee}{ \ell p }}(\mathfrak g,f)}(q,\bm x)
 =\chi_{\mathcal W_{-h^\vee+\frac{h^\vee}{ \ell   }}(\mathfrak g,f)}(q,\bm x) 
 \chi_{L_{-h^\vee+\frac{h^\vee}{ p   }}(\mathfrak g)}(q^\ell,\bm x q^{\sfh_f}).
  \label{usefulEq2}
\ee

To ensure the    admissibility and nonvanishing  of DS reduction, we require  
\be
 \ell,p\in\mathbb Z_{\geq1},\qquad
 \gcd(\ell p,h^\vee r^\vee)=1,\qquad \ell>j_{\text{max}}\equiv {\max\{j:R_j\neq0\}}.
 \label{gw-eq-ratio-hypotheses}
\ee

The proof of \eqref{usefulEq2} is simple once we use the explicit character formula. 
We consider the quotient of the characters of 
$ \mathcal W_{-h^\vee+\frac{h^\vee}{ \ell p }}(\mathfrak g,f) $ and $ \mathcal W_{-h^\vee+\frac{h^\vee}{ \ell   }}(\mathfrak g,f) $, and then take plethystic logarithms: 
\be
\PL\frac{\chi_{\mathcal W_{-h^\vee+\frac{h^\vee}{ \ell p }}(\mathfrak g,f)}}{\chi_{\mathcal W_{-h^\vee+\frac{h^\vee}{ \ell}}(\mathfrak g,f)}}
 =\left(\frac{q^\ell}{1-q^\ell}-\frac{q^{\ell p}}{1-q^{\ell p}}\right)B_f
 =\frac{q^\ell-q^{\ell p}}{(1-q^\ell)(1-q^{\ell p})}
 \chi_{\mathrm{adj}}(\bm x q^{\sfh_f}).
 \label{gw-eq-ratio-proof}
\ee

Note that  the $A_f/(1-q)$ terms cancel. The surviving term has exactly the structure of the affine character in \eqref{chiaff}, after   the substitutions $Q\to q^\ell$ and
$\bm x\to \bm x q^{\sfh_f}$.  This gives the proof of \eqref{usefulEq2}.

  The character identity \eqref{usefulEq2} translates to the following index identity among three AD theories:    
\be \label{DlpresSU}
\mathcal{I}_{D_{\ell p}(G,Y)}(q; \bm x) 
=  
\mathcal{I}_{D_\ell(G,Y)}(q;\bm{x}) 
\mathcal{I}_{D_p(G)}\left(q^\ell;\bm{x}q^{\sfh_{f_Y}}\right),
\ee
where certain conditions on the parameters are imposed implicitly as explained before.

Let us see a simple example  of  \eqref{usefulEq2} by taking $\fkg=\fks\fkl_2$ and  odd $\ell\ge5$ and odd $p\ge3$,   $f=[2]$. Then we have
\be 
\mathcal{I}_{(A_1, A_{\ell p-3})}(q) = \mathcal{I}_{(A_1, D_p)} \left(q^\ell; x=q^{1/2}\right) \mathcal{I}_{(A_1, A_{\ell-3})}(q),
\ee  
where $D_{p}(SU(2); [2])  = (A_1, A_{p-3})$ and $D_p(SU(2))  = (A_1,D_p)$.

The character identity  in \eqref{usefulEq2} involves three VOAs. It is natural to ask when we can    simplify and  establish   identities between a pair of VOAs or SCFTs.

For this purpose, we use  the following lemma. 
The following  identity holds
  \be\label{IDgsf}
\mathcal I_{D_{\ell p}(G,f)}(q,\bm x)
=
\mathcal I_{D_p(G)}
\left(q^\ell,\bm xq^{\sfh_f}\right), \qquad \ell,p\ge2,  \quad\gcd(\ell p,h^\vee r^\vee)=1 ,
\ee 
with generic surviving flavor fugacities, if and only if
\be 
R_j\cong R_{\ell-1-j},
\quad \forall j.
\label{Rjref}
\ee
Here $R_j$ is given  in the decomposition of adjoint of $\fkg$ in \eqref{gw-eq-decomposition}. 
An absent $R_j$ is understood to be zero.

The relation in \eqref{Rjref} is equivalent to 
\be\label{AArel}
A_f(q,\bm x)
=
q^{\ell+1}A_f(q^{-1},\bm x),
\ee
where $A_f$ is defined in \eqref{ABfeq}.

To prove \eqref{IDgsf}, we first note that \eqref{gw-eq-XY39} can be   rewritten as 
  \be 
\chi_{\mathcal W_{-h^\vee+\frac{h^\vee}{\ell}}(\mathfrak g,f)}(q,\bm x)
 =\PE\!\left[
 \frac{A_f(q,\bm x) -q^{\ell +1}A_f(q^{-1},\bm x)}
 {(1-q)(1-q^\ell)}\right],  
\ee 
which equals 1 if \eqref{AArel} holds.  Further applying  \eqref{usefulEq2} gives
  \be 
  \chi_{\mathcal W_{-h^\vee+\frac{h^\vee}{ \ell p }}(\mathfrak g,f)}(q,\bm x)
 = 
 \chi_{L_{-h^\vee+\frac{h^\vee}{ p   }}(\mathfrak g)}(q^\ell,\bm x q^{\sfh_f}).
\ee
This then proves the  sufficient direction. The necessary condition can be shown similarly. 
This   establishes the lemma
 \be\label{fsfsflema}
\mathcal I_{D_{\ell p}(G,f)}(q,\bm x)
=\mathcal I_{D_p(G)}
\left(q^\ell,\bm xq^{\sfh_f}\right)\qquad \Leftrightarrow \qquad
R_j\cong R_{\ell-1-j}
\quad \forall j.
\ee

\subsection{Principal case: trivial puncture}
\label{gw-subsec-principal}

For a principal  nilpotent element $f$, corresponding to the completely closed trivial regular puncture, the character is given by  \eqref{chiprin}
\be \label{chiprin2}
 \chi_{\mathcal W_{-h^\vee+h^\vee/\ell}(\mathfrak g,f_\prin)}(q )
 =\PE\!\left[
 \frac{\sum_{i=1}^r (q^{d_i}-q^{\ell+1-d_i }) }
 {(1-q)(1-q^{ \ell})}\right].
\ee
For ordered Casimir degrees $d_i$, they satisfy $d_i+d_{r+1-i}=h+2$. This means that if we choose $\ell=h+1$, then  the numerator  above vanishes.  Therefore
\be
 \chi_{\mathcal W_{-h^\vee+h^\vee/ (h+1)}(\mathfrak g,f_\prin)}(q )=1,
 \qquad
\mathcal W_{-h^\vee+h^\vee/ (h+1)}(\mathfrak g,f_\prin) \cong\mathbb C.
 \label{gw-eq-principal-seed}
\ee
This comes from the fact that the trivial VOA has only  one state given by the vacuum.  

 Combining  \eqref{gw-eq-principal-seed} with  \eqref{usefulEq2} gives the identity
\be 
  \chi_{  W_{-h^\vee+\frac{h^\vee}{ (h+1) p }}(\mathfrak g,f_\prin)}(q)
 =  \chi_{  L_{-h^\vee+\frac{h^\vee}{   p }}(\mathfrak g )}(q^{h+1},q^{\rho^\vee}),\qquad
 \gcd(p,h^\vee r^\vee)=1.
  \label{gw-eq-principal-general}
\ee

In particular, the condition in \eqref{gw-eq-ratio-hypotheses} is simplified as indicated above, because $\gcd(h+1,h^\vee r^\vee)=1$, which can be easily checked using the data in  Table~\ref{Liedata}. 

The effective central charge provides a further consistency check via  \eqref{ceffprins}. 
     \beqn
  \ceff_{    W_{-h^\vee+\frac{h^\vee}{ (h+1) p }}(\mathfrak g,f_\prin) }&=&(\frac{1}{h+1}-\frac{1}{p(h+1)})\dim \fkg
  \\&=&
  \frac{1}{(h+1)}(1-\frac{1}{p })\dim \fkg
  =  \frac{1}{(h+1)}\ceff_{  L_{-h^\vee+\frac{h^\vee}{   p }}(\mathfrak g )}.
  \eeqn

 \begin{table}
\centering
\begin{tabular}{c|c|c|c|c|c|c|l} \hline
$\fkg$ & $\dim \mathfrak{g}$ & $r$ & $h$ & $h^\vee$ & $r^\vee$ & $\ell=h+1$ & Casimir degrees $d_i$ \\\hline
$A_{N-1}$ & $N^2-1$ & $N-1$ & $N$ & $N$ & $1$ & $N+1$ & $2, 3, \dots, N$\\
$B_n$ & $2n^2+n$ & $n$ & $2n$ & $2n-1$ & $2$ & $2n+1$ & $2, 4, \dots, 2n$\\
$C_n$ & $2n^2+n$ & $n$ & $2n$ & $n+1$ & $2$ & $2n+1$ & $2, 4, \dots, 2n$\\
$D_n$ & $2n^2-n$ & $n$ & $2n-2$ & $2n-2$ & $1$ & $2n-1$ & $2, 4, \dots, 2n-2, n$\\
$E_6$ & $78$ & $6$ & $12$ & $12$ & $1$ & $13$ & $2, 5, 6, 8, 9, 12$\\
$E_7$ & $133$ & $7$ & $18$ & $18$ & $1$ & $19$ & $2, 6, 8, 10, 12, 14, 18$\\
$E_8$ & $248$ & $8$ & $30$ & $30$ & $1$ & $31$ & $2, 8, 12, 14, 18, 20, 24, 30$\\
$F_4$ & $52$ & $4$ & $12$ & $9$ & $2$ & $13$ & $2, 6, 8, 12$\\
$G_2$ & $14$ & $2$ & $6$ & $4$ & $3$ & $7$ & $2, 6$ \\\hline
\end{tabular} \caption{Lie algebra data. $\dim \mathfrak{g}$, $r$, $r^\vee$, $h$, $h^\vee$ denote the dimension, rank, lacing number, Coxeter number, and dual Coxeter number, respectively. Note that $\dim\fkg =(h+1) r$. We restrict to $A_{N-1}: N \ge 2,   B_n, C_n: n \ge 2,   D_n: n \ge 4$.}
\label{Liedata}
\end{table}

For simply laced $G$, $\rho^\vee=\rho, r^\vee=1$, \footnote{ 
The Weyl vector $\rho$ and dual Weyl vector are related via $\rho^\vee = \rho + ( r^\vee - 1)\rho_{\mathrm{short}} 
$, where $\rho_{\mathrm{short}} = \frac{1}{2} \sum_{\alpha>0,\mathrm{short}} \alpha.
$}
the character identity translates  into an index identity between a pair of untwisted
AD theories
\be
 \mathcal I_{D_p(G)}(q^{h+1},q^\rho)
 =\mathcal I_{(G,A_{(h+1)(p-1)})}(q),\qquad \gcd(p,h)=1.
 \label{gw-eq-ADE}
\ee
For non-simply-laced types, \eqref{gw-eq-principal-general} is the
unambiguous VOA statement.  A four-dimensional interpretation also
requires the twisted  puncture conventions to be specified.


\subsection{Non-principal   case: non-trivial puncture}

We now switch to the non-principal nilpotent element  case, and try to relate its character with that of an affine vertex algebra. 
 In  the 4d SCFT, this corresponds to a non-trivial regular puncture, which can contribute to the flavor symmetry.

\subsubsection{$\fks\fkl_N$}

 We start with the    case of $\fkg=\fks\fkl_N$.  This case is particularly simple, and we can represent the regular puncture using the partition of $N$. For example, $[N]$ corresponds to the principal nilpotent element that we considered in the previous subsection. Now, we want  to consider more   general   partitions.   We will not  perform a systematic classification, but only focus on   selected  families.

Our strategy is to search for cases      where one of the factors on the RHS of \eqref{usefulEq2} is trivial, such that the resulting character identity only involves two VOAs.

For this purpose, we use the result in \cite[Theorem~8.7]{Arakawa:2021ogm}:  
if $\ell\ge 2, s=1 $ or $\ell-1$, then 
\be\label{Rjref22}
\mathcal{W}_{-N+N/\ell}(\mathfrak{sl}_N, [\ell^m, s]) \cong \mathbb{C}, \qquad
\chi_{\mathcal{W}_{-N+N/\ell}(\mathfrak{sl}_N, f) }=1,
\ee
where $N=\ell m+s$. 
Here $N\ge 2, m\ge 0$, and we have $N\equiv \pm 1 \mod \ell$.
The triviality of character can be verified using \eqref{Rjref}. 

Combining \eqref{Rjref22} with  \eqref{usefulEq2} gives
  \be
 \chi_{\mathcal{W}_{-N+\frac{N }{\ell p}}(\mathfrak{sl}_N, f=[\ell^m, s]) }(q,\bm x)
 =\ 
 {\chi_{L_{-N+N/p}(\mathfrak{sl}_N)}} (q^\ell,\bm x q^{\sfh_f}). 
\ee

This gives the following identity  between the flavored indices of the two SCFTs
\be \label{Idsfs}
\mathcal{I}_{D_{\ell p}(SU(N),f=[\ell^m, s])}(q; \bm{x}) = \mathcal{I}_{D_p(SU(N))}(q^\ell; \bm{x}q^{\sfh_{f}}).
\ee
Note that the theory $D_{\ell p} (SU(N), [\ell^m,s])$ on the LHS has symmetry   $SU(m)\times U(1) $ for $m\ge 1$, which may get further enhancement.

The effective central charge provides a further consistency check.  In particular, for $s=1$, we find that 
\beqn
c_{D_{\ell p} (SU(N), [\ell^m,1])} &=& 
\frac{p-1}{12}
\left[
\ell(N^2-1)-\frac{Nm(\ell^2-1)}{p}
\right],\\[2mm]
a_{D_{\ell p} (SU(N), [\ell^m,1])} &=&c_{D_{\ell p} (SU(N), [\ell^m,1])}-\frac{(p-1)m(N+1)}{48p}.
\eeqn
Therefore the effective central charge is
\be 
\ceff_{D_{\ell p} (SU(N), [\ell^m,1])}= \frac{(p - 1)(N^2 - 1)}{ \ell p} 
=\frac{1}{\ell}\ceff_{D_{  p} (SU(N) )},
\ee
where we used \eqref{DpGceff}. For $s=\ell-1$, one can show that the effective  central charge is the same, which gives the same relation with the parent theory.

If we choose  $\ell=N+1, m=0, s=\ell-1=N$,  the nilpotent element on the LHS becomes the  principal  one, and we recover the previous identity \eqref{DpGA} between principal $W$ algebra and affine algebra. 

 Since $\ell m+s=N$ and $s=1$ or $\ell-1$, it is easy to see that $\ell |(N-1)$ or $\ell|(N+1)$, depending on the value of $s$. Therefore, there are several values of $\ell$, and we could  establish the index identity for the   family of theories $ D_{\ell p}(SU(N),f=[\ell^m, s])$ with $s\in \{1,\ell-1\}$, labelled by $\ell$. Two members in the family are related by
\be\label{Iddll}
\mathcal{I}_{D_{\ell  p}(SU(N), [\ell^m,s ])}(q; \bm x) 
=\mathcal{I}_{D_{\hat\ell p}(SU(N), [  \hat\ell  ^{\hat m},\hat s ])}(q^{ \ell/\hat\ell}; \widehat {\bm x})  , 
\qquad \ell  m+s=\hat\ell\hat  m+\hat s,
\ee
where the flavor fugacities $\widehat {\bm x}$ and $\bm x$ are related in a specific way, and the identity holds when the two sides share a compatible  specialization of flavor symmetry.

In particular, in \eqref{Iddll} we can choose $\ell=N+1, m=0, s=\ell-1=N $, which implies that the nilpotent element is principal on the LHS. 
This   gives the following index identity
 \be\label{Idds}
\mathcal{I}_{D_{n  p}(SU(N),f_\prin)}(q)
=\mathcal{I}_{D_{  p}(SU(N), [  t  ^{  n-1}, t-1 ])}(q^{ n};  \{q^{j-n/2} \}_{j=1}^{n-1}; 1),
\ee 
where $p\geq t$,   $\gcd(p,N)=1$, $tn= N+1 $ and we have also provided the explicit flavor fugacity map.  Further setting $t=1$ yields a trivial nilpotent element on the RHS, and we again recover the identity in \eqref{DpGA}.

If we set $n=2$ in \eqref{Idds}, then 
\be\label{prinsbu}
\mathcal{I}_{D_{2  p}(SU(N),f_\prin)}(q)
=\mathcal{I}_{D_{  p}(SU(N), [  \frac{N+1}{2}, \frac{N-1}{2} ])}(q^{ 2 }; 1).
\ee

In a more standard notation, this is just 
\be
\mathcal{I}_{(A_{2m}, A_{2n})}(q) = \mathcal{I}_{D_{m+n+1}(\mathfrak{sl}_{2m+1}, [m+1, m])}(q^2; 1)
= \mathcal{I}_{D_{(m+n+1)/(m+1)}(\mathfrak{sl}_{2m+1})}(q^{2m+2} ; q^{i-(m +1) }),
\ee
for $m,n\in \bZ_+$ and  $\gcd(2m+1,2n+1)=1$, where we also include the identity in \eqref{DpGA} with the extra condition   $ (m+1)\mid n$.

Since the LHS is invariant under the exchange of $m,n$, the first expression
on the right should share this symmetry. In fact,  using \eqref{gw-eq-XY39}, one can show that the symmetry persists with the flavor
 fugacity turned on:   
 \beqn
\mathcal{I}_{D_{m+n+1}(\mathfrak{sl}_{2m+1}, [m+1, m])}(q ; w)
&=&\mathcal{I}_{D_{m+n+1}(\mathfrak{sl}_{2n+1}, [n+1, n])}(q ; w)
\\&=&
\operatorname{PE} \left[ \frac{q(1 - q^m)(1 - q^n) \left[ 1 + q + (w + w^{-1})q^{1/2} \right]}{(1 - q)^2(1 - q^{m+n+1})} \right],
\qquad\qquad
\eeqn 
where $w$ is the $U(1)$ fugacity and $\gcd(2m+1,2n+1)=1$.

Similarly, if we consider $s=1$ in \eqref{Idsfs}, we can also derive
 \be
 \mathcal I_{D_{np}(SU(N),[N-1,1])}(q,w^N)=\mathcal I_{D_p(SU(N),[t^n,1])}
\Big(q^n;\{y_j=wq^{j-(n+1)/2}\}_{j=1}^n;w^{-(N-1)}\Big),
 \ee  
 where $tn=N-1$,  $p\geq t$ and $\gcd(p,N)=1$.   If $n=2$, then  $N=2t+1$ and 
  \be
\mathcal{I}_{D_{ 2 p}(SU(N),f_\text {subreg}=[N-1,1])}(q,w^N)
=\mathcal{I}_{D_{  p}(SU(N), [  t  ^{  2 },  1 ])}(q^{ 2}; wq^{-1/2},wq^{1/2};w^{-(N-1)}).
\ee

Let us be more concrete and take $N=3, p= 2, n=2,t=1, f=[2,1]$. This then gives
  \be\label{nonadmp2}
\cI_{(A_1,A_5)}(q, w^3)=\cI_{(A_1,D_4)}(q^2;w q^{\frac12},w q^{-\frac12};w  ^{-2}),
\ee  where we have used the identifications
\be
D_4(SU(3),[2,1])=(A_1,A_5),\qquad
D_2(SU(3),[1,1,1])=(A_1,D_4).
\ee
 It turns out this can be generalized further, as we will discuss in  subsection~\ref{nonadm}.

\subsubsection{Orthogonal   $\fks\fko_N$ }
 
There is another useful result proved in \cite[Theorem~9.14(1)]{Arakawa:2021ogm}: The following VOA is trivial
\be
\mathcal W_{-h^\vee+\frac{h^\vee}{ \ell   }}(\fks\fko_N,f)=\bC, 
\ee
for an odd integer $ \ell\in 2\bZ+1  $  coprime to $h^\vee$ and a nilpotent element  $f$ corresponding to partition $[\ell^m,\varepsilon ] $ with  $  \varepsilon\in\{0,1\}$. Here $N= \ell m+ \varepsilon, \ell \ge 3$, and a zero part is omitted.

This gives $  \chi_{\mathcal W_{-h^\vee+\frac{h^\vee}{ \ell   }}(\fks\fko_N,f)}=1$. Combining  this with  \eqref{usefulEq2} gives
 \be 
  \chi_{\mathcal W_{-h^\vee+\frac{h^\vee}{ \ell p }}(\fks\fko_N,f)}(q,\bm x)
 = 
 \chi_{L_{-h^\vee+\frac{h^\vee}{ p   }}(\fks\fko_N)}(q^\ell,\bm x q^{\sfh_f}).
 \label{usefulEq22}
\ee 
 
For even $ N$, this gives untwisted type-$ D$ AD index identities:
\be 
\mathcal I_{D_{\ell p}(SO(N),f)}(q,\bm x)
=
\mathcal I_{D_p(SO(N))}
\left(q^\ell,\bm xq^{\sfh_f}\right),
 \ee
for $ f=[\ell^m,\varepsilon]$ with $ \varepsilon\in \{0,1\}, \; \ell\in 2\bN+1$.   

\subsubsection{Symplectic   $\fks\fkp_N$  }

There is another useful result proved in \cite[Theorem~9.7(1)]{Arakawa:2021ogm}: The following VOA is trivial
\be
\mathcal W_{-h^\vee+\frac{h^\vee}{ \ell   }}(\fks\fkp_N,f)=\bC,  \qquad  h^\vee_{\mathfrak{sp}_N}=\frac{N}{2}+1,
\ee
for an odd integer $ \ell  $  coprime to $h^\vee_{\mathfrak{sp}_N}$ and nilpotent element  $f=[\ell^m,\varepsilon ] $ with  $  \varepsilon=0,\ell-1$. Note that both $N=\ell m+\varepsilon$ and $m$ are   even. \footnote{In this notation, $\mathfrak{sp}_N $ has rank $N/2$, so $\mathfrak{sp}_2 =\mathfrak{sl}_2$. }

This gives $  \chi_{\mathcal W_{-h^\vee+\frac{h^\vee}{ \ell   }}(\fks\fkp_N,f)}=1$. Combining  this with  \eqref{usefulEq2} gives
 \be 
  \chi_{\mathcal W_{-h^\vee+\frac{h^\vee}{ \ell p }}(\fks\fkp_N,f)}(q,\bm x)
 = 
 \chi_{L_{-h^\vee+\frac{h^\vee}{ p   }}(\fks\fkp_N)}(q^\ell,\bm x q^{\sfh_f}),
 \label{usefulEq22sp}
\ee
for $f=[\ell^m,\varepsilon ] $ with  $  \varepsilon=0,\ell-1$.
 
\subsubsection{Exceptional   $E_7$}

We also study the exceptional case, and use  
 \cite[Theorem~10.6]{Arakawa:2021ogm}: 
 \footnote{There is another relation in  \cite[Theorem~10.6]{Arakawa:2021ogm}:  $W_{-18+18/19}(E_7, E_7) \cong \mathbb{C}$, which however gives   \eqref{DpGA} with $G=E_7$. }
 \be
\mathcal{W}_{-18+18/7}(E_7, A_6) \cong \mathbb{C},\qquad
\chi_{\mathcal{W}_{-18+18/7}(E_7, A_6)}=1,
\ee
where $A_6$ denotes the Bala–Carter label of the nilpotent orbit in   $E_7$.

Applying this to \eqref{usefulEq2}, we get the character identity 
  \be
\chi_{\mathcal W_{-18+\frac{18}{ 7 p }}(E_7,A_6)}(q,\bm x)
 = 
  \chi_{L_{-18+\frac{18}{p}}(E_7)}(q^7,\bm x q^{\sfh_{  A_6}}).
\ee 
This translates to  the following index identity 
\be
\mathcal{I}_{D_{7p}(E_7, A_6)}(q; y) = \mathcal{I}_{D_p(E_7)} \left(q^7; y q^{\sfh_{A_6}}\right), \qquad \gcd(p, 18) = 1.
\ee

\subsection{A non-boundary admissible family  }\label{nonadm22}
We now consider the following pair of VOAs 
\be 
V_{t,p}=\mathcal W_{-3+t/p}(\mathfrak{sl}_3,f_{[2,1]}), 
\qquad
W_{t, p}=\mathcal W_{-3+t/(2p)}(\mathfrak{sl}_3,f_{\mathrm{prin}})=W_3(t,2p),
\ee
which correspond to the Bershadsky--Polyakov algebra and $W_3$ minimal model, respectively. We also impose the following constraints on the parameters
\be \label{tpsfsdf}
3\le t\in 2\bZ+1 ,\qquad p\ge2,\qquad \gcd(t,p)=1.
\ee 
Since $t$ can differ from 3, these $W$ algebras are admissible, but not necessarily boundary admissible.
Their central charges   are 
\be
c_{V_{t, p}}=25 - \frac{6t }{p} -\frac{ 24p}{t},
\qquad
c_{W_{t, p}}=50 - \frac{12t }{p} -\frac{ 48p}{t}.
\ee
Obviously, we have the relation 
\be
c_{W_{t, p}}=2c_{V_{t, p}}.
\ee
This naturally suggests a possible character identity  relating the two following   \eqref{ceffecL}. 
We  claim that  the following identity  holds
 \be
 \chi_{V_{t,p}}(q^2,1)=\chi_{W_{t, p}}(q).
 \ee
 This can be proved   if we use the following character formulae  \footnote{The vacuum character of the Bershadsky--Polyakov algebra can be derived from the admissible affine character formula in \cite{KacWakimoto1989}, together with the minimal-reduction character theorem in \cite{ArakawaMinimalReduction}. The corresponding principal $W$-algebra character formula is
given in \cite{FrenkelKacWakimoto1992}. Independently, we have performed extensive checks of these character formulae
through high-order $q$-expansions for a wide range of parameters, including comparisons with independent known character expressions 
 and, where applicable, checks against MLDEs and flavored MLDEs.} 
  \be  
\chi_{V_{t,p}}(q,w)
 =\frac{\mathcal N_{t,p}(q^{1/2},w)}
 {(q;q)_\infty^2(wq^{1/2};q)_\infty
 (w^{-1}q^{1/2};q)_\infty}, 
\qquad
\chi_{W_3(t,2p)}(q)
=
\frac{\mathcal N_{t ,p}(q,1)}
{(q;q)_\infty^2},
\ee 
where
\beqn
\mathcal N_{t,p}(\xi,w)
&=&
\sum_{m,n\in\mathbb Z}
\xi^{2tp(m^2+n^2-mn)-t(m+n)}w^{t(n-m)}
  \Bigl[  \xi^{2p(m+n)}-\xi^{4-2p(m+n)}
\nonumber \\ &&\qquad
-w\xi^{1+2p(-m+2n)}
-w^{-1}\xi^{1+2p(2m-n)}
 +w\xi^{3+2p(-2m+n)}
+w^{-1}\xi^{3+2p(m-2n)}
  \Bigr].\nonumber \\
    \eeqn

  Under the conditions \eqref{tpsfsdf}, the identity therefore provides a family of character identities beyond the  boundary admissible case.

    To make contact with 4d AD theories, we   specialize to the  case $t=3$ and $\gcd(3,p)=1$, which corresponds to the boundary admissible case.
We   also   take $p>2$ in order to get a negative 2d and hence a positive 4d   central charge. 
\footnote{For $p=2,t=3$, the two VOAs degenerate  to $\bC$ with trivial character 1. }

Then, the two AD theories are given by 
\be 
\mathcal T_p
=
D_p(SU(3),[2,1])
=
\bigl(A_2^{\,3}[p-3],[2,1]\bigr) \qquad\leftrightarrow \qquad 
(A_2,A_{2p-4}).
\ee
The first theory   is obtained from  $ D_p(SU(3))$ by partially closing the full puncture to $ [2,1]$.
Their  associated VOAs are
\be  
\mathbb V[\mathcal T_p]
=
\mathcal W_{-3+3/p}(\mathfrak{sl}_3,f_{[2,1]}) =V_{3,p},
\ee
and
\be  
\mathbb V[(A_2,A_{2p-4})]
=
\mathcal W_{-3+3/(2p)}(\mathfrak{sl}_3,f_{\mathrm{prin}})
=
W_3(3,2p). 
\ee

Under the $[2,1]$ embedding,  the adjoint decomposes as 
\be 
 \mathfrak{sl}_3
\cong
(V_1)_0\oplus (V_0)_0\oplus (V_{1/2})_1 \oplus (V_{1/2})_{-1},
\ee
where the outer subscripts denote  the U(1) charges. 
The boundary-admissible index formula then gives
  \be  
 {\mathcal I}_{\mathcal T_p}(q,w)
=
\operatorname{PE}\!\left[
\frac{
q+q^2+(w+w^{-1})q^{3/2}
-q^{p-1}-q^p-(w+w^{-1})q^{p-1/2}
}
{(1-q)(1-q^p)}
\right], 
\ee  
which can also be written as
  \beqn
\mathcal I_{\mathcal T_p}(q,w)
&=&\operatorname{PE}\!\left[
\frac{\bigl(q+(w+w^{-1})q^{3/2}+q^2\bigr)(1-q^{p-2})}
{(1-q)(1-q^p)}\right]\nonumber\\
&=&\operatorname{PE}\!\left[
\frac{\bigl(q+(w+w^{-1})q^{3/2}+q^2\bigr)
\sum_{j=0}^{p-3}q^j}{1-q^p}\right].
\eeqn

 Comparing with \eqref{AkANIdx}, we obtain 
\beqn\label{BPfsfsf}
\mathcal I_{D_p(\mathfrak{sl}_3,[2,1])}(q^2,1) &=& \operatorname{PE}\!\left[
\frac{
(q^2+q^3)(1-q^{2p-4})
}
{(1-q)(1-q^{2p})}
\right]
=
\mathcal I_{(A_2,A_{2p-4})}(q) .
\eeqn
 
 This actually corresponds to the  identity  \eqref{prinsbu} with $N=3$.
   
   \subsection{A  non-admissible  family}\label{nonadm}
   We now consider another  family whose natural $\fks\fkl_2$ levels are non-admissible.
    The pair of AD theories is given by 
\be   
(A_1,D_{2p}) =D_{2p}(SU(2), [1,1])\longleftrightarrow D_{4p}(SU(2),[2])=(A_1,A_{4p-3}),
\qquad p\ge2.
\ee
Since $D_{2p}(SU(2))$ and $ D_{4p}(SU(2))$ violate  the coprimality condition, their associated VOAs
are not given by the boundary-admissible $\fks\fkl_2$ formulas.

The associated VOAs are identified in    \cite{Creutzig:2017qyf,Adamovic:2020lvj}
\be 
\mathbb V[(A_1,D_{2p})]=\mathcal R^{(p)},
\qquad
\mathbb V[(A_1,A_{4p-3})]=\mathcal B^{(2p)}.
\ee
Interestingly, they can nevertheless be realized as boundary admissible $W$-algebras: 
\be 
\mathcal R^{(p)}
\cong 
\mathcal W_{-(p+1)+\frac{p+1}{p}}
\bigl(\mathfrak{sl}_{p+1},[p-1,1,1]\bigr),
\ee
and
\be 
\mathcal B^{(2p)}
\cong
\mathcal W_{-(2p-1)+\frac{ 2p-1 }{2p}}
\bigl(\mathfrak{sl}_{2p-1},[2p-2,1]\bigr).
\ee
Note the two VOAs arise from   DS reduction of different parent affine vertex algebras. 

The 4d central charges of the two AD theories are 
\beqn
a_{(A_1,D_{2p})}&=&\frac{6p-5}{12},\qquad\qquad\quad
c_{(A_1,D_{2p})}=\frac{3p-2}{6}, \\
a_{(A_1,A_{4p-3})}&=&p-\frac{29}{24}+\frac{1}{4p},\qquad
c_{(A_1,A_{4p-3})}=p-\frac76+\frac{1}{4p}.
\eeqn

The effective central charges are 
\be\label{ceffAD}
\ceff_{(A_1, D_{2p})} =4, \qquad  \ceff_{(A_1, A_{4p-3})}=2.
\ee

The flavor symmetry of  $(A_1, A_{4p-3})$ is $U(1)$ for $p\ge 2$. 
The flavor symmetry of $(A_1, D_{2p})$ is $SU(2)\times U(1)$ for $p>2$, and is further enhanced to $SU(3)$ for $p=2$, which is more manifest in terms of the equivalence $D_4(SU(2))=D_2(SU(3))=(A_1,D_4)$.
Note that the $p=2$ case has been  discussed in \eqref{nonadmp2}. 

 The vacuum characters of the two VOAs are given by  
  \be
\begin{aligned}
\mathcal I_{(A_1,D_{2p})}(Q;y,w)
&=\operatorname{PE}\Biggl[
\left(\frac{Q}{1-Q}-\frac{Q^p}{1-Q^p}\right)(y^2+2+y^{-2}) 
 +\frac{Q^{p/2}}{1-Q^p}(w+w^{-1})(y+y^{-1})\Biggr],
\end{aligned}
\ee 
 and
   \be\label{AAfsdfs}
\mathcal I_{(A_1,A_{4p-3})}(q;w)
=
\operatorname{PE}\left[
\frac{1+q}{1-q^{2p}}
\left(
\frac{q-q^{2p-1}}{1-q}
+q^{p-\frac12}(w+w^{-1})
\right)
\right],
\ee 
where $w$ is the fugacity for $ \fku(1)$ and $y$ is the  fugacity for $\fks\fku(2)$. They can be derived from the general formula \eqref{gw-eq-XY39} \cite{Xie:2019zlb}. 

It is easy to check that 
  \be\label{AAfDdfs}
\mathcal I_{(A_1,A_{4p-3})}(q;w)
=
\mathcal I_{(A_1,D_{2p})}
\left(q^2;y=q^{1/2},w\right) ,
\qquad p\ge2.
\ee

This is also consistent with the relation of effective central charges \eqref{ceffAD}. 

 \section{Graded vector space isomorphism for VOAs}\label{GVISOsec}

In the previous sections, we established numerous index identities relating pairs of AD theories with specialized fugacities. On the VOA side, these yield various character identities between affine and  $W$-algebras. Since our character identities involve a non-trivial fugacity map, one should not expect an isomorphism between the two VOAs. Instead, they are related by a graded vector space isomorphism. 

In this section, we first formalize the general aspects of graded vector space isomorphisms. As a proof of concept, we construct such a graded isomorphism for a concrete pair of AD theories. We then discuss the graded vector space isomorphism between the universal algebras of affine and principal  $W$-algebras, as well as their simple quotients up to the lowest null state.
 
\subsection{General aspects}

Let $\Gamma$ be an additive commutative monoid. A
$\Gamma$-graded vector space is a complex vector space $V$
equipped with a direct-sum decomposition
\begin{equation}
    V=\bigoplus_{\gamma\in\Gamma}V_\gamma.
\end{equation}

When $\Gamma\subset\mathbb R$, this grading can be encoded by
a diagonalizable grading operator $K$ satisfying
$     Kv=\gamma v,  v\in V_\gamma $. 
In physics, the operator $K$ corresponds to a symmetry  generator or charge.
Assuming that every graded component is finite-dimensional,
we define the formal graded character by
\begin{equation}
    \chi_{V,K}(q)
    =\operatorname{Tr}_V q^K
    =\sum_{\gamma\in\Gamma}
      \dim(V_\gamma)\,q^\gamma.
\end{equation}

A graded vector-space isomorphism (GVISO) between $V$ and $W$ is
a linear bijection $\Phi:V\xrightarrow{\sim}  W$ such that
\begin{equation}
    \Phi(V_\gamma)=W_\gamma
    \qquad\text{for every }\gamma\in\Gamma.
\end{equation}
When the gradings are encoded by $K_V$ and $K_W$, this is
equivalent to
\begin{equation}\label{PhiKVW}
    \Phi K_V=K_W\Phi.
\end{equation}
Therefore $\Phi$  acts as an intertwiner for the grading generators.
In particular, this implies the following identity  
\be\label{VWKs}
\chi_{V,K_V {   }}(  y)=\chi_{W,K _W{   }}(  y),
\ee
 or equivalently 
 \be
\operatorname{Tr}_{V}  y^{  K_V}
=\operatorname{Tr}_{W}  y^{  K_W}.
\ee

This kind of relation will be referred to as a graded vector space isomorphism (GVISO) and will be denoted as
\be
\Phi: (V,  K_V)\cong_\text{gr}(W,  K_W).
\ee

 Our character identities in the previous sections are exactly of the form in \eqref{VWKs}, where the two vector spaces are now given by  two VOAs graded by  a linear combination of conformal and flavor symmetry generators. This naturally motivates us to find an explicit map $\Phi$  between two VOAs which further  satisfies  \eqref{PhiKVW}. 
 
 Note that a GVISO is weaker than an isomorphism of VOAs, which must preserve additional algebraic structures. A GVISO identifies graded vector spaces and their characters, but
generally does not preserve OPEs or normally ordered products.

 In the rest of this section, we need to translate between operators and modes. The convention is as follows. 
For an operator $O$ with conformal weight $\Delta$, we will use the mode expansion 
\be\label{modExpO}
O(z)=\sum_{n +\Delta  \in \bZ } O_n z^{-n-\Delta}.
\ee
Note that the dimension $\Delta$ can be either integer or half-integer,   independent of the statistics.

\subsection{$(A_1,A_{4p-3})$ and $(A_1,D_{2p})$}\label{GVISOAAD}

In this subsection, we prove the GVISO between $(A_1,A_{4p-3})$ and $(A_1,D_{2p})$ explicitly. 
 
\subsubsection{Decomposition of two VOAs}

We consider the pair of AD theories $(A_1,A_{4p-3})$ and $(A_1,D_{2p})$ with $p\ge 2$.  Their associated VOAs are given by \cite{Creutzig:2017qyf,Adamovic:2020lvj} 
\be \label{VWADs}
\qquad V=\bV[(A_1,D_{2p})]=\cR ^{(p)},\qquad W=\bV[(A_1,A_{4p-3})]=\cB^{(2p)}.
\ee
Note that $(A_1,A_{4p-3})$ has  $U(1)$ flavor symmetry, and $(A_1,D_{2p})$ has flavor symmetry   $SU(2)\times U(1)$ for $p\ge 3$, but gets enhanced to $SU(3)$ at $p=2$.
 
 The central charges of the two VOAs are  
  \be
 c_V=4 - 6p, \qquad c_W=14 - 12p - \frac{3}{p}.
 \ee
 Their strong generators have conformal weights  
 \be
 V: 1, 1, 1, 1, \frac{p}{2}, \frac{p}{2},\frac{p}{2},\frac{p}{2},\qquad  \qquad W: 1, 2, p-\frac12, p-\frac12.
 \ee

 For $p=2$, the VOAs are   given by the affine vertex algebra and  BP algebra, reviewed in Appendix~\ref{BPalg}:
\be
V = \mathcal{R}^{(2)} \cong L_{-3/2}(\mathfrak{sl}_3), \qquad
W = \mathcal{B}^{(4)} \cong \mathcal{W}_{-9/4}(\mathfrak{sl}_3, f_{{\text{subreg}}}) = \mathrm{BP}_{-9/4}.
\ee

We  have established the index identity \eqref{AAfDdfs} for \eqref{VWADs}. This naturally motivates us to find its interpretation at the level of VOAs. Their different central charges already show that they cannot be isomorphic as conformal VOAs. We thus want to find a weaker isomorphism, which can be used to explain \eqref{AAfDdfs}.

Our goal in this subsection is to show that this is indeed possible, and it is realized by the  GVISO below
\be\label{VHWLJ0}
\Phi :
(V,H,a_0)\cong_\text{gr}(W,L_0^W,J_0).
\ee 

In the rest of this subsection, we  construct $\Phi$ and prove the GVISO \eqref{VHWLJ0}. We sketch our procedure below and discuss each step  in more detail in the following subsections. 
 
First, we note that the two VOAs have some useful module decompositions.  They  admit the following module  decompositions    \cite[Eqs.~(27)--(28)]{Adamovic:2020lvj}
\be
V=\bigoplus_{\substack{m\geq0\\\nu=-m,-m+2,\ldots,m}}V_{m,\nu},
\qquad
V_{m,\nu}=\mathbb V_m^{(p)}\otimes\mathcal F_\nu^a,
\label{Vdecmops}
\ee
\be
W=\bigoplus_{\substack{m\geq0\\\nu=-m,-m+2,\ldots,m}}W_{m,\nu},
\qquad
W_{m,\nu}=\mathbb M_m^{(2p)}\otimes\mathcal F_\nu^J,
\label{Wdecmops}
\ee
where $\bV,\bM, \cF$ are affine, Virasoro and Heisenberg modules, respectively, defined
below.

The module decompositions \eqref{Vdecmops} and \eqref{Wdecmops} allow  us to decompose $\Phi=\bigoplus_{m,\nu}\Phi_{m,\nu}$ above as 
\be\label{Phimu}
\Phi_{m,\nu} :
(V_{m,\nu}, H, a_0) \cong_{\mathrm{gr}} (W_{m,\nu}, L_0^W, J_0).
\qquad
\Phi_{m,\nu} a_0 = J_0 \Phi_{m,\nu}.
\qquad
\Phi_{m,\nu} H = L_0^W \Phi_{m,\nu}.
\ee
This turns out to be decomposable further into the two   factors, depending on $m$ and $\nu$, respectively: 
\be  
\varphi_\nu :\left(\mathcal{F}_\nu^a, a_0\right) \cong_{\gr} (\mathcal{F}_{\nu,\mathrm{even}}^J, J_0), \qquad \varphi_\nu a_n = J_{2n} \varphi_\nu,
\label{phinu}
\ee
and
\be
\psi_m: 
\left(\mathbb V_m^{(p)},H_{\mathfrak{sl}_2}\right)
\cong_{\gr}
\left(\mathbb M_m^{(2p)}\otimes\mathcal F^J_{0,\mathrm{odd}},
S_0+N_{\mathrm{odd}}\right).
\label{psim}
\ee
The relevant  charges  are related as follows and will be described in detail later:
\beqn
H&=&2L_0^V+\frac12h_0
=H_{\mathfrak{sl}_2}-\frac p2a_0^2
-p\sum_{n>0}a_{-n}a_n. 
\\
L_0^W&=&
S_0-\frac p2J_0^2-p\sum_{r>0}J_{-r}J_r
=
S_0+N_\text{odd}-\frac p2J_0^2-p\sum_{r\in 2\bZ_{>0}}J_{-r}J_r.
\eeqn

These decompositions make  our strategy for establishing GVISO transparent. 
We will discuss $\varphi_\nu$   \eqref{phinu} in the next subsection \ref{phinusect}, and then      $\psi_m$  \eqref{psim} in subsection~\ref{psimbij}.
Their combination to  $\Phi_{m,\nu}$ \eqref{Phimu} will be discussed in subsection~\ref{Phimusec}. The actual GVISO  $\Phi$ \eqref{VHWLJ0} and its relation to index identity  will  be discussed in subsection~\ref{PhiSec}.

\subsubsection{The  graded isomorphism $\varphi_\nu$} \label{phinusect} 
We establish \eqref{phinu} in this subsection. The two VOAs in \eqref{VWADs}  contain two $U(1)$ currents $a(z)$ and $J(z)$ with levels $-2/p$ and $-1/p$, respectively:
\begin{align*}
a(z) \in V = \mathcal{R}^{(p)}, &\qquad a(z)a(w) \sim -\frac{2/p}{(z - w)^2}, \\[2mm]
J(z) \in W = \mathcal{B}^{(2p)}, &\qquad J(z)J(w) \sim -\frac{1/p}{(z - w)^2}.
\end{align*}
Note that $ \mathcal{R}^{(p)}$ has flavor symmetry $\fks\fkl_2\oplus \fku_1$, which gets enhanced to $\fks\fkl_3$ at $p=2$, and contains affine factors  \cite{Adamovic:2020lvj} 
\be\label{LHRp}
  L_{-2+1/p}(\fks\fkl_2)\otimes \cH \subset  \mathcal{R}^{(p)} .
\ee
Therefore $a$  commutes with the $\mathfrak{sl}_2$ currents, and generates   the Heisenberg algebra $\cH$.

The modes of the two Heisenberg algebras satisfy the commutation relations
\be
[a_r,a_s]=-\frac{2r}{p}\delta_{r+s,0},
\qquad [J_r,J_s]=-\frac r p\delta_{r+s,0}.
\label{eq:heis-brackets}
\ee
For a fixed $\nu$, we define the ground states as
\be\label{Ja0nu}
a_0|\nu\rangle_a=\nu|\nu\rangle_a,
\qquad J_0|\nu\rangle_J=\nu|\nu\rangle_J,
\ee
which are also annihilated by all positive modes
\be\label{aJres}
a_r|\nu\rangle_a=0, \qquad J_r|\nu\rangle_J=0, \qquad r>0.
\ee

The two Fock spaces are then given  by
\be\label{Fnufk}
\mathcal F_\nu^a=\C[a_{-1},a_{-2},\ldots]|\nu\rangle_a,
\qquad
\mathcal F_{\nu,\mathrm{even}}^J
=\C[J_{-2},J_{-4},\ldots]|\nu\rangle_J,
\ee
We can construct a map between the two spaces
\be
\varphi_\nu\!\left(\prod_{n\geq1}a_{-n}^{b_n}|\nu\rangle_a\right)
=\prod_{n\geq1}J_{-2n}^{b_n}|\nu\rangle_J.
\label{eq:varphi}
\ee
This is a basis bijection, with inverse $J_{-2n}\mapsto a_{-n}$.
The ground charges agree, and
\be
[a_n,a_{-r}]=-\frac{2n}{p}\delta_{n,r}
=[J_{2n},J_{-2r}].
\ee
Together with \eqref{aJres}, we thus have
\be
\varphi_\nu a_n=J_{2n}\varphi_\nu\qquad  n\in\Z .
\label{eq:varphi-modes}
\ee
The corresponding oscillators both have degree $2n$ in the
full gradings defined below.

Therefore we have
\be \label{varphinus}
\varphi_\nu :\left(\mathcal{F}_\nu^a, a_0\right) \cong_\text{gr} (\mathcal{F}_{\nu,\mathrm{even}}^J, J_0), \qquad \varphi_\nu a_n = J_{2n} \varphi_\nu. 
\ee

 \subsubsection{The   graded   isomorphism $\psi_m$  }\label{psimbij}

In this subsection, we establish the GVISO
\be
\psi_m: \qquad
\left(\mathbb V_m^{(p)},H_{\mathfrak{sl}_2}\right)
\cong_{\gr}
\left(\mathbb M_m^{(2p)}\otimes\mathcal F^J_{0,\mathrm{odd}},
S_0+N_{\mathrm{odd}}\right).
\ee

\subsection*{Affine module $\mathbb V_m^{(p)}$}

We introduce $ V_{\mathfrak{sl}_2}  =L_{k_{\mathfrak{sl}_2}}(\mathfrak{sl}_2)\subset V $ with level   $k_{\mathfrak{sl}_2}=-2+1/p$, which  is the affine subsector of $\cR^{(p)}$ in \eqref{VWADs}.  The three affine currents are denoted as $e,f,h$.

  We use the conventions for ${\mathfrak{sl}_2}$:
\be
[h,e]=2e,\qquad [h,f]=-2f,\qquad [e,f]=h,
\qquad (h,h)=2,\quad(e,f)=1.
\ee
Let $E_m$ be the irreducible $(m+1)$-dimensional
$\mathfrak{sl}_2$-module. Choose its lowest-weight vector $\ket{u_m}$:
\be\label{efhu}
h_0\ket{u_m}=-m\ket{u_m},\qquad f_0\ket{u_m}=0,
\qquad E_m=\Span\{e_0^s\ket{u_m}:0\leq s\leq m\},
\ee
 subject to
\be
e_0^{m+1}\ket{u_m}=0.
\label{eq:horizontal-null}
\ee


Next, we extend the $\mathfrak{sl}_2$-module to an affine module of  $ V_{\mathfrak{sl}_2} $ by inducing from $E_m$,  on which    all strictly positive current modes act trivially. The resulting module will be denoted as $\mathbb V_m^{(p)}$, \footnote{Note that the corresponding module is simple and the PBW basis has no extra null relation \cite{Adamovic:2025efe}. } and it admits a basis 
\be
UF e_0^s\ket{u_m},\qquad 0\leq s\leq m,
\label{eq:affine-basis}
\ee
where $U$ is an ordered product of  $h_{-n},e_{-n}$ and $F$  is an ordered product of  $f_{-n}$, with   $n\geq1$.

The  $\mathfrak{sl}_2$ affine Sugawara field and its central charge are
\be
T_{\mathfrak{sl}_2}
=\frac{:ef:+:fe:+\tfrac12:hh:}{2({k_{\mathfrak{sl}_2}}+2)},
\qquad c_{\mathfrak{sl}_2}=\frac{3{k_{\mathfrak{sl}_2}}}{{k_{\mathfrak{sl}_2}}+2}=3-6p.
\ee
 
We introduce 
\be\label{Hsl2}
H_{\mathfrak{sl}_2}=2L_0^{\mathfrak{sl}_2}+\frac12h_0,
\ee
where $L_0^{\mathfrak{sl}_2}$ is the Virasoro mode of $T_{\mathfrak{sl}_2}$.

Using \eqref{efhu}, one can show 
\be
H_{\mathfrak{sl}_2}\ket{u_m}=\lambda_m \ket{u_m},
\qquad \lambda_m=\frac{p\,m(m+2)-m}{2},
\label{eq:lambda}
\ee
and
\beqn\label{sl2H1}
[H_{\mathfrak{sl}_2}, e_{-n}] &=& (2n+1)e_{-n}, \quad  
[H_{\mathfrak{sl}_2}, f_{-n}]  = (2n-1)f_{-n}, \quad  
[H_{\mathfrak{sl}_2}, h_{-n}]  = 2n\,h_{-n},\quad n\ge 1, \qquad\quad \\
{}[H_{\mathfrak{sl}_2}, e_0] &=& e_0.
\label{sl2H2}
\eeqn

\subsection*{Virasoro module
$\mathbb M_m^{(2p)}$}

The generalized BP algebra $W$ in  \eqref{VWADs} contains a stress tensor  $T$ with central charge
$c_W=14-12p-3/p$, together with  a  primary current $J$:
\be
J(z)J(w)\sim-\frac{1/p}{(z-w)^2},
\qquad
T(z)J(w)\sim\frac{J(w)}{(z-w)^2}
+\frac{\partial J(w)}{z-w}.
\label{eq:TJ}
\ee
The current $J$ generates a  Heisenberg subalgebra, whose stress tensor $T_J=-\tfrac p2:JJ:$ has central
charge $1$. One can then construct 
\be
S=T-T_J=T+\frac p2:JJ:,
\label{eq:S-construction}
\ee
which satisfies
\be
S(z)J(w)\sim0,\qquad [S_m,J_n]=0.
\label{eq:SJ}
\ee
Since $S$ also commutes with $T_J$, it is a Virasoro field with central charge
\be
c_S=c_W-1=13-12p-\frac3p
=1-\frac{6(2p-1)^2}{2p}=c_{1,2p},
\label{eq:cS}
\ee
which coincides with  the central charge of the nonrational $(1,2p)$ Virasoro algebra.

From \eqref{eq:S-construction}, we have the relation
\be
S_n=L_n^W+\frac p2\sum_{r\in\Z}:J_{n-r}J_r:,
\ee
where $S_n,L_n^W$ are the Virasoro modes of $S$ and $T$.
In particular, the zero mode of $T$ is 
\be
L_0^W= S_0 -\frac p2\sum_{r\in\Z}:J_{-r}J_r:=S_0-\frac p2J_0^2-p\sum_{r>0}J_{-r}J_r.
\label{eq:target-grading}
\ee

 The field $S$ generates the simple Virasoro algebra at central charge \eqref{eq:cS}; its universal vacuum algebra is already simple.  We study its modules generated by highest-weight states~\cite{Adamovic:2025efe}
\be 
\qquad S_n\ket{w_m}=0\ (n>0),\qquad\qquad S_0\ket{w_m}=\lambda_m\ket{w_m},
\label{eq:vir-simple}
\ee
with conformal weight 
\be\label{lambdam}
\lambda_m
=\frac{(2p(m+1)-1)^2-(2p-1)^2}{8p}=\frac{p\,m(m+2)-m}{2},
\ee
which coincides with \eqref{eq:lambda}.

 Acting with all ordered products of negative modes $S_{-n}$ with $n\geq1$ on $\ket{w_m}$ generates the Verma module.  
 Its maximal proper submodule is generated by the singular vector      $P_{m+1}(2p)\ket{w_m}$ for the  singular
polynomial  \cite{Adamovic:2025efe}  \footnote{The polynomial is known  \cite{BENOIT1988517}. For example,  
$P_1=S_{-1},  P_2=S_{-1}^2-2pS_{-2}$.}
\be\label{Prsfs}
P_r(2p)=S_{-1}^r
+\text{terms with fewer than $r$ Virasoro factors}.
\ee
After quotienting out   the corresponding  null submodules,  we obtain the irreducible Virasoro module $\mathbb M_m^{(2p)}$ at central charge $c_{1,2p}$ \cite{Adamovic:2020lvj}.

We can choose the following basis for $\mathbb M_m^{(2p)}$    
\be 
S_{-\mu}S_{-1}^s\ket{w_m},
\qquad \mu_i\geq2,\quad 0\leq s\leq m,
\label{vrsbasis}
\ee
where $S_{-\mu}=S_{-\mu_1}\cdots S_{-\mu_t}$ for the partition $\mu$ subject to 
$\mu_1\geq\cdots\geq\mu_t\geq2$, allowing the empty partition.
 Indeed, if $s=m+1$ in \eqref{vrsbasis}, we can use the null relation $P_{m+1}(2p)\ket{w_m}=0$ to rewrite $S_{-1}^{m+1}\ket{w_m}$ in terms of  
 ordered products $S_{-\nu}$ containing fewer than $m+1$ factors of $S_{-1}$. Repeating this procedure removes all powers $S_{-1}^s$ with $s>m$ and proves that \eqref{vrsbasis} spans the simple quotient.  One can   further justify that the basis is independent and complete by matching the simple module character  and the  generating function of the   spanning set \eqref{vrsbasis}.

\subsection*{Neutral odd-mode Fock space $\mathcal F^J_{0,\mathrm{odd}}$}
For the Heisenberg algebra  \eqref{eq:heis-brackets}, we constructed the even charged  sector $\cF_{\nu,\mathrm{even}}^J$  using the even modes of $J$. Similarly, we can also construct the   neutral odd-mode space and its degree operator
\be\label{OddFock}
\mathcal F^J_{0,\mathrm{odd}}
=\C[J_{-1},J_{-3},\ldots]|0\rangle_J,
\qquad
N_{\mathrm{odd}}
=-p\sum_{n\geq1}J_{-(2n-1)}J_{2n-1},
\ee
where  $ |0\rangle_J$ is annihilated by all the non-negative    modes   $J_r|0\rangle_J=0, r\ge 0$.
As a result,  $J_{-(2n-1)}$ has degree $2n-1$, namely $N_{\mathrm{odd}}J_{-(2n-1)}|0\rangle_J=(2n-1) J_{-(2n-1)}|0\rangle_J$.
 
\subsection*{The $m$-dependent bijection $\psi_m$}
\label{sec:m-map}

On  the affine side, the module $\bV_m^{(p)}$ has a basis 
$ UF e_0^s\ket{u_m} $ given in  \eqref{eq:affine-basis},
where we order the $h,e$-modes in $U$ as
$\cdots,e_{-2},h_{-2},e_{-1},h_{-1}$, followed by the ordered
$f$-modes in $F$, then $e_0^s$ with $0\leq s\leq m$. 

We then construct the following mode dictionary  
 \be 
\begin{array}{c|c|c}\centering
\text{affine mode}&\text{target mode}&\text{degree increment}\\\hline
h_{-n}&S_{-2n}&2n\\
e_{-n}&S_{-(2n+1)}&2n+1\\
f_{-n}&J_{-(2n-1)}&2n-1\\
e_0^s&S_{-1}^s&s 
\end{array} 
\label{eq:dictionary}
\ee
 where $n\geq1,\ 0\leq s\leq m$.
 Using the replacement rule in the  dictionary \eqref{eq:dictionary}, we   can map ordered products $U,F$ from the affine side to ordered products on  the target side,  which will be denoted as $U^S,F^J$. 

We can further define the map between  the affine basis and the tensor-product basis as follows:
\be
\psi_m(UF e_0^s\ket{u_m})
=U^SS_{-1}^s\ket{w_m}\otimes F^J|0\rangle_J.
\label{eq:psi}
\ee
First, it is easy to check that the dictionary  \eqref{eq:dictionary} sends the affine basis \eqref{eq:affine-basis}
bijectively to the Virasoro basis \eqref{vrsbasis} tensored
with the odd Fock basis \eqref{OddFock}.  Its inverse replaces
$S_{-2n},S_{-(2n+1)},J_{-(2n-1)}$ by
$h_{-n},e_{-n},f_{-n}$ and   $S_{-1}^s$ by $e_0^s$. So $\psi_m$ \eqref{eq:psi} is a bijection of bases. 
Second, using \eqref{sl2H1}\eqref{sl2H2}, we see that the degree increment in the last column of \eqref{eq:dictionary}  on the affine side coincides with the increment of $H_{\fks\fkl_2}$. But the degree increment in the last column of \eqref{eq:dictionary}  also equals  the  sum of the conformal weight for $S$ and the degree $N_\text{odd}$ \eqref{OddFock} for $J$. Consequently, this implies that 
\be 
\psi_mH_{\mathfrak{sl}_2}
=(S_0+N_{\mathrm{odd}})\psi_m.
\label{eq:psi-grading}
\ee

Therefore, the map $\psi_m$ is a graded vector-space isomorphism
\be
\left(\mathbb V_m^{(p)},H_{\mathfrak{sl}_2}\right)
\cong_{\gr}
\left(\mathbb M_m^{(2p)}\otimes\mathcal F^J_{0,\mathrm{odd}},
S_0+N_{\mathrm{odd}}\right).
\label{eq:m-isomorphism}
\ee

\subsubsection{The graded isomorphism $ \Phi_{m,\nu} $}\label{Phimusec}


We have just established  \eqref{eq:m-isomorphism},
where  $\mathcal F^J_{0,\mathrm{odd}}$ is the neutral odd sector  for the Heisenberg  algebra, constructed in \eqref{OddFock}, while $\bV_m^{(p)}$ and $\bM_m^{(2p)}$ are the     affine  and  Virasoro modules, respectively.

Now we can combine the odd and even sectors of $\cF_J$ together by defining a map 
\be
\Upsilon_\nu:
\mathcal F^J_{0,\mathrm{odd}}\otimes
\mathcal F^J_{\nu,\mathrm{even}}
\longrightarrow\mathcal F_\nu^J,
\ee
such that
\be
\Upsilon_\nu(P|0\rangle_J\otimes Q|\nu\rangle_J)
=PQ|\nu\rangle_J,
\label{eq:mu}
\ee
where $P$ uses odd negative $J$-modes and $Q$ uses even ones.
Its inverse separates these two sets of modes unambiguously. The charge
$\nu$ occurs once; after gluing together in the way above, the odd modes act on $|\nu\rangle_J$ as
\be
\mathcal F^J_{\nu,\mathrm{odd}}
=\C[J_{-1},J_{-3},\ldots]|\nu\rangle_J.
\ee

We can finally combine $\varphi_\nu$ in   \eqref{phinu} and     $\psi_m$  \eqref{psim}  together by considering: 
\be
\Phi_{m,\nu}
=(\id_{\mathbb M_m^{(2p)}}\otimes\Upsilon_\nu)
\circ(\psi_m\otimes\varphi_\nu).
\label{eq:Phi}
\ee
Explicitly, for $0\leq s\leq m$,   the   affine basis   \eqref{eq:affine-basis}
and the Virasoro basis \eqref{vrsbasis} are related explicitly using the dictionary in   \eqref{eq:dictionary}:
\beqn
  \Phi_{m,\nu}
\Big(UF e_0^s \ket{u_m}\otimes Q(a_{-1},a_{-2},\ldots)|\nu\rangle_a\Big)
 =U^SS_{-1}^s\ket{w_m}\otimes
F^JQ(J_{-2},J_{-4},\ldots)|\nu\rangle_J,\;\qquad\quad
\label{eq:Phi-explicit}
\eeqn  
where $0\leq s\leq m$ and  $Q$ is an arbitrary  polynomial. 
Since each factor in \eqref{eq:Phi} is bijective, the map 
$\Phi_{m,\nu}:V_{m,\nu}\to W_{m,\nu}$ is also bijective.
To establish that  $\Phi_{m,\nu}$ is a GVISO, we also need to verify it acts as  an intertwiner. 


The full   stress tensor of $\cR ^{(p)}$  is
\be
T_V=T_{\mathfrak{sl}_2}+T_a=T_{\mathfrak{sl}_2}-\frac p4:aa:, 
\qquad c_V=4-6p.
\ee
This follows because \eqref{LHRp} is a conformal embedding, so the stress tensor is a composite
of the affine and Heisenberg currents.
With this we have
\be
H=2L_0^V+\frac12h_0
=H_{\mathfrak{sl}_2}-\frac p2a_0^2
-p\sum_{n>0}a_{-n}a_n,
\label{eq:full-source-grading}
\ee
where $H_{\mathfrak{sl}_2}$  is given in \eqref{Hsl2}, and $L_0^V$ is the zero mode of $T_V$.

Combining \eqref{eq:psi-grading}   and \eqref{varphinus} gives 
\beqn
\Phi_{m,\nu}H &=&\Phi_{m,\nu} \Big(H_{\mathfrak{sl}_2}-\frac p2a_0^2
-p\sum_{n>0}a_{-n}a_n\Big) 
\\&= &
  \left(S_0+N_{\mathrm{odd}}-\frac p2J_0^2
-p\sum_{n>0}J_{-2n}J_{2n}\right)\Phi_{m,\nu}
\\&= &
 \left(S_0-\frac p2J_0^2
-p\sum_{r>0}J_{-r}J_r\right)\Phi_{m,\nu},
\eeqn
 where we used the definition of $N_{\text{odd}}$   given in \eqref{OddFock}. Using \eqref{eq:target-grading}, one realizes that the new grading is precisely $L_0^W$. As a result, we have
 \be
 \Phi_{m,\nu}H =L_0^W\Phi_{m,\nu}.\nonumber
 \ee
 
 Therefore, we have the following graded vector-space isomorphism
for   each allowed pair $(m,\nu)$:
\be
\Phi_{m,\nu}:(V_{m,\nu},H,a_0)
\cong_\text{gr}(W_{m,\nu},L_0^W,J_0),
\ee
which satisfies the further property 
$\Phi_{m,\nu}a_n=J_{2n}\Phi_{m,\nu}$ for every $n\in\Z$.

\subsubsection{Character identity from graded isomorphism   }\label{PhiSec}

 For the lowest-degree vectors $v\in V_{m,\nu}$ and $w\in W_{m,\nu}$, we have $Hv=\Delta_{m,\nu}v$ and $L_0^Ww=\Delta_{m,\nu}w$, where 
\be
\Delta_{m,\nu}
=\lambda_m-\frac p2\nu^2
=\frac{p\,m(m+2)-m-p\nu^2}{2}.
\label{eq:Delta-branch}
\ee
On the $W$ side, this follows from \eqref{lambdam}, \eqref{Ja0nu} and \eqref{eq:target-grading}. On the $V$ side, this follows from \eqref{eq:lambda},  \eqref{Ja0nu} and \eqref{eq:full-source-grading}.  
 Thus $\nu$ fixes the charge and $(m,\nu)$ fixes the lowest
weight, while descendants occupy higher weights in the same branch.

Taking the direct sum over the allowed labels gives
\be
\Phi=\bigoplus_{m,\nu}\Phi_{m,\nu}:
(V,H,a_0)\cong_\text{gr}(W,L_0^W,J_0).
\label{eq:global-Phi}
\ee
 
Since $|\nu|\leq m$, we have
$\Delta_{m,\nu}\geq(p-\tfrac12)m$; hence each total weight
receives contributions from finitely many branches.

The GVISO gives the following identity of graded characters, which can be computed
explicitly
  \beqn
&& \label{VWmnu}
\Tr_{V_{m,\nu}}q^Hw^{a_0}
=\Tr_{W_{m,\nu}}q^{L_0^W}w^{J_0}
\\&=& \label{faxsmnu}
w^\nu q^{\Delta_{m,\nu}}
\frac{1 + q + \cdots + q^m}{(q^2; q^2)_\infty^2 (q^3; q^2)_\infty (q; q^2)_\infty}
\\&=& 
w^\nu q^{\Delta_{m,\nu}}
\frac{1-q^{m+1}}{(q;q)_\infty^2},
\label{eq:branch-character}
\eeqn
where $(b;q)_\infty=\prod_{n=0}^\infty(1-bq^n)$. On the $V$ side, $ V_{m,\nu}=\mathbb V_m^{( p)}\otimes\mathcal F_\nu^a $ in \eqref{Vdecmops} admits a  basis   \eqref{eq:affine-basis} tensored with  \eqref{Fnufk}. Therefore in \eqref{faxsmnu} the numerator comes from $e_0^s$ and the denominator accounts for the contribution  of $a_{-n},h_{-n},e_{-n},f_{-n}$, respectively, after using their degrees in \eqref{eq:dictionary}.


The  flavored characters of the two VOAs are defined as
  \be
 \chi_V(Q,x,w)=\operatorname{Tr}_{V}Q^{L_0^V}x^{h_0}w^{a_0},
\qquad \chi_W(q,w)=\operatorname{Tr}_{W}q^{L_0^W}w^{J_0},
\ee 
where $x$ is the $SU(2)$ flavor fugacity and $w$ is the $U(1)$ fugacity. 

 Summing over different sectors in \eqref{VWmnu} thus gives 
\beqn
&&\chi_{V}(q^2,q^{1/2},w)=\chi_{W}(q,w)
\nonumber\\&=&
 \sum_{m,\nu}\Tr_{V_{m,\nu}}q^Hw^{a_0}
=\sum_{m,\nu}\Tr_{W_{m,\nu}}q^{L_0^W}w^{J_0}
\nonumber\\&=&
\operatorname{PE}\!\left[
\frac{q+q^2-q^{2p-1}-q^{2p}
 +(w+w^{-1})(q^{p-1/2}-q^{p+3/2})}
 {(1-q)(1-q^{2p})}\right],
\eeqn
where the last step is derived in Appendix~\ref{idfsfdsdf}.  This agrees with the index  prediction \eqref{AAfDdfs} and  \eqref{AAfsdfs}.

\subsection{$(A_{(h+1)(p-1)},G)$ and $D_p(G)$}
 
 We want to study the GVISO between the  affine VOA and   the corresponding principal
$W$-algebra, which are the associated VOAs of     $D_p(G)$ and $(A_{(h+1)(p-1)},G)$, respectively.
 
 Consider the principal boundary admissible level
\be\label{kvwpsf}
k_V=-h^\vee+\frac{h^\vee}{p},
\qquad
k_W=-h^\vee+\frac{h^\vee}{u},
\qquad
u=p(h+1),
\ee
with
\be
p\ge2,\qquad
\gcd(p,h^\vee)=\gcd(p,r^\vee)=1,
\ee
where $ r^\vee$ is the lacing number.

 Let us use   $ \widehat V,\widehat W$ for  the universal algebras, and $ V,W$ for their simple quotients, namely

\be
\begin{aligned}
\widehat V &=  V^{k_V}(\mathfrak g),
\qquad
&V &=  L_{k_V}(\mathfrak g)
=\widehat V/\mathcal N_V,
\nonumber\\
\widehat W &=  \mathcal W^{k_W}(\mathfrak g,f_{\mathrm{prin}}),
\qquad
&W &= \mathcal W_{k_W}(\mathfrak g,f_{\mathrm{prin}})
=\widehat W/\mathcal N_W,
\end{aligned}
\ee
where $\cN_V,\cN_W$ are the maximal proper ideals of $\widehat V,\widehat W$ formed by all the null states.
 
We consider the mode expansion for affine currents and $W$ generators: 
\be
J^a(z)=\sum_{n\in\mathbb Z}J^a_nz^{-n-1},
\qquad
W^i(z)=\sum_{n\in\mathbb Z}W^i_n z^{-n-d_i}.
\ee
Here $d_i=e_i+1$, where $e_i,d_i$ are the exponents and degrees of the Casimirs of $ \mathfrak g$, respectively. See Table~\ref{Liedata} for explicit values.  

The principal $W$ algebra has stress tensor given by
\be
W^1(z)=T_W(z)=\sum_{m\in\mathbb Z}L_m^W z^{-m-2}. 
\ee
Set
\be \label{Hfsfsf}
\ell=h+1,\qquad
H=\ell L_0^V-J_0^{\rho^\vee},
\ee
where $J_0^{\rho^\vee}$ is the zero mode of the affine Cartan current associated with the dual Weyl vector $\rho^\vee$.

\subsubsection{Universal algebra}
The vacuum conditions imply that the Poincaré--Birkhoff--Witt (PBW) creation modes are
\be
J^a_{-n},\qquad n\ge1, \qquad a=1, \cdots , \dim\fkg,
\ee
and
\be
W^i_{-n},\qquad n\ge d_i, \qquad i=1, \cdots, r.
\ee
 
We now organize the current labels  of $J^a$ into principal $ \mathfrak{sl}_2$ multiplets:
\be
a=(i,t),\qquad -e_i\le t\le e_i, \qquad i=1, \cdots, r,
\ee
so that
\be
[J_0^{\rho^\vee},J_n^{i,t}]=tJ_n^{i,t}.
\ee
Consequently,
\be\label{HJLW}
[H,J^{i,t}_{-n}]= (\ell n-t)J^{i,t}_{-n},
\qquad
[L^W_0,W^i_{-m}]=mW^i_{-m}.
\ee
Every current creator has positive $ H$-degree 
\be
\ell n-t\ge \ell-(h-1)=2,
\ee
because $n\ge 1$ and $t\le e_i\le h-1$.
 
All the exponents can be grouped in pairs such that $e_i +e_{\hat i}=h$. Without loss of generality, we can assume that $e_i \le e_{\hat i}$. Then we can use the following mode map between affine and principal $W$ algebra:
\be \label{JWmap1}
\widehat \Phi: \qquad J^{i,t}_{-n}\longleftrightarrow W^{\hat i}_{-(\ell n-t)},
\qquad  -e_i\le t\le e_i,
\ee
and
\be 
\widehat \Phi: \qquad  J^{\hat i,t}_{-n}\longleftrightarrow
\begin{cases}\label{JWmap2}
W^i_{-( \ell n-t)},&-e_{  i}\le t\le e_{ \hat i},\\
W^{\hat i}_{-(\ell n-t)},&-e_{ \hat i}\le t\le-e_{   i}-1.
\end{cases} 
\ee
 For an unpaired self-dual exponent $ e_i= e_{\hat i}=h/2$ with $i=\hat i$, the rule reduces to 
\be\label{JWmap3}
\widehat \Phi: \qquad  J^{i,r}_{-n}\longleftrightarrow W^i_{-(\ell n-r)},\qquad  -e_i\le r\le e_i.
\ee
This is bijective for a simple reason.  For    fixed $n$ and a pair   $( i , \hat  i)$, as   $t$ varies, the map in \eqref{JWmap2} sends  $J_{-n} $ to  $ W^i_{-k}$ with mode numbers
\be
k: \quad\ell n-e_{ \hat i},\ldots,\ell n+e_{  i}.
\ee
This is a consecutive block of length $ \ell $, because $e_i+e_{  \hat i}+1=h+1=\ell$. At $ n=1$, it starts at
$
\ell -e_{ \hat i} =d_i.
$
Increasing $ n$ by one gives the next consecutive block. Thus every allowed creator
\be
W^i_{-d_i},W^i_{-(d_i+1)},\cdots,
\ee 
appears exactly once.  For $ W^{\hat i}$,   the map in  \eqref{JWmap1}\eqref{JWmap2}  sends  $J_{-n} $ to  $ W^{\hat i}_{-k}$ with mode numbers
\be
k: \quad\ell n-e_{   i},\ldots,\ell n+e_{\hat  i}.
\ee
Again, this is a consecutive block of length $ \ell $. For $n=1$ it starts at $\ell -e_{   i} =d_{ \hat i}$, and it is easy to see that every mode of $ W^{\hat i}_{-k}$ appears exactly once for $k\ge d_{\hat i}$. Using the PBW basis, the map  \eqref{JWmap1}\eqref{JWmap2}\eqref{JWmap3} further allows us to establish a bijection between the two vector spaces   of the affine and $W$ algebras. 

Using the commutator in \eqref{HJLW}, one can also see that \eqref{JWmap1}\eqref{JWmap2}\eqref{JWmap3} preserves the grading, namely 
\be
\widehat\Phi H=   L_0^W\widehat  \Phi.
\ee
Therefore, we have proved the  GVISO for the universal algebras of affine and principal $W$ algebras: 
\be 
\widehat\Phi:
(\widehat V,H)\cong_{\mathrm{gr}}
(\widehat W,L^W_0).
\ee
This is consistent with the following character identity between affine vertex and principal $W$ algebras: 
\be
\operatorname{Tr}_{\widehat V}q^H=
\PE\Big[\frac{\sum_{i=1}^r \sum_{t=-e_i}^{e_i}q^{\ell-t}}{1-q^\ell}\Big]=\PE\Big[ \frac{\sum_{i=1}^r q^{d_i}}{1-q}\Big]
=\operatorname{Tr}_{\widehat W}q^{L_0^W},
\ee
where the $H$-grading on the left can be seen from \eqref{HJLW}.

\subsubsection{Simple quotient  }


Now we want to generalize and establish the GVISO for the simple quotients $V,W$
\be \label{Phifsvs}
 \Phi:
(  V,H)\cong_{\mathrm{gr}}
(  W,L^W_0).
\ee
For this purpose,    we need to account for the null states. The previous map $\widehat \Phi$ defined in \eqref{JWmap1}\eqref{JWmap2}\eqref{JWmap3} holds for the universal algebras where  the  ordered creation-mode monomials
form PBW bases.  Therefore, we need to modify and construct a new $\widehat \Phi^\prime$ such that it preserves the null states, in the sense that
\be\label{fdsfsfHVP}
\widehat\Phi^{\prime}:
(\widehat V,H)\cong _{\mathrm{gr}}(\widehat W,L^W_0),
\qquad
\widehat\Phi^{\prime}(\mathcal N_V)=\mathcal N_W.
\ee
 Linearity implies that 
\be
\widehat\Phi^{\prime}(v+\cN_V)=\widehat\Phi^{\prime}(v)+\mathcal N_W.
\ee
 
Let us  denote by $[v]\in V$ the image of $v\in \widehat V$ under the quotient map, and similarly for $W$. Then we can define
\be\label{Phisfsfsf}
\Phi([v])=[\widehat\Phi^{\prime}(v)]\in   W .
\ee
This map is injective because
\be
\Phi([v])=0
\Longrightarrow
\widehat\Phi^{\prime}(v)\in\mathcal N_W
\Longrightarrow
v\in\mathcal N_V
\Longrightarrow
[v]=0.
\ee
Surjectivity follows from that of $\widehat\Phi^{\prime}$, and grading preservation descends to the quotient. This gives the GVISO \eqref{Phifsvs}.
 The key in this construction is thus the map $\widehat\Phi'$ in \eqref{fdsfsfHVP} for the universal algebras
\be\label{widhefafsffsd}
 \widehat\Phi': \cN_V\to \cN_W,\qquad \dim (\cN_V)_d=\dim (\cN_W)_d,\qquad \forall d,
\ee
where the subscript denotes the degree-$d$ component of the respective null ideal under the $H$- and $L_0^W$-gradings.

The map $\widehat \Phi'$ is guaranteed to exist if the character identity holds, but constructing this explicitly is not straightforward, because a concrete construction requires control of the full
null spaces. We are not able to construct the desired  $\widehat \Phi'$ universally. Nevertheless, below we   show this indeed holds for the lowest null state for all $\fkg$. 
 
On the affine side, the first  null state takes the following form
\be 
\psi_V=
\sum_{a_i,n_i } C^{a_1\cdots a_s}_{n_1\cdots n_s}
J^{a_1}_{-n_1}\cdots J^{a_s}_{-n_s}|0\rangle,
\qquad n_j>0, \qquad
\sum_jn_j=p,
\ee
where $p$ is the denominator in the boundary admissible level \eqref{kvwpsf}. 
This state is a vacuum descendant and, as the generating affine singular vector at this boundary level, is also an affine primary whose horizontal highest weight is the highest root $\theta$ \cite{Arakawa_2014}. It generates an adjoint representation under the affine current zero modes.    
Therefore~\footnote{In the current boundary admissible case, the first null state is associated with the highest weight of the adjoint representation, labelled by $\theta$.  Therefore we have 
\be
L_0^V\psi_V=h_\theta\psi_V, \qquad
J_0^{\rho^\vee}\psi_V
=\langle\theta,\rho^\vee\rangle\psi_V.
\ee
Here 
\be\nonumber
h_\theta=\frac{(\theta, \theta + 2\rho)}{2(k_V + h^\vee)} = \frac{h^\vee}{k_V + h^\vee} = p, \qquad
\langle\theta,\rho^\vee\rangle=h-1. 
\ee
where we used $(\rho, \theta) = \langle \rho, \theta^\vee \rangle = h^\vee - 1,(\theta, \theta) = 2$. 
 }
 \be
L_0^V\psi_V=p\psi_V,
\qquad
J_0^{\rho^\vee}\psi_V=(h-1)\psi_V,
\ee
and
\be \label{Hd0}
H\psi_V=d_0\psi_V,\qquad
d_0=p(h+1)-(h-1)=u-h+1,
\ee
where $u$ is given in \eqref{kvwpsf}.

 Finite lowering-current zero modes  $J_0^{-\alpha}$ generate the adjoint multiplet from $\psi_V$; they preserve the $L_0^V$ weight and increase the $H$ weight. Therefore $\psi_V$ spans the null space of lowest $H$-degree, given by \eqref{Hd0}.

On the $W$ algebra side, the first   null state     has the structure
\footnote{See  \cite[Lemma~3.3]{Creutzig:2020zaj}, where the authors provide  the conformal weights for the lowest null state in both affine and principal $W$ algebras, whose structure exactly agrees with our $L_0^W \leftrightarrow H$ following \eqref{Hfsfsf}.}
\be 
\psi_W=
\sum_{i_j,m_j} B^{i_1\cdots i_s}_{m_1\cdots m_s}
W^{i_1}_{-m_1}\cdots W^{i_s}_{-m_s}|0\rangle,
\qquad
\sum_jm_j=d_0. 
\ee
As a singular vector, this state is annihilated by all positive modes  
$
W^i_m\psi_W=0,   m>0 
$,
which implies that $ \psi_W$ is also  Virasoro primary. 
The state $\psi_W$ has conformal weight $d_0$ and spans the null space at this weight
\be
L_0^W \psi_W=d_0  \psi_W.
\ee

As a result, on both sides  the lowest-degree null spaces are one-dimensional and occur in degree $d_0$
\be
\dim(\mathcal N_V)_{d_0}
=
\dim(\mathcal N_W)_{d_0}
=1.
\ee
This is thus consistent with \eqref{widhefafsffsd}, and gives a natural map up to this degree such that $\widehat \Phi':\psi_V\mapsto \psi_W$.
  
The null spaces become increasingly complicated at higher degrees, so we will not try to establish \eqref{widhefafsffsd} in general.

To be more concrete, let us briefly look at  the simplest example of $\fkg=\fks\fkl_2$.  The two VOAs are given by the $L_{-2+2/p}(\fks\fkl_2)$ affine vertex algebra and Virasoro minimal model $\text{Vir}(2,3p)$. Our   map   in \eqref{JWmap1}\eqref{JWmap2}\eqref{JWmap3} then gives 
\be\label{Jwsl2}
\widehat \Phi:\qquad
J^e_{-n}\longleftrightarrow L^W_{-(3n-1)},\qquad
J^h_{-n}\longleftrightarrow L^W_{-3n},\qquad
J^f_{-n}\longleftrightarrow L^W_{-(3n+1)}.
\ee
 This provides the GVISO for universal algebras. For the simple quotient, $\widehat \Phi$ requires  modification to preserve the null ideals. 
 \footnote{To illustrate, take $p=3$. Then the two null states are 
 \beqn\nonumber
\psi_V&=&\Big[
 (J^h_{-1})^2J^e_{-1}
+4J^f_{-1}(J^e_{-1})^2
+\frac{22}{3}J^h_{-2}J^e_{-1} -\frac{16}{3}J^e_{-2}J^h_{-1}
-\frac{88}{9}J^e_{-3}
\Big] \bm 1_V,
\\ \nonumber
\psi_W&=&\left[
 (L^W_{-2})^4
-\frac{4}{9}(L^W_{-3})^2 L^W_{-2}
-\frac{26}{9}L^W_{-4}(L^W_{-2})^2
+\frac{7}{9}(L^W_{-4})^2 
+\frac{4}{27}L^W_{-5}L^W_{-3}
-\frac{88}{27}L^W_{-6}L^W_{-2}
-\frac{278}{81}L^W_{-8}
\right] \bm 1_W .
\eeqn
There is an immediate obstruction to the literal PBW substitution \eqref{Jwsl2}: $ \psi_V$ contains at most three current factors, whereas the unique target null contains the nonzero four-Virasoro term. Thus the ordered PBW replacement rule  \eqref{Jwsl2} would not  map $ \psi_V$ to a nonzero multiple of $ \psi_W$, and therefore  fails to preserve the null ideals.
}
   In   \cite{Adamovic:2025efe}, the authors studied  the GVISO  of the simple quotients in detail and proved \eqref{Phifsvs} using the  character formula.

 \subsection{$(A_1,A_5)$ and $(A_2,A_4)$}
Our final example concerns 
\be
(A_1,A_5)\longleftrightarrow(A_2,A_4),
\ee
which is the  special case of subsection~\ref{nonadm22} with $p=4$.   
Their associated VOAs and central charges are 
\be
V=\bV[(A_1,A_5)]= {\text{BP}}_{-9/4}
=
\mathcal W_{-9/4}(\mathfrak{sl}_3,f_{\mathrm{subreg}}), \qquad
c_V=-\frac{23}{2},
\ee
and
\be
W =\bV[(A_2,A_4)]=\mathcal W_{-21/8}(\mathfrak{sl}_3,f_{\mathrm{prin}})
=W_3(3,8), \qquad c_W=-23,
\ee
corresponding to the Bershadsky--Polyakov  algebra and the $W_3$ minimal model.

The Bershadsky--Polyakov  algebra $V$ has four bosonic strong generators  $ J,G^\pm,T$ of  conformal weights $ 1,\frac32,\frac32,2$, whose   OPEs      are given in  Appendix~\ref{BPalg}. The $W_3 $ minimal model has strong generators $T_W,U$ of weight $2,3$, where $T_W$ is the stress tensor. Their mode expansions follow the general equation \eqref{modExpO}.

As shown in \eqref{BPfsfsf}, we have the following identity
\be
\chi_V(q^2,1)= {\cI}_{(A_1,A_5)}(q^2,1)= {\cI}_{(A_2,A_4)}(q)=\chi_W(q)
= \mathrm{PE} \left[\frac{q^2+q^3}{(1 - q) (1 + q^4)} \right].
\ee

 This character identity suggests a graded vector-space isomorphism
\be\label{VWFDfsd}
\Phi:(V,2L_0)\cong_{\mathrm{gr}}(W,L^W_0).
\ee

Let us first look at the universal algebras and forget about the null relations. Then the vacuum module   of $\widehat V$ is generated by 
acting with $J_{-n},G^+_{-n-\frac12},G^-_{-n-\frac12},L_{-n}$ on the vacuum state $\bm 1_V$. Similarly,  the vacuum module  of $\widehat W$ is generated by 
acting with $L^W_{-m},U_{-n}$ on the vacuum state $\bm 1_W$.  We can then establish a one-to-one correspondence $\widehat \Phi$ between them via
\beqn
J_{-n} &\longmapsto& L^W_{-2n},\qquad n\ge1,
\nonumber\\
G^+_{-n-\frac12} &\longmapsto& L^W_{-(2n+1)},\qquad n\ge1,
\nonumber\\
G^-_{-n-\frac12} &\longmapsto& U_{-(2n+1)},\qquad n\ge1,
\nonumber\\
L_{-n} &\longmapsto& U_{-2n},\qquad n\ge2,
\label{Phimap}
\eeqn
 together with $\bm 1_V \mapsto \bm 1_W$. 
   This can also be understood from the character identity  between   two universal algebras $\widehat V$ and $\widehat W$
\be
\chi_{ \widehat V}(q^2,1)=\operatorname{PE}\!\left[
\frac{q^2+2q^3+q^4}{1-q^2}
\right] =
\operatorname{PE}\!\left[
\frac{q^2+q^3}{1-q}
\right]= \chi _{\widehat W}(q).
\ee

To establish the   correspondence between   the simple  quotients $V$ and $W$, we need to account for the    null relations. 

On the BP side, the first   vacuum null has conformal weight $ 3$:
\beqn
&& \psi_V=\Bigl( G^+_{-\frac32}G^-_{-\frac32}
+J_{-1}^{\,3}
-3J_{-2}J_{-1}
+\frac54J_{-3} 
 -\frac32J_{-1}L_{-2}
+\frac38L_{-3}
\Bigr)\mathbf1_V.
\label{NullVs}
\eeqn

On the $W_3 $ side,  one   similarly finds a null state at weight 6
\footnote{Here we use the standard normalization $U(z)U(w)\sim\frac{c_W/3}{(z-w)^6}
+\frac{2T_W(w)}{(z-w)^4}
+\frac{\partial T_W(w)}{(z-w)^3} +\cdots$.}
\be 
\psi_W=\frac{1}{558} 
\Bigl(
558U_{-3}^{\,2}
-32(L^W_{-2})^3
+51(L^W_{-3})^2
+168L^W_{-4}L^W_{-2}
-216L^W_{-6}
\Bigr)\mathbf1_W.
 \ee

We can then apply the map \eqref{Phimap} to  \eqref{NullVs} 
\be 
\widehat\Phi(\psi_V)=\Bigl(
 L^W_{-3}U_{-3}
+(L^W_{-2})^3
-3L^W_{-4}L^W_{-2}
+\frac{5}{4}L^W_{-6} 
-\frac{3}{2}L^W_{-2}U_{-4}
+\frac{3}{8}U_{-6}
\Bigr)\mathbf1_W .
\ee
Therefore the resulting state $\widehat \Phi (\psi_V) \not \propto \psi_W$, hence is not null. 
So we need to modify the map $\widehat \Phi $ to a new one $\widehat \Phi ^\prime$ at this order. One choice is given by  
\beqn
\widehat \Phi ^\prime(v) &=&\widehat \Phi (v) , \qquad v\in \text{Span} \Big\{J_{-1}^{\,3} \mathbf{1}_V,   J_{-2}J_{-1} \mathbf{1}_V,   J_{-3} \mathbf{1}_V,   J_{-1}L_{-2} \mathbf{1}_V,   L_{-3} \mathbf{1}_V \Big\},\qquad\qquad\quad
\\
\widehat \Phi ^\prime( G^+_{-\frac32}G^+_{-\frac32}\mathbf1_V) &=&(L_{-3}^W)^2 \mathbf1_W,\qquad
\widehat \Phi ^\prime( G^-_{-\frac32}G^-_{-\frac32}\mathbf1_V) =L_{-3}^W U_{-3}  \mathbf1_W,
\\
\widehat \Phi ^\prime( G^+_{-\frac32}G^-_{-\frac32}\mathbf1_V)
&=&\psi_W -\widehat \Phi \Big( \psi_V- G^+_{-\frac32}G^-_{-\frac32}\mathbf1_V \Big) .
\eeqn
The corrected map satisfies $\widehat\Phi^{\prime}(\psi_V)=\psi_W$. It is also easy to show that this map is invertible. Thus the map identifies the one-dimensional null spaces and induces the claimed GVISO through degree six, by \eqref{Phisfsfsf}. 

We do not give an explicit formula at higher degrees. Existence follows by choosing bases of the null spaces and their complements at each degree: equality of both universal and simple characters implies equality of the corresponding dimensions.
   
\section{Character identities for  modules     } \label{charmodules}
 
 In this section, we generalize the previous discussion of vacuum character identities to non-vacuum modules. On the SCFT side, non-vacuum modules correspond to the insertion of certain BPS surface defects \cite{Cordova:2017mhb}. We will first discuss the module identities in a concrete example involving three AD theories in the next subsection, and then formulate a more general conjecture.
  
  \subsection{A detailed example: $(A_1,A_5)$, $(A_1,D_4)$, $(A_2,A_4)$ } 
  We start by considering the index identities  previously established  for a triple of AD theories. The three AD theories under consideration are $(A_1,A_5)$, $(A_1,D_4)$, $(A_2,A_4)$, and the corresponding index identities are:   
   \footnote{For convenience, in this section, we use a different set of fugacities $(\mathsf{x}_1,\mathsf{x}_2,\mathsf{x}_3)$ for $\fks\fkl_3$, which are related to the previous ones by   $\mathsf{x}_1=x_2/x_3$, $\mathsf{x}_2=x_3/x_1$, and $\mathsf{x}_3=x_1/x_2$. The fundamental character is $\chi_{\mathbf3}=x_1+x_2+x_3$.}
  \beqn\label{AAcase}
   \cI_{(A_2,A_4)}(q) &=& \cI_{(A_1,A_5)}(q^2;1)   ,
\\
  \cI_{(A_1,A_5)}(q;w)&=&  \cI_{(A_1,D_4)}(q^2; w q^{-1/2}, \ w^{-1}q^{-1/2}, q  ), 
  \label{A1a5d4}
  \eeqn 
Together these identities imply 
  \be
  \mathcal I_{(A_2,A_4)}(q)
=\mathcal I_{(A_1,D_4)}(q^4;q^{-1},q^{-1},q^2).
  \ee

  We will use the shorthand notation   $\fks\fkl_3$, BP,  and $W_3$   to denote their associated VOAs:
  \beqn
  \fks\fkl_3 & \equiv& L_{-3/2}(\fks\fkl_3)=\bV[(A_1,D_4)], \\
     {\text{BP}}& \equiv& W_{-9/4}(\fks\fkl_3,f_ {\text{subreg}})=\bV[(A_1,A_5)], \\
  W_3& \equiv& W_{-21/8}(\fks\fkl_3,f_\prin)=W_3(3,8)=\bV[(A_2,A_4)].
  \eeqn
   
  The central charges are given by 
  \be\label{cw3bp}
 c_{W_3}=-23, \qquad   c_{ \text{BP}}=-23/2, \qquad  c_{\fks\fkl_3}=-8.
  \ee
 Note that the simple relation $c_{W_3}=2 c_{ \text{BP}}$   implies that  \eqref{AAcase} can be further  promoted to     an identity for the full characters with the factor of $q^{-c/24}$: $ \widehat \chi^{     \text{BP}}(q^2,1)= \widehat \chi^{    W_3}(q)$. From \eqref{A1a5d4} and using \eqref{cCsJ}, we  find
  \be
\cC_{\mathfrak{sl}_3}
=c_{\mathfrak{sl}_3}-12k_{\mathrm{aff}}(\xi,\xi)
=-8+\frac94=-\frac{23}{4},\qquad
\cC_{\mathrm{BP}}=c_{\mathrm{BP}}=-\frac{23}{2},
 \ee 
where we have used $k_{\mathrm{aff}}=-\frac32$ and    the  following embedding of $U(1)$ into $\fks\fkl_3$ 
\be
\xi=\frac12 \times \frac12(e_1-e_2),\qquad (\xi,\xi)=\frac18.
\ee
 Therefore we have the simple relation 
 \be
 \cC_{\mathrm{BP}}=2\cC_{\mathfrak{sl}_3},
 \ee
in agreement with \eqref{LRcc}.
 
  The effective central charges are
    \be
 \ceff_{W_3}=1, \qquad   \ceff_{\text{BP}}=2, \qquad\ceff_{\fks\fkl_3}=4,
  \ee
 which can be obtained from the 4d central charges using \eqref{ceffac}.    They are   consistent with  {\eqref{LRcc}}, see   \eqref{AAcase} and  \eqref{A1a5d4}.

 %
%
  
\subsubsection{$\fks\fkl_3$ affine modules}
 
 The simple affine vertex algebra $L_{-3/2}(\fks\fkl_3)$ has four  admissible irreducible highest-weight modules. Their finite $\fks\fkl_3$ highest weights are 
\be 
0,\quad-\frac32\omega_1,\quad-\frac32\omega_2,\quad-\frac12\rho,
\ee
where $ \omega_1,\omega_2$ are the   fundamental weights of $\fks\fkl_3$ and $\rho=\omega_1+\omega_2$.
  The conformal weights are $0,-\frac12,-\frac12,-\frac12$, respectively.  

It is convenient  to introduce four vectors as follows:
\be 
 \mathcal{K} = \{(0,0,0), (1,0,-1), (0,1,-1), (1,1,0)\}.
\ee
Then their  corresponding highest weights and conformal weights are given by 
\footnote{Note that we have the relation $
\omega_1=\frac{2\alpha_1+\alpha_2}{3},
 \omega_2=\frac{\alpha_1+2\alpha_2}{3},
 \rho=\alpha_1+\alpha_2
$, and $\alpha_1=e_2-e_3,  \alpha_2=e_3-e_1, \alpha_3=-\alpha_1-\alpha_2=e_1-e_2 $.
} 
\be 
\lambda_{\bm v} = -\frac{1}{2}\sum_{i=1}^3 v_i\alpha_i, \qquad h^{\fks\fkl_3}_{\bm v} = -\frac{1}{4}\sum_i v_i^2,
\ee
 for any $\bm v\in \cK$.

The affine characters of the four modules are given by  \footnote{Note that the flavor fugacity $\sfx$ here uses a different convention. }
\be \label{sl3affc}
\widehat\chi_{\bm v}^{\fks\fkl_3}(q,\boldsymbol{\mathsf{x}})
=q^{\frac13-\frac14\sum_i v_i^2}\prod_{i=1}^3\mathsf{x}_i^{-v_i/2}
\operatorname{PE}\left[\frac{P^{\fks\fkl_3}_{\bm v}(q,\boldsymbol{\mathsf{x}})}{1-q^2}\right],
\ee
where $\mathsf{x}_1\mathsf{x}_2\mathsf{x}_3=1$ and
\be 
P^{\fks\fkl_3}_{\bm v}(q,\boldsymbol{\mathsf{x}})
=2q+\sum_{i=1}^3\left(q^{1+v_i}\mathsf{x}_i+q^{1-v_i}\mathsf{x}_i^{-1}\right).
\ee 
  These characters can be derived from the general  boundary admissible formulae  \cite{KacWakimotoBoundary2017}.

 The unrefined vacuum character also satisfies an MLDE, which is given  by  \cite{Beem:2017ooy} 
  \be
\mathscr L_2^{\mathfrak{sl}_3}\widehat\chi_{(0,0,0)}(q,1)=0,
\qquad \mathscr L_2^{\mathfrak{sl}_3}=D_q^{(2)}-40\mathbb{E}_4.
\ee 
The relevant conventions are collected in Appendix~\ref{MLDEconvention}. 

We can also  find other solutions to the MLDE. In particular, we can make the ansatz $\widehat\chi=q^\alpha(1+a_1 q+\cdots )$ and substitute it into $\mathscr L_2^{\mathfrak{sl}_3}\widehat\chi=0$, whose leading-order  term gives the indicial equation 
\be
(\alpha-\frac13)(\alpha+\frac16)=0.
\ee
The roots of the indicial equation above are $\{ \frac13, -\frac16 \}$, which correspond  to   
  conformal  weights
\be
h_{  \fks\fkl_3}=\left\{ 0, -\frac12
\right\},
\ee
where we used the fact that $h=\alpha+c/24$. 

It turns out the   non-vacuum solution is given by 
\beqn
\widehat \chi_\text{non-vacuum}&=&
-8   \lim_{\boldsymbol{\sfx} \to \mathbf{1}} \left( \widehat{\chi}_{(1,0,-1)}(q, \boldsymbol{\sfx}) + \widehat{\chi}_{(0,1,-1)}(q, \boldsymbol{\sfx}) - \widehat{\chi}_{(1,1,0)}(q, \boldsymbol{\sfx}) \right)
\\&=&
q^{-\frac16} \Big(1+28q+134q^2+568q^3+1809q^4 +5316q^5+13990q^6
+O(q^7)\Big).\qquad\qquad
\eeqn
   Note that in the unflavored limit,  the   vacuum character is finite, but the   three   non-vacuum characters are actually divergent. Nevertheless, taking the special linear combination above also gives a finite series in the unflavored limit.  

  \subsubsection{BP modules}

  The module structures of BP algebra have been studied in detail   in \cite{FKR2021Review}. 
   Here we restrict to the ordinary untwisted, or Neveu--Schwarz (NS), highest-weight modules.   
   \footnote{Other modules, like the twisted   Ramond modules, also exist, but will not be considered here. See \cite{FKR2021Review} for details. }
There are   six simple ordinary untwisted modules which can be labelled  by
\be 
 (a,b),\qquad a,b\in\mathbb Z_{\geq0},\qquad a+b\leq2.
\ee
Their highest-weight eigenvalues are
\be 
j^{ \text{BP}}_{(a,b)}=\frac{b-a}{4},
\qquad
h^{ \text{BP}}_{(a,b)}
=
\frac{2(a^2+ab+b^2)-5(a+b)}8.
\ee
Thus their $ (J_0,L_0)$ highest weights are
\be\label{jlist}
(0,0),\qquad
\left(\pm\frac12,-\frac14\right),\qquad
\left(0,-\frac12\right),\qquad
\left(\pm\frac14,-\frac38\right).
\ee

 The corresponding characters can again  be derived from \cite{KacWakimotoBoundary2017}     
 \be \label{BPcharcsfs}
 {\widehat\chi}^{ \text{BP}}_{(a,b)}(q,w)
=
q^{h^{ \text{BP}}_{(a,b)}+23/48}w^{(b-a)/4}
\operatorname{PE}\left[
\frac{P^{ \text{BP}}_{(a,b)}(q,w)}{1-q^4}
\right], 
\ee 
where
 \beqn
P^{ \text{BP}}_{a,b}(q,w) &=& 2(q+q^2+q^3)-q^{a+b+1}-q^{3-a-b} +(w+w^{-1})(q^{1/2}+q^{3/2}+q^{5/2}+q^{7/2})
\nonumber\\
&& -w^{-1}q^{a+1/2}-wq^{7/2-a} -wq^{b+1/2}-w^{-1}q^{7/2-b}.
\eeqn 
 
Independently, we can study their unrefined characters and    modules using the MLDE. In this case, the character is a series in $q^{1/2}$, and  it is  described by a fourth-order $\Gamma^0(2)$ differential operator given  in~\cite{Beem:2017ooy}
 \beqn\label{BPMLD}
\mathscr L_4^{\mathrm{BP}}&=&D_q^{(4)}
+\frac5{24}\mathbb{X}D_q^{(3)}-25\mathbb{E}_4D_q^{(2)}
+\left(-\frac{115}{18432}\mathbb{X}^3
+\frac{625}{96}\mathbb{X}\mathbb{E}_4\right)D_q^{(1)}
\nonumber\\
&&
-\frac{3065}{5308416}\mathbb{X}^4
+\frac{1325}{1536}\mathbb{X}^2\mathbb{E}_4
-\frac{7125}{16}\mathbb{E}_4^2,
\eeqn 
which satisfies
\be
\mathscr{L}^{ \text{BP}}_4   {\widehat\chi}^{ \text{BP}}_{(0,0)}(q,1)=0.
\ee

The roots of the indicial equation are given by 
\be 
\Big\{
\frac{23}{48},\quad
\frac{11}{48},\quad
\frac5{48},\quad
-\frac1{48} 
\Big\},
\ee
corresponding to conformal weights 
\be\label{hBP}
h_{ \text{BP}}=\Big\{ 0,\quad-\frac14,\quad-\frac38,\quad-\frac12 \Big\}.
\ee
 This is consistent with \eqref{jlist}. Note that in the unrefined limit  $w=1$, the character formulae in \eqref{BPcharcsfs} are invariant under the exchange of $a,b$, which flips the sign of $j$ but keeps $h$. Therefore the six ordinary modules only give four distinct unrefined characters above. Keeping the flavor fugacity distinguishes the charge-conjugate pairs, which could be detected by flavored MLDEs.
   
\subsubsection{$W_3(3,8)$  modules}
The modules of the $ W_3(3,8)$ minimal model are labeled by
  \be 
(r,s), \qquad (r,s)\sim(5-r-s,r)\sim(s,5-r-s),\qquad r,s\in\mathbb Z_{\geq0}, \qquad r+s\leq5,
\ee 
modulo cyclic permutation $(r,s,5-r-s)$.
In total, there are seven inequivalent classes, which will be chosen as  $(0,0),(0,1),(1,0),(0,2),(2,0),(1,1),(1,2)$.  Their conformal weights are
\be 
h^{W_3}_{(r,s)}
=
\frac{r^2+rs+s^2-5r-5s}{8}.
\ee
 The characters can be obtained by principal Drinfel'd--Sokolov reduction of admissible affine characters \cite{FrenkelKacWakimoto1992,KacWakimotoBoundary2017}. 
\be 
\widehat\chi^ {W_3}_{(r,s)}(q)
=
q^{h^{W_3}_ {(r,s)}+23/24}
\operatorname{PE}\left[
\frac{P^{W_3}_{(r,s)}(q)}{1-q^8}
\right],
\ee
where
\be 
P^{W_3}_{(r,s)}(q)
=
2\sum_{n=1}^{7}q^n
-
\sum_{d\in\{r+1,s+1,r+s+2\}}
\left(q^d+q^{8-d}\right).
\ee
Note that  $\{r+1,s+1,r+s+2\}$ is counted with multiplicity.

The  modules can also  be understood using the MLDE \cite{Jiang:2024baj}:  
 \be\label{W3MLD}
\mathscr L_5^{W_3}=D_q^{(5)}-125\mathbb{E}_4D_q^{(3)}
-\frac{6825}{8}\mathbb{E}_6D_q^{(2)}
-\frac{24125}{16}\mathbb{E}_4^2D_q^{(1)}
-\frac{221375}{32}\mathbb{E}_4\mathbb{E}_6,
\ee 
which annihilates the vacuum character $\widehat\chi^ {W_3}_{(0,0)}(q)$.
The roots of the indicial equation are
\be
\left\{\frac{23}{24}, \quad \frac{11}{24}, \quad \frac{5}{24}, \quad -\frac{1}{24}, \quad \frac{1}{12}\right\},
\ee
corresponding to conformal weights
\be\label{hW3}
h_{ W_3}=\left\{ 0, \ -\frac{1}{2}, \ -\frac{3}{4}, \ -1, \ -\frac{7}{8} \right\}.
\ee
The seven modules only give rise to five distinct characters, as   $(0,s)$ and $(s,0)$ are different modules for $s=1,2$, but  share the same character.

We see that except for the last entry, the module conformal weights in  \eqref{hW3} are twice   those in  \eqref{hBP}.

\subsubsection{Module character identities }
 The first relation between module characters  is:
\be\label{w3BP}
h^{W_3}_{(2a,2b)} 
=2h^{ \text{BP}}_{(a,b)} , \qquad
 {\widehat\chi}^{W_3}_{(2a,2b)}(q)
= {\widehat\chi}^{ \text{BP}}_{a,b}(q^2,1),  \qquad a,b\in\mathbb Z_{\geq0},\qquad a+b\leq2,
\ee
which is the generalization of the vacuum character identity in \eqref{AAcase}. 

 To generalize \eqref{A1a5d4},  we can apply the same   fugacity map $q\to q^2$, $\mathsf{x}_1=wq^{-1/2}$,  $\mathsf{x}_2=w^{-1}q^{-1/2}$, $\mathsf{x}_3=q$. Then the affine character becomes 
  \beqn
&& q^{-3/16}  \widehat\chi_{\bm v}^{\mathfrak{sl}_3} \left(q^2; wq^{-1/2}, w^{-1}q^{-1/2},q\right) 
= q^{23/48 - (v_1+v_2)/4} w^{(v_2-v_1)/2} \operatorname{PE} \left[ \frac{ \widetilde P^{\fks\fkl_3}_{\bm v}(q,w) }{1 - q^4} \right],
\qquad\qquad
\eeqn 
where 
  \be
\widetilde P^{\fks\fkl_3}_{\bm v}(q,w)  = q + 2q^2 + q^3   + w \left( q^{3/2+2v_1} + q^{5/2-2v_2} \right) 
 + w^{-1} \left( q^{5/2-2v_1} + q^{3/2+2v_2} \right).
\ee 

One can  explicitly check that 
  \be
  h^{\mathrm{BP}}_{a,b} 
=2h^{\fks\fkl_3}_\lambda-\frac12(\lambda,\rho)
=2h^{\fks\fkl_3}_\lambda+\frac{v_1+v_2-2v_3}{4}, \qquad 
\widetilde P_{(v_1,v_2,v_3)}^{\fks\fkl_3} (q,w) =  P_{(a,b)}^{ \text{BP}} (q,w)    ,
\ee 
  if we identify $(a,b)=\bigl(v_1(2-v_2),v_2(2-v_1)\bigr)$ for $\bm v\in\mathcal K$.
Therefore we have the character identity
  \be
q^{-3/16}\widehat\chi^{\mathfrak{sl}_3}_{\bm v}
\left(q^2;wq^{-1/2},w^{-1}q^{-1/2},q\right)
=\widehat\chi^{ \text{BP}}_{a,b}(q,w), 
\ee
where the non-trivial prefactor is $q^{( 2c_{\fks\fkl_3}-c_\text{BP})/24}$, with the central charges given in \eqref{cw3bp}.

 We collect the relevant relations in Table~\ref{tabmodules}, which  gives the character correspondences within the selected classes: four affine admissible highest-weight modules, six ordinary BP NS modules, and seven simple $W_3(3,8)$ modules.  The (1,1) module of $W_3$ algebra, with conformal weight $h_W=-7/8$,  has no counterpart among the modules paired by the two fixed
regradings in Table~\ref{tabmodules}.  Nevertheless, we find the following identity
\be\label{W3missing}
\widehat\chi^{W_3}_{(1,1)}(q)
=\widehat\chi^{\mathfrak{sl}_3}_{\text{vac}}(q^{1/4};i,i,-1), 
\ee
 where  the vacuum character of the  affine algebra  $ \widehat{\chi}_{\text{vac}}^{\mathfrak{sl}_3}( {q}; \mathsf{x}_1, \mathsf{x}_2, \mathsf{x}_3) $ is  given in     \eqref{sl3affc} subject to $\sfx_1\sfx_2\sfx_3=1$. The identity \eqref{W3missing} is very different from other character identities in the rest of the paper.  It would be interesting to understand whether it is a coincidence or points towards another   relation between VOAs.  Nevertheless, we find  that it can be generalized to other $W_3$ modules:
  \begin{align}
\widehat{\chi}^{W_3}_{(0,0)}(q)
&=
q^{-3/8}\,
\widehat{\chi}^{\mathfrak{sl}_3}_{\mathrm{vac}}
\left(q^4;\,q^{-1},q^{-1},q^2\right),
\\[4pt]
\widehat{\chi}^{W_3}_{(0,1)}(q)
=
\widehat{\chi}^{W_3}_{(1,0)}(q)
&=
q^{-7/8}\,
\widehat{\chi}^{\mathfrak{sl}_3}_{\mathrm{vac}}
\left(q^4;\,q,q^2,q^{-3}\right),
\\[4pt]
\widehat{\chi}^{W_3}_{(2,0)}(q)
=
\widehat{\chi}^{W_3}_{(0,2)}(q)
&=
q^{-1/8}\,
\widehat{\chi}^{\mathfrak{sl}_3}_{\mathrm{vac}}
\left(q;\,-1,-q^{-1/2},q^{1/2}\right), \\
\widehat{\chi}^{W_3}_{(2,1)}(q)
=
\widehat{\chi}^{W_3}_{(2,2)}(q)
&=
-q^{-19/8}\,
\widehat{\chi}^{\mathfrak{sl}_3}_{\mathrm{vac}}
\left(q^4;\,q^{-5},q^3,q^2\right).
\end{align}
 These identities relate the    modules of  $W_3$ algebra to the  vacuum module of  the affine vertex algebra. Note that the last one has an overall sign  and its fugacity specialization requires meromorphic continuation.

 \begin{table}[t]
\centering
\begingroup
\renewcommand{\arraystretch}{1.7}
\setlength{\arraycolsep}{10pt}
$
\begin{array}{ccc|cc|cc}
\multicolumn{3}{c|}{L_{-3/2}(\mathfrak{sl}_3)}
&
\multicolumn{2}{c|}{\mathrm{BP}_{-9/4}}
&
\multicolumn{2}{c}{W_3(3,8)}
\\[3pt] \hline
\boldsymbol{v} & \lambda & h_\lambda
& (a,b) & (j,h)
& (r,s) & h_W
\\ \hline
(0,0,0) & 0 & 0
& (0,0) & (0,0)
& (0,0) & 0
\\
(1,0,-1) & -\frac32\omega_1 & -\frac12
& (2,0) & \left(-\frac12,-\frac14\right)
& (4,0)\sim(0,1) & -\frac12
\\
(0,1,-1) & -\frac32\omega_2 & -\frac12
& (0,2) & \left(\frac12,-\frac14\right)
& (0,4)\sim(1,0) & -\frac12
\\
(1,1,0) & -\frac12\rho & -\frac12
& (1,1) & \left(0,-\frac12\right)
& (2,2)\sim(1,2) & -1
\\
\text{---} & \text{---} & \text{---}
& (1,0) & \left(-\frac14,-\frac38\right)
& (2,0) & -\frac34
\\
\text{---} & \text{---} & \text{---}
& (0,1) & \left(\frac14,-\frac38\right)
& (0,2) & -\frac34
\\
\text{---} & \text{---} & \text{---}
& \text{---} & \text{---}
& (1,1) & -\frac78
\end{array} 
$
\endgroup
\caption{Modules and conformal weights  for $L_{-3/2}(\mathfrak{sl}_3)$,
$\mathrm{BP}_{-9/4}$, and $W_3(3,8)$.  Non-empty entries in the same row satisfy the character identities, including the indicated regrading and affine prefactor.    }
\label{tabmodules}
\end{table}

 The simple relation in \eqref{w3BP} also  allows us to establish a connection between the    two modular linear differential operators \eqref{W3MLD} and \eqref{BPMLD}.
 
 For this purpose, we introduce the scaling operator $R_{1/2}$ which acts as
\be
(R_{1/2}f)(q)=f(q^{1/2} ).
\ee
  
The covariant derivative transforms as
 \be
\frac12R_{1/2}\partial_{(k)}R_{1/2}^{-1}
=\partial_{(k)}+\frac{k\mathbb{X}}{24}.
\ee 
This  can be checked  using \eqref{e246res} -  \eqref{serdfdsf}:
  \beqn
\frac12(\partial_{(k)}f)(q^{1/2} )
&=&q\partial_q[f(q^{1/2} )]
+\frac{k}{2}\mathbb{E}_2(q^{1/2})f(q^{1/2} )\\
&=&q\partial_q[f(q^{1/2} )]
+k\mathbb{E}_2(q)f(q^{1/2} )
+\frac{k}{24}\mathbb{X}(q)f(q^{1/2} ).
\eeqn

We can then define  
\be
\widehat {\mathscr L}^{W_3}_5 
=
2^{-5}R_{1/2}\mathscr L^{W_3}_5 R_{1/2}^{-1}.
\ee
This operator can be rewritten as
 \be\label{DDks}
\widehat{\mathscr L}^{W_3}_5
=\left(\partial_{(8)}+\frac58\mathbb{X}\right)\circ\mathscr L_4^{\mathrm{BP}},
\ee 
by using \eqref{e246res} -  \eqref{serdfdsf} and replacing $\partial_{(k)}$ with
$\overline\partial_{(k)}=\partial_{(k)}+k\mathbb{X}/24$. 
The factorization in \eqref{DDks}   gives  an inclusion of solution
spaces after rescaling, and  thus provides a natural  explanation  for   \eqref{w3BP}.
 
  In section \ref{GVISOAAD}, we established the GVISO between $\fks\fkl_3$ and BP in the vacuum module.  
It is then natural to ask whether this construction can be generalized to non-vacuum modules.
This requires a deeper understanding of the modules and is beyond the scope of this paper.   Nevertheless, we believe the GVISO construction in the vacuum case should be generalizable to  a special class of modules, which are obtained from the vacuum module by spectral flow.  Note that the two VOAs have   U(1) symmetries which can be used to generate spectral flows. We leave a detailed study to the future. 
  
\subsection{General conjecture}

The concrete example above suggests that the non-vacuum module character identity is a general feature. We therefore conjecture the following character identity between  an affine vertex algebra and a principal  $W$-algebra:
\be\label{VMgsfsdf}
\Tr_{\cM (L_k(\fkg),\lambda)}q^{(h+1)L_0- J_0^{\rho^\vee}}=\Tr_{\cM (W_{k'}(\fkg,f_\prin),\mu)}q^{  L_0 }
\ee
  for general $\fkg$ at general admissible levels  subject to $(k+h^\vee) =(k'+h^\vee) (h+1)$. 
  \footnote{We   require both  
levels to be admissible before cancellation, namely 
$k=-h^\vee+n/p$ and $k'=-h^\vee+n/[p(h+1)]$ with
$p,n\geq1$ and $\gcd(n,p)=\gcd(n,h+1)=1$, in addition to the other conditions in footnote~\ref{admiscond}.} 
We expect the conjecture holds for a  selected class  of modules with a suitable map $\lambda\mapsto\mu$. This generalizes the vacuum-module identity in the boundary admissible case that we discussed extensively in the previous sections. 
We conjecture that similar module character identities  also hold beyond the affine/principal setting.

We defer the detailed discussions to \cite{Jiang20272}, but present the key module character formulae in the boundary admissible case with  $\fkg=\fks\fkl_N$, which are particularly relevant for 4d SCFTs and provide a starting point for proving the above conjecture in this case. 

 The two relevant VOAs are given by
\be
W_N(N,U):=\mathcal W_{-N+\frac N U}(\mathfrak{sl}_N,f_{\mathrm{prin}}), \qquad U>N+1, \quad \gcd(N,U)=1,
\ee
and
\be
L_{-N+\frac Nu}(\mathfrak{sl}_N), \qquad u> 1,\qquad \gcd(N,u)=1.
\ee
 For $N\ge2$, the former is the $W_N$ minimal model associated with $(A_{N-1},A_{U-N-1})$, and the latter is the affine algebra associated with $D_u(SU(N))$ when $u>1$. Note that the endpoints $u=1$ and $U=N+1$ give    trivial VOAs at  zero central charges. 

 Their central charges   are
\be
c_{\mathrm{aff}}  = (1 - u)(N^2 - 1), \qquad c _W= (N - 1) - \frac{(N^2 - 1)(U - N)^2}{U}.
\ee
Their modules are specified by   weights 
\footnote{We use the standard type-$A_{N-1}$ normalization
$\alpha_i=e_i-e_{i+1}$ and $(\alpha_i,\alpha_i)=2$.
The fundamental weights satisfy $(\omega_i,\alpha_j)=\delta_{ij}$, and the Weyl vector is
$
\rho 
=\sum_{i=1}^{N-1}\omega_i 
$.
All weight pairings below use the inverse Cartan matrix $(A^{-1})_{ij}=\min(i,j) - \frac{ij}{N}$. In particular,
$
(\lambda,\rho)=\frac12\sum_{i=1}^{N-1}i(N-i)\lambda_i,
  (\rho,\rho)=\frac{N(N^2-1)}{12} 
$. } 
\be \label{almbdamu}
\lambda=\sum_{i=1}^{N-1}\lambda_i\omega_i,
\qquad
\mu=\sum_{i=1}^{N-1}\mu_i\omega_i.
\ee
 
 On the $W$-algebra side, all simple ordinary   modules are labelled by nonnegative tuples
\be 
(\mu_0, \mu_1, \dots, \mu_{N-1}), \qquad \sum _{i=0}^{N-1}  \mu_i = U - N, \quad \mu_i \in \bZ_{\ge 0},
\ee
 modulo cyclic rotation. The conformal weight is 
 \be
 h^W_\mu = \frac{N}{2U}(\mu, \mu + 2\rho) - (\rho, \mu).
 \ee
 The character is given by 
\be  \label{Wmdch}
\widehat\chi^W_{[\mu]}(q) = q^{h^W_\mu - c_W/24} \operatorname{PE} \left[ \frac{A_\mu(q)}{1 - q^U} \right],
 \qquad
\ee
where 
\be  
A_\mu(q) = (N - 1) \sum_{d=1}^{U-1} q^d - \sum_{1\le i < j \le N } \left(q^{d_{ij}} + q^{U-d_{ij}}\right), \qquad
d_{ij} =  \sum_{a=i}^{j-1} (\mu_a + 1).   
\ee

 On the affine side, the irreducible   admissible highest-weight modules can be labelled by $\lambda_i$ subject to 
\be\label{lambdai}
S_{ij} 
:=\sum_{s=i}^{j-1}(\lambda_s+1)\in\mathcal P_{N,u},
\qquad  \forall 1\le i<j\le N,\qquad S_{ll}=0,\qquad  \forall 1\le l \le N
\ee
where
\be\label{eq:typeA-allowed-pairings}
\mathcal P_{N,u}
=\left\{m-\frac{Nn}{u}\;\middle|\;
m,n\in\mathbb Z,\quad 1\le m\le N-1,\quad 0\le n\le u-1\right\}.
\ee
 Note that the  affine Dynkin label   is given by $\lambda_0=k -\sum_{i=1}^{N-1} \lambda_i$ for $k =-N+\frac Nu$.

The conformal weight of the module is given by
\be\label{eq:typeA-affine-label-weight}
h^{\mathrm{aff}}_\lambda=\frac{u}{2N}(\lambda,\lambda+2\rho).
\ee
The module character is 
 \be \label{affmdch}
\widehat\chi_{\lambda}^{\mathrm{aff}}(Q,\bm x) = Q^{h^{\mathrm{aff}}_{\lambda} - c_{\mathrm{aff}}/24} \bm x^{\bm \lambda} \operatorname{PE} \left[ \frac{ {N}_{\lambda}(Q,\bm x)}{1 - Q^u} \right],
\ee
where 
 \be
 {N}_{\lambda}(Q,\bm x)  = (N - 1)(B_u(Q) - 1) 
 + \sum_{1 \le i < j \le N} \left[ (Q B_u(Q) - Q^{u - \nu_{ij}}) \frac{x_i}{x_j} + (B_u(Q) - Q^{\nu_{ij}}) \frac{x_j}{x_i} \right],
\ee 
and
  \be
   \nu_{ij}=[-a\,uS_{ij}]_u , \qquad 
\bm x^{\bm \lambda}   = \prod_{a=1}^{N-1} (x_1 \cdots x_a)^{\lambda_a},\qquad 
B_u(Q) = 1 + Q + \cdots + Q^{u-1}.
\ee 
Here $a\equiv N^{-1} \mod u$, and  $[m]_u \in \{0, \cdots, u-1\}$ denotes the mod $u$ reduction of $m$. Note that $x_1\cdots x_N=1$.

 The character formulae  \eqref{Wmdch} and  \eqref{affmdch} can be derived from   \cite{KacWakimotoBoundary2017}.  \footnote{We have also tested them extensively using MLDE and flavored MLDE over   a broad range of parameters. }
Their simple PE form is related to the boundary admissibility of the levels. One can verify that they reduce to the    formulae in 
\eqref{chiaff} and \eqref{chiprin} for  the vacuum module.  

Using these character formulae, one can show that the following character identity holds
\be\label{WVidensf}
 \widehat\chi^W_{[\mu]}(q) =q^{\,kN(N-1)/24}\,
\widehat\chi^{\mathrm{aff}}_\lambda
\left(q^{  N+1 } ; \ x_i=q^{\,i- (N+1)/2}\right),
\ee
under   identification $  U=(N+1)u  $ and the  module map: $ \mu_i=p_{ i+1}-p_i-1  $ for $i=1, \cdots, N-1$, where    the ordered set  $0=p_1< p_2\cdots <p_N <U$ is just a permutation of  $b_i:= i-1+(N+1) \nu_{1i} $, namely  $p_i=b_{\sigma(i)}$ for a permutation $\sigma \in S_N$.

  We will derive the character formulae  \eqref{Wmdch} and  \eqref{affmdch}, discuss the   proof of  \eqref{WVidensf}, and explore further generalization  to modules  for other Lie algebras at general        admissible levels   in  detail in \cite{Jiang20272}.

\section{Character identities    beyond   SCFT/VOA}\label{IsingnoSCFT}
 
Our discussions so far have been motivated by the SCFT/VOA correspondence, where the VOAs feature quasi-lisse properties and non-unitarity. In this section, we generalize the character identity and GVISO beyond the realm of the SCFT/VOA correspondence. 
 
 \subsection{Affine and Ising    from Majorana  }

  We begin by discussing this relation in a simple unitary setting involving the Ising VOA and the level-one affine VOA.

  For $n\ge2$,  \footnote{The case of $n=1$  
was studied explicitly  in  \cite{Adamovic:2025efe}, where $L_1(\mathfrak{so}_{3})$ is written as $L_2(\fks\fkl_2)$. We restrict to $n\ge 2$ just to ensure a  uniform set of  notations later.} we now consider the following pair of VOAs 
 \be
 V=L_1(\mathfrak{so}_{2n+1}),\qquad W=\text{Vir}(3,4),
 \ee
 which correspond  to the level one affine vertex algebra  and the Ising VOA, respectively.  Their central charges are
 \be
 c_V=\frac{2n+1}{2}, \qquad c_W=\frac12.
 \ee
 
 We will show that there is a GVISO between them 
  \be\label{IsingSO}
 \left(L_1(\mathfrak{so}_{2n+1}),
H\right)
\cong_{\mathrm{gr}}
\left(\operatorname{Vir}(3,4),S_0\right),
\ee
where
\be\label{Hising}
H=\ell L_0^V-J_0^{\rho^\vee},\qquad\qquad  \ell=2n+1.
\ee

The two VOAs admit  free-fermion realizations  
\be\label{VWF}
V\cong(F^{\otimes(2n+1)})_{\overline0},
\qquad
W\cong F_{\overline0},
\ee
where $ F$ is the free Majorana fermion and the subscript  ${\overline0}$ implies that we only consider the even fermion parity sector. 

On the $W$ side, we choose one Majorana  $\chi$, with modes $\chi_b$, $b \in \bZ+\frac12$ with 
\be
\{\chi_{b_1},\chi_{b_2}\}
= \delta_{b_1+b_2,0}.
\ee

On the $V$ side, we choose $2n+1$ Majorana  fermions and linearly combine them such that
\be\label{psimode}
\psi^r_a,\qquad
r=-n,\ldots,n,\qquad a\in\mathbb Z+\frac12,
\ee
with
\be
\{\psi^r_{a_1},\psi^s_{a_2}\}
=\delta_{r+s,0}\delta_{a_1+a_2,0},
\qquad
[J_0^{\rho^\vee},\psi^r_a]=r\psi^r_a,
\label{J0rho-charge}
\ee
where the explicit form of the current $J_0^{\rho^\vee}$ is given below in \eqref{J0rho}.

 The two Fock vacua are defined as 
 \be\label{Fpsichi}
 \ \chi_b\bm 1_W=0, \qquad b>0;
 \qquad\quad    \psi_ a^r\bm 1_V=0, \qquad a>0, \quad\forall r .
 \ee
 
We   introduce the following intertwiner map:
  \be \label{PhiF}
\Phi_F \bm 1_V=\bm 1_W, 
\qquad
\Phi_F\psi^r_a=\chi_{\ell a+r}\Phi_F.
 \ee
It has the following four   properties:
 
 1) The indices are in bijection. Indeed, one  can introduce $a'=a-\tfrac12$, $b'=b-\tfrac12$ and
 $r'=r+\tfrac \ell 2-\tfrac12=r+n$, so that the map between $(a,r)$ and $b$ reduces to  $b'=\ell a'+r'$ with  $0\le r'<\ell$. This gives an  explicit  bijection of indices between $\bZ$ and $\bZ\times \bZ_\ell$.

 2)   Creation and annihilation modes remain creation and annihilation modes. 
 Indeed, if $a>0$, then  $a'=a-1/2\ge 0$, and   $b'=\ell a'+r' \ge 0$. If $a<0$, then  $\ell a+r<0$. Similarly, if $a<0$, then $a'=a-1/2\le -1$ and   $b'=\ell a'+r' \le  - \ell+r' < 0$  as    $r'\in \{0, \cdots, \ell-1\}$. Therefore, the creation and annihilation modes on the $V$ side  are mapped to   creation and annihilation modes on the $W$ side.
 
  3) The anticommutators agree.  This follows from 
  \be
   \ell(a_1+a_2)+r+s=0
   \iff {a_1}+a_2=0,\quad r+s=0,
   \ee 
 due to   $ |r+s|\le \ell-1$.
 
 4) The  fermion parity is preserved.

  These properties   guarantee  that \eqref{PhiF} is indeed a bijection  between the two Fock-spaces.  It provides a complete specification of the linear map on the two fermion Fock spaces 
 \be 
\Phi_F \left( \psi_{-a_1}^{r_1} \cdots \psi_{-a_s}^{r_s} \mathbf{1} \right) = \chi_{-\ell a_1+r_1} \cdots \chi_{-\ell a_s+r_s} \mathbf{1}, \qquad a_j > 0,
\ee

Moreover, using  \eqref{Hising}  and \eqref{J0rho}, one can see that
  \be
[H,\psi^r_a]=-(\ell a+r)\psi^r_a,\qquad [S_0, \chi_b]=-b\chi_b.
\ee 
Combining with \eqref{PhiF}, we thus have
\be
\Phi_FH=S_0\Phi_F.
\ee

Since our two VOAs are the even fermion-parity sectors of the corresponding free Majorana algebras \eqref{VWF}, we  thus have the following graded vector space isomorphism
\be 
\left(L_1(\mathfrak{so}_{2n+1}),(2n+1)L_0^V-J_0^{\rho^\vee}\right)
\cong_{\mathrm{gr}}
\left(\text{Vir}(3,4),S_0\right). 
\ee

This realization also gives further correspondences. 
The affine currents in $L_1(\mathfrak{so}_{2n+1})$ admit  the following description 
\be
J^{rs}_m
=
\sum_{a\in\mathbb Z+1/2}
:\psi^r_a\psi^s_{m-a}:.
\ee
Under  \eqref{PhiF}, these currents are mapped to  
  \be 
\widetilde J^{rs}_m
=
\sum_{a\in\mathbb Z+1/2}
:\chi_{\ell a+r}\chi_{\ell(m-a)+s}:,
\qquad
\Phi_FJ^{rs}_m=\widetilde J^{rs}_m\Phi_F.
\ee 
Therefore, for every current relation, we have
\be
P(J)|0\rangle=0
\quad\Longleftrightarrow\quad
P(\widetilde J)|0\rangle=0,
\ee
where $P(J),P(\widetilde J)$ are the same finite noncommutative polynomial in the respective
current modes.     This resolves the question of all null relations in this case: the invertible
oscillator map transports every relation.

The   Virasoro modes on the  Ising VOA are
\be
S_m=\frac14\sum_{t+v=m}(v-t):\chi_t\chi_v:.
\ee
Under the inverse  of   \eqref{PhiF}, these modes are  mapped to
  \be 
S^V_m=
\frac14
\sum_{\substack{
a,b\in\mathbb Z+\frac12,\ -n\le r,s\le n\\
\ell(a+b)+r+s=m}}
\bigl[\ell(b-a)+s-r\bigr]:\psi^r_a\psi^s_b:.
\ee 
They satisfy the Virasoro algebra  
\be
[S^V_m,S^V_j]
=(m-j)  S^V_{m+j}
+\frac{m(m^2-1)}{24}\delta_{m+j,0},
\ee  
with central charge $1/2$. This should be distinguished from the original Sugawara stress tensor in the affine algebra. 
In addition, we have
 \beqn
 S^V_0 &=& -\frac12
 \sum_{a\in\mathbb Z+\frac12}\sum_{r=-n}^n
 (\ell a+r):\psi^r_a\psi^{-r}_{-a}:
 \label{eq:fermion-S0-full}\\
 &=&-\sum_{\substack{a>0\\a\in\mathbb Z+\frac12}}\sum_{r=-n}^n
 (\ell a+r):\psi^r_a\psi^{-r}_{-a}:
 =\ell L_0^V-J_0^{\rho^\vee}=H.
 \label{eq:fermion-S0-half}
\eeqn 
Here we used
\be\label{J0rho}
J_0^{\rho^\vee}
=\sum_{\substack{a>0\\a\in\mathbb Z+\frac12}}\sum_{r=-n}^n
r:\psi^r_a\psi^{-r}_{-a}:
=\frac12\sum_{r=-n}^n rJ^{r,-r}_0.
\ee

 The universal Virasoro vacuum algebra at $c=1/2$ contains the nonzero weight-six singular vector
\be\label{eq:ising-weight-six-singular}
\mathcal N_{\rm Ising}=
\left(
S_{-2}^3
-\frac{33}{8}S_{-4}S_{-2}
+\frac{93}{64}S_{-3}^2
-\frac{27}{16}S_{-6}
\right)|0\rangle.
\ee
Its image is zero in the simple Ising VOA and in its fermion realization. Therefore replacing every $S_m$ by $S_m^V$ gives zero in the source fermion realization as well, by conjugation with $\Phi_F$. 
 
We now extend this discussion to modules. First, the affine algebra $L_1(\mathfrak{so}_{2n+1})$ has three affine primaries,   denoted as $0, \omega_1, \omega_n$, corresponding to the trivial, vector and spinor representations.  Similarly, the Ising VOA has three  Virasoro primaries, $1, \epsilon, \sigma$. They are related as follows:  
  \be
\begin{array}{c|c|c|c|c}
W\text{ module}&V\text{ module}&\text{lowest }L_0^V
&\text{max. ground }J_0^{\rho^\vee}&\text{lowest }H=\ell L_0^V-J_0^{\rho^\vee}\\\hline
I_0=L^{\mathrm{Vir}}(\frac12,0)&M_0=L_1(\mathfrak{so}_{2n+1};0)&0&0&0\\
I_\varepsilon=L^{\mathrm{Vir}}(\frac12,\frac12)&M_v=L_1(\mathfrak{so}_{2n+1};\omega_1)&\frac12&n&\frac12\\
I_\sigma=L^{\mathrm{Vir}}(\frac12,\frac1{16})&M_s=L_1(\mathfrak{so}_{2n+1};\omega_n)&\frac{2n+1}{16}&\frac{n(n+1)}4&\frac1{16}
\end{array}
\ee 

 The modules of the Ising VOA also admit free field realizations
\beqn  
I_0
&=&(\mathcal F_\chi^{\mathrm{NS}})_{\mathrm{even}}
=\operatorname{span}\left\{
\chi_{-a_1}\cdots\chi_{-a_s}|0\rangle:
 a_1>\cdots>a_s>0,\; a_i\in\mathbb Z+\tfrac12, \; s\in 2\bN
\right\},
\qquad\qquad \qquad\\  
I_\varepsilon
&=&(\mathcal F_\chi^{\mathrm{NS}})_{\mathrm{odd}}
=\operatorname{span}\left\{
\chi_{-a_1}\cdots\chi_{-a_s}|0\rangle:
 a_1>\cdots>a_s>0 ,\; a_i\in\mathbb Z+\tfrac12, \; s\in 2\bN+1 
 \right\},
   \\  
 I_\sigma
&=&(\mathcal F_\chi^{\mathrm{R}}) 
=\operatorname{span}\left\{
\chi_{-a_1}\cdots\chi_{-a_s}|\sigma\rangle:
 a_1>\cdots>a_s>0,\; a_i\in\mathbb Z , \; s\in  \bN 
 \right\},
 \eeqn  
 where the ground state $\ket{\sigma}$  satisfies
\be
\chi_a|\sigma\rangle=0\quad(a>0),\qquad
\chi_0|\sigma\rangle=\frac1{\sqrt2}|\sigma\rangle,\qquad
S_0 \ket{\sigma}=\frac{1}{16}\ket{\sigma}.
\ee
The ground state in $I_\epsilon $ is $\ket{\epsilon}  =\chi_{-\frac12}\ket{0}$ which has conformal weight $1/2$.
A similar construction also applies for level 1 affine algebra $L_1(\mathfrak{so}_{2n+1})$  \cite{Ginsparg:1988ui}.  See e.g. \cite{Bockenhauer:1995hx}.

The Ising characters  are given in  \cite{Ginsparg:1988ui}
\be\label{eq:Isingcharacters} 
\widehat\chi_{I_0}(q)=\frac{q^{-1/48}}2\left[ \cP_++\cP_-\right],\qquad
\widehat\chi_{I_\varepsilon}(q)=\frac{q^{-1/48}}2\left[ \cP_+-\cP_-\right], \qquad
\widehat\chi_{I_\sigma}(q) =q^{1/24}\cP_R,
\ee
where 
\be
\mathcal P_\pm(q) =(\mp q^{1/2};q)_\infty,\qquad
\mathcal P_{\mathrm R}(q)=(-q;q)_\infty, \qquad (a;q)_\infty =\prod_{i=0}^\infty(1-a q^i).
\ee

 The affine   characters of the three modules are given by  \cite{Delmastro:2021otj}
\be\label{eq:Bn-level-one-flavored-characters}
\begin{aligned}
\widehat\chi_{M_0}(Q,\bm x)
&=\frac{Q^{-\ell/48}}{2}\bigl[\mathcal A_+(Q,\bm x)+\mathcal A_-(Q,\bm x)\bigr],\\
\widehat\chi_{M_v}(Q,\bm x)
&=\frac{Q^{-\ell/48}}{2}\bigl[\mathcal A_+(Q,\bm x)-\mathcal A_-(Q,\bm x)\bigr],\\
\widehat\chi_{M_s}(Q,\bm x)&=Q^{\ell/24}\mathcal B(Q,\bm x),
\end{aligned}
\ee
where
\be\label{eq:Bn-flavored-fermion-products}
\begin{aligned}
\mathcal A_\pm(Q,\bm x)
&=(\mp Q^{1/2};Q)_\infty
\prod_{r=1}^{n}(\mp x_rQ^{1/2};Q)_\infty (\mp x_r^{-1}Q^{1/2};Q)_\infty,\\
\mathcal B(Q,\bm x)
&=(-Q;Q)_\infty\prod_{r=1}^{n}
(x_r^{1/2}+x_r^{-1/2})(-x_rQ;Q)_\infty(-x_r^{-1}Q;Q)_\infty.
\end{aligned}
\ee
 
Setting $Q=q^\ell$ and $x_r=q^{-r}$, we have
\be 
\mathcal A_\pm(q^\ell,x_r=q^{-r}) =\mathcal P_\pm(q), \qquad
\mathcal B(q^\ell,x_r=q^{-r}) =q^{-n(n+1)/4}\mathcal P_{\mathrm R}(q).
\ee
As a result,  we obtain the character identities  
\be\label{eq:Bn-Ising-normalized-character-identity}
\widehat\chi_{I_\alpha}(q)
 =q^{(\ell c_V-c_W)/24}
\widehat\chi_{M_\alpha}(q^\ell,x_r=q^{-r}),\qquad \frac{\ell c_V-c_W}{24} =\frac{\ell^2-1}{48},
\ee
for the  three pairs of modules $(M_0,I_0),(M_v,I_\varepsilon),(M_s,I_\sigma)$. 
Equivalently, we have the equality  $\operatorname{Tr}_{M_\alpha}q^H=\operatorname{Tr}_{I_\alpha}q^{S_0}$.  The product formulae also show that all eigenspaces of these regraded
operators are finite-dimensional and that their spectra are bounded below.   
 
These character identities imply the following graded vector-space isomorphisms for 
the modules
\beqn
&& (L_1(\mathfrak{so}_{2n+1},0),H)\cong_{\rm gr}(I_0,S_0),
\nonumber\\
&& (L_1(\mathfrak{so}_{2n+1},\omega_1),H)\cong_{\rm gr}(I_\varepsilon,S_0),
\nonumber\\
&& (L_1(\mathfrak{so}_{2n+1},\omega_n),H)\cong_{\rm gr}(I_\sigma,S_0).
\eeqn

\subsection{Further generalization}

It turns out the  preceding example admits broad generalizations, which will be explored in more detail in  \cite{Jiang2027}. Below we present several interesting   results. 

 First, for $m \ge 2, \gcd(m+2,3)=1$, we claim that 
\beqn\label{SpVirsfs}
 \left(L_1(\mathfrak{sp}_{2m}),
(2m+1)L_0^V-J_0^{\rho^\vee}\right)
& \cong_{\mathrm{gr}}&
\left(\operatorname{Vir}(3,m+2),S_0\right). 
\eeqn
This   follows from the specialized character formula in \cite[Theorem~7.14]{ButoracWakimoto2024}, which gives
\beqn
\operatorname{Tr}_{L_1(\mathfrak{sp}_{2m})} q^{(2m+1)L^V_0 - J_0^{\rho^\vee}} 
&=& \frac{(q, q^{2m+3}, q^{2m+4}; q^{2m+4})_\infty (q^{2m+2}, q^{2m+6}; q^{4m+8})_\infty}{(q; q)_\infty} \label{Virseresi}
\\ &=&\label{SpVirsfs2}
 \frac{1}{(q; q)_\infty}\sum_{n \in \mathbb{Z}} \left( q^{3(m+2)n^2 + (m-1)n} - q^{3(m+2)n^2 + (m+5)n + 1} \right)\qquad\qquad
\\ &=& \operatorname{Tr}_{\operatorname{Vir}(3,m+2)} q^{L_0},
\eeqn
where $(a_1, \cdots,a_s;q)_\infty =(a_1;q)_\infty \cdots (a_s;q)_\infty$ and we used the quintuple product identity in the second equality.~\footnote{The   quintuple product identity states that $\sum_{n=-\infty}^{\infty} (-1)^n Q^{n(3n-1)/2} z^{3n} (1 + zQ^n) = (Q, -z, -Q/z; Q)_\infty (Qz^2, Q/z^2; Q^2)_\infty $, and we have used $Q=q^{2m+4},z=-q$ in our application.}
 The resulting infinite sum in \eqref{SpVirsfs2} is exactly the vacuum character of the Virasoro minimal model. 
 This thus establishes \eqref{SpVirsfs}.   Setting $m=2$  in \eqref{SpVirsfs} reduces to \eqref{IsingSO} with $n=2$, due to the isomorphism $\mathfrak{sp}_{4}\cong \mathfrak{so} _5$.

Under the same conditions $m \ge 2$ and $\gcd(m+2,3)=1$, there is also a GVISO between $\mathfrak{sl}_2$ and the Virasoro algebra \cite{Adamovic:2025efe}: 
\be\label{afsfdw}
\left(L_m(\mathfrak{sl}_2 ),
3L_0-\frac12J^h_0\right)
\cong_{\mathrm{gr}}
\Big(
 {\mathrm{Vir}}\!\left(
 {3,m+2} 
\right), S_0\Big).
\ee

Combining \eqref{SpVirsfs}  and \eqref{afsfdw}  gives
\be 
\left(L_1(\mathfrak{sp}_{2m} ),
(2m+1)L_0-J_0^{\rho^\vee}\right)
\cong_{\mathrm{gr}}
\left(L_m(\mathfrak{sl}_2 ),
3L_0-\frac12J^h_0\right)
=\left(L_m(\mathfrak{sp}_2 ),
3L_0-J_0^{\rho^\vee}\right).
\ee
This admits a further generalization to modules. It turns out that this is just the special case of the following general family of relations:
\be \label{speclevelrank}
 \left(L_m(\mathfrak{sp}_{2n}),
(2n+1)L_0-J_0^{\rho_n^\vee}\right)  \cong_{\mathrm{gr}}
\left(L_n(\mathfrak{sp}_{2m}),
(2m+1)L_0-J_0^{\rho_m^\vee}\right). 
\ee
Actually, the formula in \cite[Theorem~7.14]{ButoracWakimoto2024} allows us to compute the graded character explicitly
\be \label{speclevelrank2}  
\Tr_{L_m(\mathfrak{sp}_{2n})} q^{(2n+1)L_0-J_0^{\rho_n^\vee}}=
\operatorname{PE}\!\left[
\frac{q^2(1-q^{2m})(1-q^{2n})(1+q^{2m+2n+3})}
{(1-q)(1-q^2)(1-q^{4m+4n+4})}
\right],
\ee 
which is obviously symmetric in $m$ and $n$, thus proving the GVISO \eqref{speclevelrank}. 
Alternatively, \eqref{speclevelrank} can be understood by combining  the VOA isomorphism \cite[Theorem~9.3(2)]{Kanade:2018qut} 
\be 
\mathcal W_{-(n+1)+(m+n+1)/(2n+1)}(\mathfrak{sp}_{2n})
\cong
\mathcal W_{-(m+1)+(m+n+1)/(2m+1)}(\mathfrak{sp}_{2m}), 
\ee 
 together with a    generalization of affine/principal-$W$ character identity in \eqref{gw-eq-principal-general} beyond  the boundary admissible regime. 
  
   Eq.~\eqref{speclevelrank} gives rise to the character identity between two affine algebras whose rank and level are exchanged.  Because of this property, this kind of relation will be referred to as \emph{spectral level-rank duality}. This is  reminiscent of the  standard level-rank duality in WZW and CS theory, which shall be  referred to as \emph{topological level-rank duality}. Detailed derivations of \eqref{speclevelrank} and \eqref{speclevelrank2},  and further generalizations   to other pairs of VOAs and modules will be presented in \cite{Jiang2027}.

 \section{Conclusion}\label{conssec}
 
 In this paper, we establish infinite families of index identities among AD theories and, equivalently, character identities for the associated affine and $W$-algebras. At the VOA level, we construct an explicit graded vector space isomorphism (GVISO) for a concrete pair of VOAs. We also establish the correspondence for universal affine and principal $W$-algebras and test its compatibility with their simple quotients by matching the lowest-degree null states. We extend our discussion to non-vacuum modules and to examples beyond the SCFT/VOA correspondence. In the latter setting, free Majorana fermion realizations provide a direct explanation of identities relating level-one affine and Virasoro algebras. These examples finally lead us to a precise and new  proposal of \emph{spectral level-rank duality}.

Taken together, these results suggest a broader conjectural framework in which distinct VOAs and selected representations are related by graded vector space isomorphisms after suitable regradings. Such correspondences identify graded state spaces without requiring preservation of the OPEs and are therefore weaker than VOA isomorphisms. They would account for identities between appropriately specialized characters of selected modules, while allowing the underlying operator algebras to remain distinct.

Our main constructions concern boundary admissible affine and $W$-algebras associated with four-dimensional AD theories. A central family relates affine algebras to principal $W$-algebras. Non-principal $W$-algebras also appear, but our examples are far from exhaustive. It would be valuable to develop systematic tools for constructing further graded isomorphisms between affine and more general $W$-algebras. From a complementary perspective, one may seek to classify the specializations of flavored affine characters that yield series with nice modular transformation properties. Character identities with $W$-algebras  provide one mechanism for obtaining such specializations, thus naturally ensuring the requisite modular behavior. The new  perspective allows one to investigate the modular structure of affine characters directly, without first identifying a corresponding $W$-algebra. Such an investigation is of independent interest and may also guide the discovery of new correspondences.~\footnote{In particular, non-vacuum module character formulas of affine vertex algebras may diverge in the unflavored limit; therefore, specializing them to other meaningful values is expected to open a new avenue for studying modular properties.}

Another important direction is to extend these graded isomorphisms beyond boundary admissible levels, encompassing non-boundary admissible and even non-admissible algebras and their modules. We will address some of these questions in \cite{Jiang20272} and \cite{Jiang2027}. Such extensions may reveal deeper structural principles underlying the identities studied here. In particular, spectral level-rank duality may provide a framework for exploring further relations with WZW models and Chern--Simons theories, including possible connections to topological quantum field theory and its categorical structures. A central question is whether, and under what additional conditions, these spectral correspondences admit a categorical interpretation.

The high-temperature limits of the index identities constrain combinations of central charges and flavor levels, together with the modular data governing these asymptotics. These relations provide useful criteria for identifying candidate GVISOs and guiding the search for new examples. It would be interesting to determine whether these asymptotic constraints, supplemented by non-vacuum character identities, yield direct relations between the corresponding modular $S$ and $T$ matrices, as well as between the (flavored) MLDEs satisfied by the characters. More broadly, one may ask whether the SCFTs connected by these index identities also exhibit relations in other sectors, in particular their Coulomb branches. Such connections could clarify the physical origin of the index identities and
may help build more fundamental connections   between the underlying SCFTs.

Finally, the Macdonald index     provides another useful and  natural direction to explore, as it incorporates more refined  information beyond  the Schur index. Although a universal intrinsic prescription for recovering the Macdonald index from the VOA alone is not available, a distinguished non-Schur limit does admit an intrinsic VOA interpretation \cite{Jiang:2026ffv}. It would be interesting to establish analogous identities for these indices and to determine whether the corresponding GVISOs can be refined to respect the additional gradings or filtrations involved. Such refinements could reveal further structure in the relations between the underlying state spaces of VOAs.

We hope to report further progress on these questions    in   future work.

\acknowledgments
   This work was supported by the startup grant at SIMIS and the Shanghai Pujiang Program (No. 25PJA128).
We acknowledge the assistance of AI tools with computations and language improvement.
   
\appendix

\section{Bershadsky--Polyakov algebra}\label{BPalg}

The Bershadsky--Polyakov algebra    has  strong generators   given by $T, G^+,G^-,J$, with conformal weights $2,\frac32,\frac32,1$. They are all bosonic and obey  the OPEs  \cite{Beem:2017ooy}
  \beqn
T(z)T(w) &\sim& \frac{\frac{k(14-9k)}{4(2+k)}}{(z-w)^4} + \frac{2T(w)}{(z-w)^2} + \frac{T'(w)}{z-w} ,
\nonumber\\
T(z)G^+(w) &\sim& \frac{\frac{3}{2}G^+(w)}{(z-w)^2} + \frac{{G^+}'(w)}{z-w} ,
\nonumber\\
T(z)G^-(w) &\sim& \frac{\frac{3}{2}G^-(w)}{(z-w)^2} + \frac{{G^-}'(w)}{z-w} ,
\nonumber\\
T(z)J(w) &\sim& \frac{ J(w)}{(z-w)^2} + \frac{{J}'(w)}{z-w} ,
\nonumber\\
J(z)J(w) &\sim& \frac{k/2}{(z-w)^2} ,
\nonumber\\
J(z)G^+(w) &\sim& \frac{G^+(w)}{(z-w)} ,
\nonumber\\
J(z)G^-(w) &\sim& \frac{-G^-(w)}{(z-w)} ,
\nonumber\\
G^+(z)G^-(w) &\sim& \frac{\frac{3}{8}(3k^2-2k)}{(z-w)^3} + \frac{\frac{3}{4}(3k-2)J(w)}{(z-w)^2}
+ \frac{3(JJ)(w) + \frac{3}{8}(3k-2)J'(w) - \frac{3}{4}(k+2)T(w)}{z-w} ,
\nonumber\\
G^\pm(z)G^\pm(w) &\sim& 0.\nonumber
\eeqn

The Bershadsky-Polyakov algebra can be obtained as the    subregular Drinfel\textquoteright{}d-Sokolov reduction of $\fks\fkl_3$, namely
\be
 \text{BP}_{ \tilde  k}=\cW_{\tilde k}(\fks\fkl_3,f_\text{subreg}). 
\ee
 Here $k$ is twice the $\mathfrak u(1)$ level. We assume $k\ne-2$, equivalently $\tilde k\ne-3$, so that the conformal vector is defined. The parameter $k$ and central charge $c$ are related to the parent affine $\mathfrak{sl}_3$ level $\tilde k$ by  
\be
 \qquad k = \frac{2}{3}(2\tilde{k}+3), \qquad
 c =\frac{k(14-9k)}{2(2+k)}=7-6 \tilde k-\frac{24}{\tilde k+3}. 
\ee

 \section{Modular forms and derivatives}\label{MLDEconvention}
 
The Eisenstein series are given by 
  \be\label{Dwmld}
\mathbb{E}_{2k}(q)
=-\frac{B_{2k}}{(2k)!}
+\frac{2}{(2k-1)!}\sum_{n\ge1}\frac{n^{2k-1}q^n}{1-q^n},
\qquad q=e^{2\pi i\tau},
\ee
where $B_{2k}$ are Bernoulli numbers. 
We also introduce
\be
\mathbb{X}(q) 
=12\left[\mathbb{E}_2(q^{1/2})-2\mathbb{E}_2(q)\right]=\theta_2(q)^4+\theta_3(q)^4,
\ee
where   the theta functions are defined by
\be
\theta_2(q)=\sum_{n\in\mathbb Z}q^{(n+1/2)^2/2},\qquad
\theta_3(q)=\sum_{n\in\mathbb Z}q^{n^2/2}.
\ee 

We also have the following identities: 
\beqn\label{e246res}
\mathbb{E}_2(q^{1/2})&=&2\mathbb{E}_2(q)+\frac{\mathbb{X}(q)}{12},\\
\mathbb{E}_4(q^{1/2})&=&\frac{\mathbb{X}(q)^2}{144}-4\mathbb{E}_4(q), \label{e246res1}\\
\mathbb{E}_6(q^{1/2})&=&\frac{11\mathbb{X}(q)^3}{30240}
-\frac27\mathbb{X}(q)\mathbb{E}_4(q).\label{e246res2}
\eeqn

The Serre derivatives are 
\be\label{serdfdsf}
\partial_{(w)}=q\frac{d}{dq}+w\mathbb{E}_2,
\qquad D_q^{(r)}=\partial_{(2r-2)}\circ\cdots\circ\partial_{(2)}\circ\partial_{(0)},
\qquad D_q^{(0)}=1.
\ee

\section{A character  summation formula     }\label{idfsfdsdf}

Our goal in this appendix is to evaluate the following sum:
 \be\label{chiqw}
\chi(q,w)= \sum_{m\geq0}\ \sum_{\substack{|\nu|\leq m\\\nu - m\in 2\bZ }}w^\nu q^{\Delta_{m,\nu}}
\frac{1-q^{m+1}}{(q;q)_\infty^2}  ,
 \ee
 where $\Delta_{m,\nu}$ is given in \eqref{eq:Delta-branch}. 
 
Set
\be
a=(m+ \nu)/2 ,\qquad b=(m-\nu)/2, \qquad
Z=q^{2p},\qquad
u=wq^{p-\frac12},\qquad
v=w^{-1}q^{p-\frac12},
\qquad uv=\frac{Z}{q}.
\ee
The numerator in \eqref{chiqw} becomes
\begin{align}
 (q;q)_\infty^2 \; \chi(q,w)
 =\sum_{a,b\geq0}u^a v^b Z^{ab}(1-q^{a+b+1})
& =\sum_{a\geq0}\frac{u^a}{1-vZ^a}
  -q\sum_{b\geq0}\frac{(qv)^b}{1-quZ^b}
\nonumber\\
&=\sum_{n\in\mathbb Z}\frac{u^n}{1-vZ^n}.
\label{chifsf2}
\end{align}
For the last equality, write $n=-b-1$ in the negative part
of the bilateral sum and use $Z/u=qv$ and $Z/v=qu$:
\be
\frac{u^{-b-1}}{1-vZ^{-b-1}}
=-\frac{Z}{uv}\frac{(Z/u)^b}{1-(Z/v)Z^b}
=-q\frac{(qv)^b}{1-quZ^b}.
\ee
 
Next we   use   Kronecker's identity:
\begin{equation}\label{kronide}
\sum_{n\in\mathbb Z}\frac{u^n}{1-vZ^n}
=
\frac{(Z;Z)_\infty^2(uv;Z)_\infty(Z/(uv);Z)_\infty}
 {(u;Z)_\infty(Z/u;Z)_\infty
  (v;Z)_\infty(Z/v;Z)_\infty},
\end{equation}
which follows by setting $(A,B,\xi)=(v,Zv,u)$ in Ramanujan's ${}_1\psi_1$ summation
\cite{warnaar2012ramanujans1psi1summation}:
\begin{equation}
\sum_{n\in\mathbb Z}\frac{(A;Z)_n}{(B;Z)_n}\,\xi^n
=
\frac{
 (Z;Z)_\infty(B/A;Z)_\infty
 (A\xi;Z)_\infty(Z/(A\xi);Z)_\infty
}{
 (B;Z)_\infty(Z/A;Z)_\infty
 (\xi;Z)_\infty(B/(A\xi);Z)_\infty
}.
\end{equation}

Applying the identity \eqref{kronide} to  \eqref{chifsf2} gives  
\be
(q;q)_\infty^2 \, \chi(q,w) = \frac{(Z;Z)_\infty^2 \, (uv;Z)_\infty \, (Z/(uv);Z)_\infty}{(u;Z)_\infty \, (Z/u;Z)_\infty \, (v;Z)_\infty \, (Z/v;Z)_\infty}.
\ee

As a result, we obtain
\begin{equation} \label{chifsdfsf}
\chi (q,w)=
\frac{(q^{2p};q^{2p})_\infty^2
      (q;q^{2p})_\infty(q^{2p-1};q^{2p})_\infty}
 {(q;q)_\infty^2
  \displaystyle\prod_{\epsilon=\pm1}
    (w^\epsilon q^{p-\frac12};q^{2p})_\infty
    (w^\epsilon q^{p+\frac12};q^{2p})_\infty}.
\end{equation}
Using
\be
\log(t;Z)_\infty
=-\sum_{r\geq1}\frac{t^r}{r(1-Z^r)},
\ee
we can write \eqref{chifsdfsf} as 
\begin{equation}  
\chi (q,w)
=
\operatorname{PE}\!\left[
\frac{
 q+q^2-q^{2p-1}-q^{2p}
 +(w+w^{-1})(q^{p-\frac12}-q^{p+\frac32})
}{
 (1-q)(1-q^{2p})
}
\right].
\end{equation}

\bibliographystyle{JHEP} 
 \bibliography{ref}

\end{document}